\documentclass[10pt,aps,twocolumn,showpacs,preprintnumbers,pra,superscriptaddress,hidelinks]{revtex4-1}

\usepackage{graphicx}
\usepackage[normalem]{ulem}
\usepackage{booktabs}
\usepackage{soul}
\usepackage{color}
\usepackage{amssymb}
\usepackage{amsmath}
\usepackage{mathrsfs}
\usepackage{empheq}
\usepackage{tikz}
\usepackage{bm}
\usepackage{mathtools}
\usepackage{microtype}

\newcommand{\ssection}{Section }
\newcommand{\ssections}{Sections }
\newcommand{\sectionp}{Sec. }
\newcommand{\eph}{\textit{e-ph} }
\newcommand{\ephh}{\textit{e-ph}}
\newcommand{\gww}{\textit{GW }}
\newcommand{\gw}{\textit{GW}}
\newcommand{\xc}{\textit{xc} }
\newcommand{\xcc}{\textit{xc}}
\newcommand{\ibcs}{IBCs}
\newcommand{\ibccs}{IBCs }

\newcommand{\ibcc}{IBC }
\newcommand{\abinitio}{\textit{ab-initio} }
\newcommand{\ifc}{IFCs }

\newcommand{\GG}{\mathbf{\Gamma}}
\newcommand{\suppinfo}{Supplemental Material}
\newcommand{\A}{\text{A}}
\newcommand{\BZ}{\text{BZ}}

\newcommand{\rot}{\mathbf{S}}
\newcommand{\invrot}{\mathbf{S}^{-1}}

\newcommand{\R}{\mathbf{r}}
\newcommand{\X}{\mathbf{x}}

\newcommand{\RR}{\mathbf{R}}
\newcommand{\Q}{\mathbf{q}}
\newcommand{\K}{\mathbf{k}}
\newcommand{\G}{\mathbf{G}}

\newcommand{\TT}{\mathcal{T}}
\newcommand{\KK}{\mathbf{K}}

\newcommand{\abs}[1]{\lvert{#1}\rvert}
\newcommand{\ub}[1]{\underline{\bm{#1}}}
\newcommand{\uub}[1]{\underline{\underline{\bm{#1}}}}

\newcommand{\elmat}[3]{\bra{#1} #2 \ket{#3}}

\newcommand{\be}{\begin{equation}}
\newcommand{\ee}{\end{equation}}
\newcommand{\bea}{\begin{eqnarray}}
\newcommand{\eea}{\end{eqnarray}}
\newcommand{\bean}{\begin{eqnarray*}}
\newcommand{\eean}{\end{eqnarray*}}

\def\bra#1{\mathinner{\langle{#1}|}}
\def\ket#1{\mathinner{|{#1}\rangle}}
\def\braket#1{\mathinner{\langle{#1}\rangle}}

\begin{document}

\date{\today}

\title{Electron-Phonon Coupling in
Many-Body Perturbation Theory: Developments within the
Quasiparticle Self-Consistent GW approximation and LMTO Formalism}

\author{Savio Laricchia}
\affiliation{Centro S3, CNR-Istituto Nanoscienze, 41125 Modena, Italy}
\altaffiliation{Current affiliation: Istituto di Struttura della Materia-CNR (ISM-CNR), Area della Ricerca di Roma 1, Monterotondo Scalo, Italy}
\email[corresponding author: ]{savio.laricchia@mlib.ism.cnr.it}
\affiliation{Department of Physics, King's College London, Strand,
  London WC2R 2LS, United Kingdom}
\author{Casey Eichstaedt}
\affiliation{National Renewable Energy Laboratory, Golden, CO 80401,USA}
\author{Dimitar Pashov}
\affiliation{Department of Physics, King's College London, Strand,
  London WC2R 2LS, United Kingdom}
\author{Mark van Schilfgaarde}
\affiliation{National Renewable Energy Laboratory, Golden, CO 80401,USA}

\begin{abstract}

The calculation of electron-phonon (\ephh) coupling from first principles
is a topic of great interest in materials science, offering a robust,
non-empirical framework to understand and predict a wide range of physical phenomena.
While significant progress has been made using the Kohn-Sham framework
of density functional theory (KS-DFT), it is increasingly evident that
standard approximations in KS-DFT often fall short of providing accurate
results. These shortcomings are frequently linked to the non-local nature
of the exchange-correlation potential, prompting the development of advanced
methodologies within DFT and many-body perturbation theory.
Despite these efforts, a highly reliable and efficient first-principles
approach to accurately capture \eph interactions remains elusive.
To address this challenge, we introduce a novel field-theoretical methodology
that integrates the foundational work of Baym and Hedin with the
Quasiparticle Self-Consistent \gw (QS\gw) approximation, implemented
within the Questaal electronic structure suite. Our approach, based
on a response function framework, ensures that Pulay-like
incomplete-basis-set corrections
are not required to account for changes in basis functions, paving
the way for a high-fidelity description of \eph coupling.
\end{abstract}

\maketitle

\section{Introduction}
\label{sec:intro}
The interaction between electrons and a vibrating lattice (phonons) gives
rise to a diverse range of material properties. For instance, it determines
the temperature dependence of electrical transport coefficients in metals
and semiconductors~\cite{pizzi14}, enables optical transitions in
indirect-gap semiconductors~\cite{louie-elph-2012}, renormalizes the
effective mass of charge carriers, and plays a central role in the
thermalization of hot carriers. This thermalization process critically
affects the performance of electronic, optoelectronic, photovoltaic,
and plasmonic devices~\cite{sjakste07}. Furthermore,
electron-phonon (\ephh) coupling governs the lifetimes of electron spins
in perfect crystals or at defect sites~\cite{el-ph-lifetime}, which is
essential for spintronics and quantum information technologies.
In conventional superconductors, \eph coupling drives the formation of Cooper
pairs in the superconducting condensate. However, its role in unconventional
superconductors, such as copper oxides, remains
a subject of ongoing debate~\cite{1d-doped}.

Owing to its fundamental importance in so many phenomena,
accurately and reliably
determining the \eph interaction is a major focus in physics, chemistry,
electrical engineering, materials science, and mechanical engineering.
The first \abinitio framework for modeling \eph coupling, based on the
Kohn-Sham formulation of density functional theory (KS-DFT), was developed
in the early 1990s~\cite{baroni87,gonze92,savrasov92}.
This approach remains the most widely used method for studying \eph
interactions and lattice-related properties~\cite{froz-phon,DFPT},
typically employing the local density approximation (LDA) or the
generalized gradient approximation (GGA) for the
exchange-correlation (\xcc) potential.
These methods rely on density-functional perturbation theory (DFPT), a
perturbative extension of KS-DFT. DFPT is an efficient tool for calculating
phonon modes at any wave vector in the Brillouin zone (BZ) and the \eph
matrix elements coupling
two KS electronic states. While DFPT is
broadly applicable, it often underestimates \eph coupling
strength~\cite{shang2023}. This discrepancy arises partly because KS
states derive from a fictitious auxiliary Hamiltonian,
limiting their direct comparability to experimental results.
This limitation is particularly significant in the
calculation of \eph coupling matrix elements, where
non-local correlations and many-body effects are critical
and highly sensitive to the accuracy of the
computed excitation energies.
Gr\"uning et al. have shown that, at least for some systems,
the primary source of error does not stem from
approximations in the \xc energy $E_{xc}$,
but rather from the fictitious Kohn-Sham
eigenfunctions $\psi_i$ and their corresponding Lagrange
multipliers $\varepsilon_i$~\cite{Gruning06}.
These Lagrange multipliers are almost always interpreted as
excitation energies. For example, KS-DFT is well-known for
underestimating band gaps.
Similar underlying factors likely
contribute to inaccuracies in DFPT calculations of \eph coupling strength.

KS-DFT often fails to accurately describe the \eph coupling,
even in simple \textit{sp}-bonded compounds such as graphene and diamond.
In graphene and graphite, the non-local, long-range nature of the
Coulomb interaction enhances the coupling of electrons
to the intervalley $A'_1$ optical phonon.
This enhancement is particularly noticeable when incorporating leading
logarithmic corrections via the Renormalization Group
approach~\cite{basko08}, which are not captured
by LDA and GGA density functional approximations.
Lazzeri et al.~\cite{lazzeri08} demonstrated that a non-local
$G_0W_0$ self-energy approach applied within a
frozen-phonon scheme---which
implicitly includes \eph vertex corrections---results
in a ${\sim}40\%$ increase in
the intraband \eph matrix element and improved phonon dispersions at
the $\KK$ point compared to KS-DFT methods.
For diamond, Antonius et al.~\cite{antonius2014}
reported a ${\sim}40\%$ increase in the renormalized band gap,
attributed to the zero-point motion of the lattice, by computing
\eph self-energy
through varations
in the $G_0W_0$ band structure within a frozen-phonon framework.

Li et al.~\cite{GWPT} developed a first-principles
linear-response method, the \gw perturbation theory ({\it GW}PT),
which models the \eph coupling as the
interaction of a true quasiparticle with phonons within the
\gw approximation. This approach accounts for non-local
many-electron correlation and dynamical self-energy effects,
going beyond the limitations of traditional DFPT by
replacing the local \xc contribution
with the first-order variation of the \gw self-energy
induced by a phonon perturbation.
This methodology extends beyond DFPT while avoiding
limitations of the frozen-phonon technique.
\textit{GW}PT provides a significant enhancement of \eph interactions
for states near the Fermi surface. This was notably demonstrated for the
bismuthate superconductor Ba$_{1-x}$K$_x$BiO$_3$, where \textit{GW}PT
explains the material's high superconductivity transition
temperature of 32 K~\cite{GWPT}, with a threefold enhancement compared to DFPT.
Another key application of this method is to understand
the kink observed around 70 meV in the
energy-momentum dispersion of high-$T_c$ cuprates,
such as single-copper-oxygen-layer cuprate
La$_{2-x}$Sr$_x$CuO$_4$ (LSCO) \cite{valla99,lanzara01}.
Previous DFPT calculations have shown a threefold underestimate
 in the magnitude of the kink observed in angle-resolved photoemission
spectroscopy (ARPES)~\cite{giustino08,heid08}.
By contrast, including the \gww band structure
and non-local self-energy effects in the evaluation of the \eph matrix
elements through \textit{GW}PT
significantly enhances the phonon-induced component of the self-energy
by a factor of 2-3~\cite{li2021}.

KS-DFT does not directly address the single-particle excitation spectrum
of materials, often leading to
coupling between excited quasiparticles and other elementary excitations
being significantly underestimated
in certain materials~\cite{giustino17}. Recent advancements in
modeling \eph coupling highlight the potential of \gw and related
techniques to accurately capture \eph scattering effects in materials
where non-local correlations play a critical role and are inadequately
addressed by KS-DFT. Nonetheless, a systematic, reliable, and efficient
first-principles approach to model \eph coupling remains an open challenge.

In this paper, we introduce a novel field-theoretical methodology designed
for predicting electronic quasiparticles and their interactions with lattice
vibrations, using Green’s functions as the foundational framework.
\ssection \ref{sec:theory} outlines the formal theory underlying our approach.
In \sectionp \ref{sec:phonons}, we provide an overview of the vibrating crystal
within the Born-Oppenheimer and harmonic approximations.
\ssection \ref{sec:field_theoretic} summarizes the Green’s function
treatment of the coupled \eph system, following the framework and
general notation of Ref. \onlinecite{giustino17}.
The impact of Pulay-like incomplete-basis-set corrections in the
evaluation of phonon dispersions and \eph coupling within a
field-theoretic formalism is addressed in \sectionp \ref{sec:pulay}.
\ssection \ref{sec:qsgw} details the Quasiparticle Self-Consistent \gw
(QS$GW$) approximation as implemented in
the \texttt{Questaal} electronic structure suite.

\ssection \ref{sec:implementation} discusses the formal implementation of
this methodology using \texttt{Questaal}’s optimized linearized muffin-tin
orbital (LMTO) basis functions combined with the mixed product basis.
This approach enables the efficient computation of many-body quantities,
including excitonic effects. \ssections \ref{sec:eph_mpb}-\ref{sec:dynmat}
delve into the specifics of computing \eph matrix elements within the
\texttt{Questaal} framework,
with additional technical details provided in the \suppinfo.

Finally, in \sectionp \ref{sec:results}, we demonstrate the capability of
our field-theoretic approach by presenting results that show excellent
agreement for graphene with the experimentally derived Fermi
surface-averaged \eph matrix elements at $\Q=\bm{\Gamma},\KK$.
These results are extracted from the slope of the Kohn
anomaly for the highest optical phonon mode at $\Q=\bm{\Gamma}$, and
under the assumption of negligible coupling between electrons and
multiple phonons at $\Q=\KK$.

\section{Theory}
\label{sec:theory}

In this section, we outline the fundamental formalism used to compute
\eph interactions within a field-theoretic framework. Starting
with the Hamiltonian of a vibrating crystal, we adopt the
Born-Oppenheimer (BO) approximation in the adiabatic regime.
This Hamiltonian is derived from potential energy surfaces obtained via
the electronic averaging over quasiparticle states~\cite{gross1991many}.

We then present a concise derivation and summary of the established
field-theoretic methodology for calculating \eph matrix elements, following
the conventions detailed in Refs. \onlinecite{hedin65,baym61} and further
refined in the modern review presented in Ref. \onlinecite{giustino17}.
This approach treats electrons and phonons as a unified system
rather than distinct subsystems, enabling a comprehensive theoretical
treatment of their interactions, as described in
\sectionp \ref{sec:field_theoretic}.

It is worth noting that a novel framework for modeling coupled \eph
systems was recently introduced in Ref. \onlinecite{stefanucci23}.
While this framework shows promise, its application to extended systems
poses conceptual challenges that are beyond the scope of this work.
A thorough examination of this alternative method is deferred to future studies.
\subsection{Lattice vibration in crystals}
\label{sec:phonons}

Lattice dynamics in crystals have been extensively explored by various
authors~\cite{born_huang54,ziman60,kittel63,kittel76,ashcroft_mermin76,
maradudin1971theory}. For the purposes of this paper, we introduce specific
notations and summarize essential concepts to facilitate subsequent
discussions. Consider a system consisting of $N$ nuclei within each
unit cell, where each nucleus has mass $m_r$, proton number $Z_r$,
and an equilibrium position $\bm{\tau}^0_r$, free of external forces.
The out-of-equilibrium nuclear position of the $r$-th atom in the unit
cell is denoted by the vector $\bm{\tau}_r$, with its Cartesian components
represented as $\tau_{r\alpha}$.

The unit cells span a Born-von Kármán (BvK) macrocrystal comprising
$N_{\text{BvK}}$ units under periodic boundary conditions. Each unit cell
is identified by a direct lattice vector $\textbf{R}_l$ (where
$l=1,2,...,N_{\text{BvK}}$) and has a volume $\Omega_0$. The
position of the $r$-th nucleus in the $l$-th unit cell is given
by $\bm{\tau}_{rl}=\bm{\tau}_{r}+\textbf{R}_l$. The total number of
unit cells in the BvK macrocrystal equals the number of Bloch wave
vectors $\K$ on a uniform grid in the reducible BZ, such that
$N_\K = N_{\text{BvK}}$. The volume of the BvK macrocrystal is thus
defined as $\Omega = N_\K\Omega_0$.

Determining the lattice vibrations involves calculating
the total potential energy $U(\{\bm{\tau}_{rl}\})$
arising from systems of interacting electrons
(in their ground state) and nuclei, with the latter being
treated as \textit{classical particles} located at
$\bm{\tau}_{rl}$. Here, $\{\bm{\tau}_{rl}\}$ represents
the set of all nuclear positions. For small displacements
$\Delta \bm{\tau}_{r l}$ from equilibrium, the harmonic approximation expands
the potential energy to second order
\be\label{eq:full_potential_energy}
U(\{\bm{\tau}_{rl}\}) \approx U_0 +\tfrac{1}{2}
\sum_{r\alpha l}\sum_{s\beta l'}
C_{rs}^{\alpha\beta}(\RR_l,\RR_{l'})   \Delta \tau_{rl\alpha}
   \Delta \tau_{sl'\beta} ,
\ee
where $U_0$ is the potential energy at equilibrium, and
$\{C_{rs}^{\alpha\beta}(\RR_l,\RR_{l'})\}$ are the interatomic
force constants (IFCs), defined as
\be\label{eq:force_constants_classical_def}
C_{rs}^{\alpha\beta}(\RR_l,\RR_{l'}) \equiv
\left.\frac{\partial^2 U(\{\bm{\tau}_{rl}\})}
{\partial\tau_{rl\alpha }\partial\tau_{sl'\beta }}\right|_{\{\bm{\tau}^0_{rl}\}}.
\ee
Here, Greek letters $\alpha$ and $\beta$ label
the Cartesian components, the Latin letters
$r$, $s$ denote different atoms in the unit cell,
and the derivatives are evaluated at the equilibrium lattice configuration.
The translational symmetry implies that the \ifc  depend only on the
difference $\RR_l-\RR_{l'}$ i.e., $C_{rs}^{\alpha\beta}(\RR_l,\RR_{l'}) =
C_{rs}^{\alpha\beta}(\RR_l-\RR_{l'},\textbf{0})$.
The nuclear displacements $\Delta \bm{\tau}_{r l}$
are measured from their equilibrium positions $\bm{\tau}^0_{r l}$, such that
\bea\label{eq:vector_position}
\bm{\tau}_{r l} &=& \bm{\tau}_{r l}^0 +
\Delta \bm{\tau}_{r l}\nonumber \\
&=& \bm{\tau}_{r}^0 + \RR_l +
\Delta \bm{\tau}_{r l} \quad \quad l=1,\dots,N_{\K} \, .
\eea
Within the Born-Oppenheimer approximation, the quantum mechanical treatment
of lattice vibrations leads to the Hamiltonian of the harmonic
vibrating crystal
\be\label{eq:nuclear_hamiltonian}
\hat{H}_n = \hat{T}_n + \tfrac{1}{2} \sum_{r\alpha l}\sum_{s\beta l'}
C_{rs}^{\alpha\beta}(\RR_l,\RR_{l'}) \Delta \hat{\tau}_{rl\alpha}
     \Delta \hat{\tau}_{sl'\beta}
\ee
where the nuclear kinetic energy operator is
\be\label{eq:tn_operator}
\hat{T}_n = \sum_{rl\alpha} \frac{\hat{p}_{rl\alpha }
     \hat{p}_{r l\alpha}}{2 m_r} .
\ee
Displacement operators $\Delta \hat{\mathbf{\tau}}_{r l\alpha}$ and
momentum operators $\hat{p}_{rl \alpha}$ can be expanded in a basis
of phonon modes
\be\label{eq:displacement}
\Delta \hat{\mathbf{\tau}}_{r l\alpha } =
\frac{1}{\sqrt{N_{\K}}}\sum_{\Q \nu}
\sqrt{\frac{\hbar}{2 m_r \omega_{\Q\nu}}}
e^{i\Q\cdot\RR_l} e_{r\alpha,\nu}(\Q) \bigl(\hat{a}_{\Q\nu} +
\hat{a}^\dagger_{-\Q\nu}\bigr) \,
\ee
and
\be\label{eq:nucl-mom}
\hat{p}_{r l \alpha} = \frac{i}{\sqrt{N_{\K}}}\sum_{\Q \nu}
\sqrt{\frac{\hbar m_r \omega_{\Q\nu}}{2 }}
e^{-i\Q\cdot\RR_l} e^*_{r\alpha,\nu}(\Q) \bigl(\hat{a}_{\Q\nu} -
\hat{a}^\dagger_{-\Q\nu}\bigr) \, .
\ee
In the theory of vibrating crystals, the displacement
vector $\Delta \bm{\tau}_{rl}$
corresponds to the expectation value of the nuclear displacement operator
$\Delta\hat{\bm{\tau}}_{rl}$ of Eq. (\ref{eq:displacement}).
The orthonormal polarization vectors $\bm{e}_{r,\nu}(\Q)$
and associated phonon frequencies $\omega_{\Q\nu}$,
for a given $\nu$-th phonon mode and phonon wave vector $\Q$,
are respectively the eigenvectors and the square root of the eigenvalues
of the Hermitian dynamical matrix $D_{rs}^{\alpha\beta}(\Q)$, defined as Bloch
transform of the \ifc
\be\label{eq:dynmat}
D_{rs}^{\alpha\beta}(\Q) = \frac{1}{\sqrt{m_rm_s}}
\sum_p C_{rs}^{\alpha\beta}(0,\RR_p) e^{i\Q\cdot\RR_p} \, .
\ee
The bosonic operators $\hat{a}^\dagger_{\Q\nu}$ and
$\hat{a}_{\Q\nu}$ are respectively the creation and destruction operators for
a phonon in a state $\textbf{e}_{r,\nu}(\Q)$ and energy $\hbar\omega_{\Q\nu}$.
These bosonic operators are introduced to conveniently describe the classical
nuclear dynamics in terms of quanta of lattice vibration (phonons) and
obey the canonical commutation relations
\bea\label{eq:commutation_a_operator}
&& [\hat{a}_{\Q\nu},\hat{a}^\dagger_{\Q'\nu'}] =\delta_{\nu\nu'}\delta_{\Q,\Q'}
\nonumber \\
&& [\hat{a}_{\Q\nu},\hat{a}_{\Q'\nu'}] =
[\hat{a}^\dagger_{\Q\nu},\hat{a}^\dagger_{\Q'\nu'}] = 0 \, .
\eea
By combining
Eqs. (\ref{eq:nuclear_hamiltonian})-(\ref{eq:commutation_a_operator}),
the BO nuclear Hamiltonian can be written in its spectral representation
of phonons in terms of $3 N N_{\K}$ independent harmonic oscillators
\be\label{eq:harmonic_oscillator_ham}
\hat{H}_n = \sum_{\Q\nu} \hbar\omega_{\Q\nu}(\hat{a}^\dagger_{\Q\nu}
\hat{a}_{\Q\nu} + \tfrac{1}{2}) \, .
\ee
The ground state nuclear eigenfunction is the product of Gaussians and
all additional states can be generated by applying the operators
$\hat{a}^\dagger_{\Q\nu}$ to it.

In the long-wavelength limit $\abs{\Q}=0$,
solving the eigenvalue problem for the
dynamical matrix in Eq. (\ref{eq:dynmat}) yields three acoustic normal modes,
which correspond to the rigid translation of the entire crystal.
These modes have frequencies $\omega_{\mathbf{0}\nu}=0$.
However, for these acoustic modes, the expectation value of the nuclear
displacement operator in Eq. (\ref{eq:displacement}) becomes ill-defined
in the long-wavelength limit, potentially posing challenges in
modeling \eph coupling.

Despite this, the \textit{acoustic sum rules}
ensure that the \eph matrix elements
associated with acoustic phonon modes in the long-wavelength limit are
identically zero for degenerate electronic states~\cite{suppinfo}.
Consequently, these modes can be excluded
from numerical solutions of expressions involving
summations over all phonon modes and the entire BZ.
In contrast, for non-degenerate electronic states, the \eph matrix elements
are not guaranteed to vanish and may even
diverge under these conditions~\cite{suppinfo}.
Nonetheless, this divergence does not present significant issues in practical
applications. This is because the long-wavelength contributions of acoustic
phonons to the Fan-Migdal and Debye-Waller \eph
self-energies cancel each other out~\cite{giustino17}, effectively mitigating
any problematic behavior. For further details, the reader can refer to
\sectionp S.6 of the \suppinfo.

We conclude this section by briefly discussing the primary approaches
available for computing the \ifc.
The most widely adopted first-principles method is based on
DFPT~\cite{baroni87,gonze92,savrasov92}, which involves solving a
Sternheimer equation~\cite{sternheimer54} for the first-order
lattice-periodic variation of the local mean-field KS potential, i.e.,
\be\label{eq:delta_vKS_dfpt}
\partial_{r\alpha,\Q} v^{\text{KS}} = \sum_p e^{-i\Q\cdot(\R-\RR_p)}
\left. \frac{\partial v^{\text{KS}}(\R-\RR_p)}
     {\partial \tau_{rp\alpha}}\right|_{\bm{\tau}^0_{rp}} \, ,
\ee
to evaluate Eq. (\ref{eq:force_constants_classical_def}). DFPT circumvents the
primary limitation of the frozen-phonon algorithms, where the supercell
size can become impractically large for computing the dynamical matrix,
Eq. (\ref{eq:dynmat}).

In addition to DFPT, an alternative method for computing the screened
perturbation $\partial_{r\alpha,\Q} v^{\text{KS}}$ is the
\textit{dielectric approach}~\cite{pick_cohen_martin70,quong_klein92}.
Although less widely used, this method provides a valuable connection between
DFT calculations of phonon dispersions and \eph matrix elements within the
field-theoretic formulation presented in \sectionp \ref{sec:field_theoretic}.
Within the dielectric approach, the first-order lattice-periodic variation
of the electron density, $\partial_{r\alpha,\Q} n_e$, is related to
$\partial_{r\alpha,\Q} v^{\text{KS}}$ by:
\be
\partial_{r\alpha,\Q} n_e(\R) = \int_{\Omega_0} d\R' \chi^\Q_{\text{KS}}(\R,\R')
\partial_{r\alpha,\Q} v^{\text{KS}}(\R') \, ,
\ee
where $\chi^\Q_{\text{KS}}$ is the lattice-periodic independent-particle
electron polarizability, projected onto the phonon
wave vector $\Q$ in the BZ. This dielectric formalism complements
the field-theoretic approach detailed in \sectionp \ref{sec:Wph}.

Both the dielectric and field-theoretic methods converge to the same
expression for the IFCs under the harmonic and adiabatic approximations.
The IFCs, $C^{\alpha\beta}_{rs}(\RR_l,\RR_{l'})$, can be derived as:
\begin{widetext}
\be\label{eq:ifc_manybody}
C^{\alpha\beta}_{rs}(\RR_l,\RR_{l'}) = \sum_{tp}\biggl( \delta_{lp}\delta_{rt} -
\delta_{ll'}\delta_{rs}\biggr) \int_\Omega\int_\Omega\int_\Omega
\frac{\partial V^{(0)}_{sl'}(\R)}{\partial r_\beta}  v^{-1}(\R-\R_1)
\varepsilon^{-1}_{e,\text{TDDFT}}(\R_1,\R_2)
\frac{\partial V^{(0)}_{tp}(\R_2)}{\partial r_{2\alpha}}
\,d\R\,d\R_1\,d\R_2 \, .
\ee
\end{widetext}
Here, $\varepsilon^{-1}_{e,\text{TDDFT}}(\R_1,\R_2)$ represents the static
limit of the inverse dielectric function for the electrons, defined through
the density response function of time-dependent density functional theory
(TDDFT). In symbolic notation, the electron dielectric function can
be expressed as
\be\label{eq:eps_tddft}
\varepsilon_{e,\text{TDDFT}} = 1-(v + f_{xc})\chi_{\text{KS}},
\ee
where $\chi_{\text{KS}}$ denotes the KS electron density response function to
perturbations arising from both the classical electrostatic Hartree
kernel $v$ and quantum effects captured by the \xc kernel $f_{xc}$.
The \xc kernel is defined as the second functional derivative of the
\xc energy functional within the KS-DFT framework~\cite{petersilka96}.
Neglecting the \xc kernel reduces Eq. (\ref{eq:eps_tddft}) to a
\textit{test charge} static dielectric function within the
random-phase approximation (RPA) of TDDFT.

In Eq. (\ref{eq:ifc_manybody}), $v^{-1}(\R_1-\R_2)$ acts as the
Green’s function for the long-range Coulomb interaction
$v(\R_1-\R_2)=e^2\abs{\R_1-\R_2}^{-1}$, satisfying the identity
\be\label{eq:v_vinv}
\int_\Omega \, d\R_1 \,v^{-1}(\R-\R_1) v(\R_1-\R_2) = \delta(\R-\R_2) \, .
\ee
The term $V^{(0)}_{rl}(\R) = -Z_r e^2\abs{\R-\bm{\tau}^0_{rl}}^{-1}$ represents
the bare nuclear potential for the $r$-th atom in the $l$-th unit cell at
its equilibrium position $\bm{\tau}_r^0$, with nuclear charge $Z_re$.

The second term in Eq. (\ref{eq:ifc_manybody}) ensures the satisfaction of
the acoustic sum rule, which maintains translational invariance of the
IFCs and conserves total momentum:
\be\label{eq:acoustic_sum_rule}
\sum_{sl'} C^{\alpha\beta}_{rs}(\RR_l,\RR_{l'}) = 0 \, .
\ee
This constraint arises from the invariance of the lattice potential under
a global translation of all atoms, ensuring zero net forces when the
atoms are in equilibrium.

\subsection{Field theoretic approach to the electron-phonon interaction}
\label{sec:field_theoretic}

In this section, we provide a concise overview of the field-theoretic framework
for \eph interaction, which is
widely regarded as the most comprehensive theory of
the \eph problem to date. In this framework, electrons are treated as quantum
particles, while ions are treated as classical particles. Initially
introduced by Baym~\cite{baym61} and further developed by Hedin and
Lundqvist~\cite{hedin_lundqvist69}, this approach offers a fully general and
robust formulation of the \eph interaction problem.

By employing the field-theoretic approach, it is possible to address the
limitations inherent in the approximations within the KS-DFT framework.
Unlike KS-DFT, which relies on the assumption of an effective mean-field KS
potential and is sensitive to the choice of the \xc functional, the
field-theoretic method provides a more physically grounded framework.
This reduces the sensitivity of the \eph matrix elements to the specific
\xc functional employed in KS-DFT calculations. Moreover, KS-DFT relies
on the BO approximation, which can be insufficiently accurate for
certain systems, particularly metals and narrow-gap semiconductors.
Addressing these limitations requires incorporating retardation effects
into the evaluation of \eph scattering.

Additionally, the \eph interaction induces renormalizations in both the
electronic structure and the lattice dynamics of solids,
thereby coupling electrons and phonons.
Thus, a comprehensive understanding of the \eph
interaction necessitates a self-consistent treatment, which is achievable
within the rigorous and general field-theoretic framework for interacting
electrons and phonons in solids.

The starting point for studying the \eph interaction within a field-theoretic
context is to define the Fock space and the relevant operators for electrons
and nuclei. The representation of many-body electronic states using Slater
determinants is straightforward, as electrons are indistinguishable particles.
Their behavior is conveniently described
using second-quantized electronic field operators $\hat{\psi}$.
Conversely, nuclei are distinguishable entities, and their
dynamics are described using first-quantized operators for nuclear
momenta ($\hat{\textbf{p}}$) and nuclear displacements from equilibrium
($\Delta\bm{\hat{\tau}}$), as introduced in Eqs. (\ref{eq:nucl-mom}) and
(\ref{eq:displacement}), respectively.
Phonons, which result from quantizing nuclear
displacements, are treated as indistinguishable particles.
For this discussion, we focus on equilibrium Green’s functions at zero
temperature. As a result, all expectation values
are evaluated for the electron-nuclei ground state $|0\rangle$, such that
$\langle\ldots\rangle\equiv \langle 0|\ldots |0\rangle$.

In second quantization, the electronic field operators $\hat{\psi}(\X)$ and
$\hat{\psi}^\dagger(\X)$ respectively create or annihilate an electron at
$\X=\{\R,\sigma\}$, where $\R$ is the spatial position and
$\sigma$ is the spin projection. These operators obey the anticommutation
relations~\cite{merzbacher98}
\bea\label{eq:anticommutation_psi_operator}
&&\{\hat{\psi}(\X),\hat{\psi}(\X')\} =
\{\hat{\psi}^\dagger(\X),\hat{\psi}^\dagger(\X')\} = 0 \nonumber \\
&&\{\hat{\psi}(\X),\hat{\psi}^\dagger(\X')\} = \delta(\X-\X') \, .
\eea
The general non-relativistic Hamiltonian for an unperturbed system of coupled
electrons and nuclei is
\be\label{eq:many_body_H}
\hat{H}_0 = \hat{T}_e + \hat{T}_n + \hat{U}_{ee} + \hat{U}_{en} +
\hat{U}_{nn},
\ee
where $\hat{T}_n$ is the nuclear kinetic energy operator defined in
Eq. (\ref{eq:tn_operator}), and $\hat{T}_e$ is the electronic kinetic
energy operator
\be
\hat{T}_e = -\frac{\hbar^2}{2 m_e} \int d\X\, \hat{\psi}^\dagger(\X) \nabla^2
\hat{\psi}(\X) \, ,
\ee
with $m_e$ the electron mass and where the integral
$\int d\X\equiv \sum_\sigma \int_\Omega d^3 \R$ denotes
the sum over spin states and the integration over the BvK macrocrystal.
The electron-electron interaction term is
\be\label{eq:ee-int}
\hat{U}_{ee} = \tfrac{1}{2} \int_\Omega d\R \int_\Omega d\R'\, \hat{n}_e(\R)
[\hat{n}_e(\R') - \delta(\R-\R')] v(\R-\R') \, ,
\ee
where
the Dirac delta function is used here to remove the
unphysical self-interaction, and the electron density operator is defined as
\be\label{eq:ne_operator}
\hat{n}_e(\R) = \sum_\sigma \hat{\psi}^\dagger(\R\sigma)\hat{\psi}(\R\sigma) \,.
\ee
The electron-nuclear interaction is
\be\label{eq:Uen}
\hat{U}_{en} = \int_\Omega  d\R \int_\Omega d\R' \hat{n}_e(\R) \hat{n}_n(\R')
v(\R-\R')
\ee
with the nuclear density operator given by
\be\label{eq:nuclear_density_operator}
\hat{n}_n(\R) = -\sum_{rl} Z_r  \delta(\R-\bm{\tau}_{rl}^0 -
\Delta\hat{\bm{\tau}}_{rl}) \, .
\ee
The nuclear-nuclear interaction is instead expressed as
\be\label{eq:Unn}
\hat{U}_{nn} = \tfrac{1}{2} \sum_{rl\ne sl'} Z_rZ_s
v(\bm{\tau}^0_{rl} + \Delta\hat{\bm{\tau}}_{rl},
\bm{\tau}^0_{sl'} + \Delta\hat{\bm{\tau}}_{sl'}).
\ee
By combining Eqs. (\ref{eq:many_body_H})-(\ref{eq:Unn}),
the Hamiltonian becomes
\begin{align}\label{eq:many_body_H_2}
&\hat{H}_0 = \hat{T}_n + \hat{U}_{nn} + \int d\X\,\hat{\psi}^\dagger(\X)\biggl[
  -\frac{\hbar^2}{2 m_e} \nabla^2  + \hat{V}_n(\R) \biggr]
\hat{\psi}(\X) + \nonumber \\
& \,\,\,\,\,\,\,+\tfrac{1}{2}\int\int d\X d\X'\, \hat{\psi}^\dagger(\X)
\hat{\psi}^\dagger(\X') v(\R-\R') \hat{\psi}(\X') \hat{\psi}(\X) \, ,
\end{align}
where the nuclear potential operator $\hat{V}_n(\R)$ is defined as
\begin{align}\label{eq:nuclear_potential}
\hat{V}_n(\R) &= \int_\Omega d\R' v(\R-\R') \hat{n}_n(\R') \nonumber \\
&= \sum_{rl} \hat{V}_{rl}(\R)
\end{align}
with $\hat{V}_{rl}(\R)$ the out-of-equilibrium bare electron-nuclear potential
$-Z_re^2\abs{\R-\bm{\tau}^{0}_{rl}-  \Delta\hat{\bm{\tau}}_{rl}}^{-1}$ or
its ionic pseudopotential.
A detailed analysis of how the core electrons screen the bare potential
$\hat{V}_{rl}(\R)$ is given in \sectionp \ref{sec:core} and the \suppinfo.

\subsubsection{The electron Green's function for a system of coupled
electrons and phonons}
\label{sec:green_function_eph_coupling}

We briefly review the derivation of a set of self-consistent equations
that describe a system of coupled electrons and phonons entirely from
first principles. Since these equations merge insights from
two foundational approaches, the Hedin's formulation~\cite{hedin65},
describing interacting electrons in the potential of clamped nuclei,
and the Baym's earlier scheme~\cite{baym61}, modeling interacting
nuclei in the presence of an effective nuclear-nuclear bare interaction.
the complete set has been recently
referred to as the {\it Hedin-Baym equations}~\cite{giustino17}.
The set of Hedin-Baym equations provides a nonperturbative framework for
the coupled \eph system through a set of nonlinear, self-consistent
equations. Solving these equations yields the time-ordered
one-electron Green’s function, $G(\X t,\X' t')$, essential for
exploring the excitation spectrum of the corresponding many-body
Hamiltonian~\cite{kato_kobayashi_namiki60,fetter_walecka03} defined
in Eq. (\ref{eq:many_body_H}). Instead of relying on a diagrammatic
approach, this formalism employs functional differentiation
techniques~\cite{schwinger51} to derive the equations governing
the exact propagator.

In this formalism, an external potential $\phi(\R t)$ that
couples to the \textit{total} charge density adds an additional perturbative
term to the Hamiltonian (\ref{eq:many_body_H}),
i.e. $\hat{H}(t) = \hat{H}_0 + \hat{H}_1(t)$ with
$\hat{H}_1(t) = \int_\Omega d\R \, \hat{n}(\R) \phi(\R t)$.
Such a perturbative potential is introduced with the
aim of exploiting the Schwinger’s functional derivative technique
in finding a mathematical framework for the evaluation
of the time-ordered one-electron Green's function
in the Heisenberg picture~\cite{hedin_lundqvist69}
\be\label{eq:green}
G(\X t,\X' t') \equiv -\frac{i}{\hbar}
\frac{\braket{\hat{T} \hat{U}_0^\dagger(\TT,-\TT)
    \hat{U}(\TT,-\TT) \hat{\psi}(\X t)
    \hat{\psi}^\dagger(\X' t')}}{\braket{\hat{U}_0^\dagger(\TT,-\TT)
    \hat{U}(\TT,-\TT)}} \, .
\ee
In Eq. (\ref{eq:green}) $\TT$ tends to $\infty$,
$\hat{T}$ is the Wick's time-ordering operator for fermions,
$\hat{U}_0(t,t_0) = \exp[-i\hbar^{-1}\hat{H}_0(t-t_0)]$
is the evolution operator corresponding to the
unperturbed Hamiltonian (\ref{eq:many_body_H}), and
\be\label{eq:evolution_operator}
\hat{U}(t,t_0) = \hat{U}_0(t,t_0) - \frac{i}{\hbar} \int_{t_0}^t \hat{U}_0(t,t')
  \hat{H}_1(t') \hat{U}(t',t_0) dt'
\ee
is the time evolution operator for the solution of the Schr\"odinger equation
for the perturbed Hamiltonian $\hat{H}(t)$,
i.e. $\ket{t} = \hat{U}(t,t') \ket{t'}$, with $\ket{t'}$ the perturbed
electron-nuclei ground state at time $t'$. By treating the problem
through functional differentiation rather than perturbative
expansions, this approach ensures a rigorous
and self-consistent description of the interaction.

To develop a mathematical formalism
enabling a systematic evaluation of the Green’s function
(\ref{eq:green}) and incorporating the full dynamics of the
coupled \eph system,
it is essential to have knowledge of the time-dependence of the field operators.
In the Heisenberg picture~\cite{fetter_walecka03,hedin_lundqvist69},
the evolution operator (\ref{eq:evolution_operator})
generates the time evolution as
\be\label{eq:psi_time_evolution}
\hat{\psi}(\X t) = \hat{U}^\dagger(t,-\TT) \hat{\psi}(\X)
\hat{U}(t,-\TT) \, .
\ee
We can then write the equation of motion for the creation field operator as
\begin{align}\label{eq:equation_motion_psi}
& i\hbar \frac{\partial}{\partial t} \hat{\psi}^\dagger(\X t) = \bigl[\hat{H},
  \hat{\psi}^\dagger(\X)  \bigr](t) \nonumber \\
& \,\,\,\,\,\,\,\, =  \biggl[ - \frac{\hbar^2}{2 m_e}\nabla^2 +
  \int_\Omega d\R' v(\R-\R') \hat{n}(\R't) + \phi(\R t) \biggr]
\hat{\psi}^\dagger(\X t)
\end{align}
Multiplying Eq. (\ref{eq:equation_motion_psi}) on the left by the Hermitian
adjoint $\hat{\psi}$, applying the Wick time-ordering and the
evolution operators as in Eq. (\ref{eq:green}), evaluating
the expectation value over the unperturbed electronic ground state,
and combining Eqs. (\ref{eq:anticommutation_psi_operator}) and
(\ref{eq:equation_motion_psi}),
we obtain the equation of motion for the
time-ordered one-electron Green’s function
\begin{align}\label{eq:equation_motion_psi2}
& \biggl[i\hbar \frac{\partial}{\partial t} + \frac{\hbar^2}{2 m_e}\nabla^2
  - \phi(\R t) \biggr] G(\X t,\X' t') = \delta(\X t - \X'' t'') -
\nonumber \\ & \,\, -
\frac{i}{\hbar} \int_\Omega d\R'' \int dt'' v(\R t - \R'' t'')
\langle\langle \hat{n}(\R'' t'') \hat{\psi}(\X t)
\hat{\psi}^\dagger(\X' t') \rangle\rangle \, , \nonumber \\
\end{align}
where  $v(\R t - \R'' t'') \equiv v(\R -\R'')\delta(t-t'')$.
$\delta(\X t - \X'' t'') \equiv \delta(\X -\X'')\delta(t-t'')$, and where
$\langle\langle \dots \rangle\rangle = \braket{\hat{T}
  \hat{U}_0^\dagger \hat{U} \dots}/\braket{\hat{U}_0^\dagger \hat{U}}$.
Applying the general result that relates the functional differentiation of
the expectation value of time-ordered products in the Heisenberg
representation to correlation functions
\be\label{eq:kato_formula}
\frac{\delta \langle\langle\hat{\mathcal{O}}_1(t_1)
  \hat{\mathcal{O}}_2(t_2) \dots\rangle\rangle}{\delta \phi(\R t)} =
-\tfrac{i}{\hbar} \langle\langle\delta\hat{n}(\R t)\hat{\mathcal{O}}_1(t_1)
      \hat{\mathcal{O}}_2(t_2) \dots\rangle\rangle \, ,
\ee
with the induced charge density operator defined as
$\delta\hat{n}(\R t) = \hat{n}(\R t) - \braket{\hat{n}(\R t)}$, we arrive at
\be\label{eq:kato_formula_on_G}
\hbar^2 \frac{\delta G(\X t, \X' t')}{\delta \phi(\R'' t'')} =
- \langle\langle\delta\hat{n}(\R'' t'')
    \hat{\psi}(\X t) \hat{\psi}^\dagger(\X' t')\rangle\rangle \, .
\ee
Using the compact notation $(\X t)$ or $(\R t) \to
1$ and $(\R, t+\eta) \to 1^+$, where $\eta$ is a positive infinitesimal
arising from time ordering, the equation of motion for the
one-electron Green’s function can be rewritten as
\begin{align}\label{eq:equation_motion_psi3}
& \biggl[i\hbar\frac{\partial}{\partial t} + \frac{\hbar^2}{2m_e}\nabla^2_1
  - V_{tot}(1) - \nonumber \\ & \,\,\,\,\,
  -i\hbar\int d3\, v(1^+-3)\frac{\delta}{\delta\phi(3)}
  \biggr]G(12) = \delta(1-2) \, .
\end{align}
Here, the total potential $V_{tot}(1)$ is the sum of the external
potential and the Hartree mean-field potential
\be\label{eq:Vtot}
V_{tot}(1) = \int d2\, v(1-2)\braket{\hat{n}(2)} + \phi(1) \, .
\ee

A set of self-consistent equations coupling electrons and phonons
can be derived by eliminating the functional derivative with respect
to the external field, which is set to zero at the conclusion
of the Schwinger’s functional differentiation. To achieve this,
we use the variation of the identity $\delta\left(G^{-1}G\right)$ and
apply the chain rule for functional differentiation.
The functional derivative of the Green’s function can then be rewritten as
\be\label{eq:dGdphi}
\frac{\delta G(12)}{\delta\phi(3)} =
-\int d(456) G(14) \frac{\delta G^{-1}(45)}{\delta V_{tot}(6)}
\frac{\delta V_{tot}(6)}{\delta \phi(3)} G(52) \, ,
\ee
which can be expressed further as
\be\label{eq:dGdphi2}
\frac{\delta G(12)}{\delta\phi(3)}
= \int d(456) G(14) \Gamma(45,6) \varepsilon^{-1}(63) G(52).
\ee
In Eq. (\ref{eq:dGdphi2}) we introduce the three-point vertex
$\Gamma(45,6) \equiv - \delta G^{-1}(45)/\delta V_{tot}(6)$ and the inverse
dielectric function $\varepsilon^{-1}(12)$, with the latter defined
through the variation of the total potential $V_{tot}$ with respect to the
external perturbation
\bea\label{eq:eps_expansion}
\varepsilon^{-1}(12) &=& \frac{\delta V_{tot}(1)}{\delta \phi(2)} =
\delta(1-2) + \int d3 \,
v(1-3)\frac{\delta \braket{\hat{n}(3)}}{\delta\phi(2)} \nonumber  \,
\eea
where we use the definition given by Eq. (\ref{eq:Vtot})
for the total potential $V_{tot}(1)$.
By applying a chain rule on the total charge
density, we achieve a \textit{Dyson equation for the inverse dielectric
function}
\bea\label{eq:eps_expansion2}
\varepsilon^{-1}(12)
&=& \delta(1-2) + \int d(34) v(1-3)
\frac{\delta \braket{\hat{n}(3)}}{\delta V_{tot}(4)} \varepsilon^{-1}(42) \, .
\nonumber \\
\eea
Substituting Eq. (\ref{eq:dGdphi2}) into the equation of motion
(\ref{eq:equation_motion_psi3}), we obtain
\bea
\label{eq:equation_motion_psi4}
&&\biggl[i\hbar\frac{\partial}{\partial t} + \frac{\hbar^2}{2m_e}\nabla^2_1
  - V_{tot}(1) \biggr]G(12) - \nonumber \\
&& \,\,\,\,\,\,\,\,\,\,\,\,\,\,\,\,\,\,\,\,\,\,\,\,\,
- \int d3\,\Sigma(13) G(32) = \delta(1-2) \, ,
\eea
where we introduce the {\it self-energy} $\Sigma (12)$ for a
coupled electrons-nuclei system
\be\label{eq:sigma_el_ph}
\Sigma(12) = i\hbar \int d(34) G(13)\Gamma(32,4) W(41^+) \, ,
\ee
with the \textit{screened Coulomb interaction} defined as
\be\label{eq:W_el_ph}
W(12)= \int d3 \, \varepsilon^{-1}(13) v(32)\, ,
\ee
and the integral equation for the vertex function
\begin{align}\label{eq:vertex}
& \Gamma(12,3) = \delta(12)\delta(13) + \nonumber \\
& \,\,\,\,\,\,\,\,\,\,\, +
\int d(4567) \frac{\delta\Sigma(12)}{\delta G(45)} G(46) G(75) \Gamma(67,3) \, .
\end{align}
A Dyson equation for the screened Coulomb interaction can be derived
by combining Eqs. (\ref{eq:W_el_ph}) and (\ref{eq:eps_expansion})
\be\label{eq:W_dyson_def}
W(12) = v(1-2) + \int d(45) v(1-4) P(45) W(52) \, ,
\ee
where $P(12)\equiv\delta \langle\hat{n}(1)\rangle/\delta V_{tot}(2)$
separates into electronic and nuclear polarizabilities.

\subsubsection{The phonon contribution to the screened Coulomb interaction}
\label{sec:Wph}

A decomposition of the screened Coulomb interaction into a purely electronic
contribution and a term describing the effect of the \eph scattering
can be obtained by separating the total charge density into nuclear
and electronic components. Consequently, Eq. (\ref{eq:W_dyson_def})
can be rewritten as:
\begin{align}\label{eq:W_decomposition}
& W(12) = v(1-2) + \int d(34) v(1-3) P_e(34) W(42) + \nonumber \\
& \,\,\,\,\,\,\,\,\,\,\,\,\,\,\,\,\,\,\,\,\,\,\,\,\,\,\,\,\, + \int d(34)
\,v(1-3) \frac{\delta \braket{\hat{n}_n(3)}}{\delta \phi(4)} v(4-2) \, ,
\end{align}
where the electronic polarizability $P_e(34)$ is defined as:
\begin{align}\label{eq:electron_pol}
& P_e(34) = \frac{\delta \braket{\hat{n}_e(3)}}{ \delta V_{tot}(4)} \nonumber \\
& \,\,\,\,\,\,\,\,\,\,\,\,\,\,\,\,\,
= -i\hbar \sum_{\sigma}\int d(56) G^{\sigma}(35) G^{\sigma}(63^+) \Gamma(56,4)
\end{align}
representing the polarization propagator associated with
the electronic response to the total potential.
The term $\delta\langle\hat{n}_n(3)\rangle/\delta \phi(4)$
describes the nuclear charge density response to the external potential.

Omitting the last term in Eq. (\ref{eq:W_decomposition}) recovers the effect
of electronic screening on the bare Coulomb interaction, as derived in
Hedin's seminal work~\cite{hedin65}, $W_e(12) = \int d3 \,\varepsilon_e^{-1}(13)
v(3-2)$, where the purely electronic dielectric function $\varepsilon_e$ is
\be\label{eq:electron_eps}
\varepsilon_e(12) = \delta(1-2) - \int d(3) v(1-3) P_e(32)
\ee
To rewrite the \eph contribution to the screened Coulomb interaction,
Baym~\cite{baym61} introduced a perturbative, time-dependent external
scalar field $J(\R t)$, coupling exclusively to the nuclei,
i.e. $\hat{H}_2(t) = \int_\Omega d\R \hat{n}_n(\R) J(\R t)$.
By exploiting the general relation (\ref{eq:kato_formula}),
the equivalence $\delta\braket{\hat{n}_n(1)}/\delta \phi(2) =
\delta\braket{\hat{n}(2)}/\delta J(1)$ holds within this formulation.
Using functional differentiation and the
decomposition of the total charge density
($\hat{n}=\hat{n}_{e}+\hat{n}_n$), we derive
\begin{align}
&\frac{\delta\braket{\hat{n}(2)}}{\delta J(1)} =
\int d(34) \frac{\delta\braket{\hat{n}_e(2)}}{\delta V_{tot}(3)}
\frac{\delta V_{tot}(3)}{\delta\braket{\hat{n}(4)}}
\frac{\delta\braket{\hat{n}(4)}}{\delta J(1)} +
\frac{\delta\braket{\hat{n}_n(2)}}{\delta J(1)} \nonumber \\
& \,\,\,\,\,\,\,\,\,\,\,\,\,\,\,\,\,\,\,\,\,
= \int d(34) P_e(23) v(3-4) \frac{\delta\braket{\hat{n}(4)}}{\delta J(1)} +
\frac{\delta\braket{\hat{n}_n(2)}}{\delta J(1)} \, ,
\nonumber \\
\end{align}
where the definition (\ref{eq:electron_pol}) of the irreducible
electronic polarizability was used, together with the fact that
$\delta V_{tot}(3)/\delta\braket{\hat{n}(4)} = v(3-4)$,
which follows from the definition of the total potential
in Eq. (\ref{eq:Vtot}).
Solving for $\delta\braket{\hat{n}(2)}/\delta J(1)$ yields
\be\label{eq:dndJ}
\frac{\delta \braket{\hat{n}_n(1)}}{\delta \phi(2)}  =
\frac{\delta \braket{\hat{n}(2)}}{\delta J(1)}       = \int d3 \,
\varepsilon^{-1}_e(23) \frac{\delta\braket{\hat{n}_n(3)}}{\delta J(1)} \,
\ee
In Eq. (\ref{eq:dndJ}) it is convenient to define the
(time-ordered) {\it nuclear density-density response function}
\begin{align}\label{eq:den_den_correl_func}
D(31)& \equiv\frac{\delta\braket{\hat{n}_n(3)}}{\delta J(1)}
\nonumber \\ & =
-\tfrac{i}{\hbar}\langle\langle \hat{n}_n(3)
\hat{n}_n(1) \rangle\rangle+\tfrac{i}{\hbar}
\braket{\hat{n}_n(3)}\braket{\hat{n}_n(1)} \, .
\end{align}
By combining Eqs. (\ref{eq:W_decomposition}), (\ref{eq:dndJ}), and
(\ref{eq:den_den_correl_func}) and solving for $W$, we can rewrite
the screened Coulomb interaction $W$ as
\be\label{eq:Wfinal}
W(12) = W_e(12) + W_{ph}(12) \, ,
\ee
where the contribution from the \eph scattering to the screened Coulomb
interaction~\cite{hedin_lundqvist69}
\be\label{eq:Wph}
W_{ph}(12) = \int d(34) W_e(13) D(34) W_e(24),
\ee
describes the dynamic polarization effects of the lattice.

To account for small displacements of the nuclei, $\Delta\bm{\tau}_{rl}$,
from their equilibrium positions, $\bm{\tau}^0_{rl}$, we expand the
time-dependent nuclear density operator using a Taylor series within
the harmonic approximation~\cite{baym61,maksimov76} yielding
\begin{align}\label{eq:nuclear_density_operator_expansion}
& \hat{n}_n(\R t) = n^0_n(\R) + \sum_{rl\alpha}
Z_r\Delta\hat{\tau}_{rl\alpha}(t)
\frac{\partial \delta(\R-\bm{\tau}^0_{rl})}{\partial r_\alpha} - \nonumber \\
& \,
- \tfrac{1}{2} \sum_{rl}\sum_{\alpha\beta} Z_r\Delta\hat{\tau}_{rl\alpha}(t)
\frac{\partial^2\delta(\R-\bm{\tau}^0_{rl})}{\partial r_\alpha \partial r_\beta}
 \Delta\hat{\tau}_{rl\beta}(t).
\end{align}
Here, $n_n^0(\R) = -\sum_{rl} Z_r  \delta(\R-\bm{\tau}_{rl}^0)$ represents the
density of nuclear point charges at the clamped nuclear equilibrium positions,
$\bm{\tau}^0_{rl}$. The nuclear density-density response function from
Eq. (\ref{eq:den_den_correl_func}) can then be expressed, to second order
in nuclear displacements, as:
\begin{align}\label{eq:den_den_correl_func2}
& D(12) = \sum_{rl\alpha}\sum_{sl'\beta}
Z_r\frac{\partial\delta(\R_1 -
\bm{\tau}_{rl}^0)}{\partial r_{1\alpha}} \times \nonumber \\
& \,\,\,\,\,\,\,\,\,\,\,\,\,\,\,\,\,\,\,\,\,\,\,\,\,\, \times
D_{rl\alpha,sl'\beta}(t_1t_2)\,  Z_s
\frac{\partial\delta(\R_2 - \bm{\tau}_{sl'}^0)}{\partial r_{2\beta}} \, ,
\end{align}
where we have utilized the fact that
$\braket{\Delta\hat{\tau}_{r l \alpha}}=0$ at equilibrium and introduced the
{\it displacement-displacement correlation function} for the nuclei
\be\label{eq:displ_displ_corr_function}
D_{rl\alpha,sl'\beta}(t_1t_2) \equiv -\tfrac{i}{\hbar}
\langle\langle\Delta\hat{\tau}_{r l \alpha}(t_1)\Delta\hat{\tau}_{sl'\beta}(t_2)
\rangle\rangle\, .
\ee
By inserting Eq. (\ref{eq:den_den_correl_func2}) into the \eph
contribution (\ref{eq:Wph}) to the screened Coulomb interaction
and taking the Fourier transform over time, we get in the frequency domain
\bea\label{eq:Wph2}
&& W_{ph}(\R_1,\R_2;\omega) = \sum_{r\alpha l}\sum_{s\beta l'}
\int_\Omega d\R_3 \int_\Omega d\R_4\,
\varepsilon_e^{-1}(\R_1,\R_3;\omega) \times
\nonumber \\ && \, \times \frac{\partial
  V^{(0)}_{rl}(\R_3)}{\partial r_{3\alpha}}D_{rl\alpha,sl'\beta}(\omega) \,
\varepsilon_e^{-1}(\R_2,\R_4;\omega)
\frac{\partial V^{(0)}_{sl'}(\R_4)}{\partial r_{4\beta}} \, ,
\nonumber \\
\eea
where $V^{(0)}_{rl}$ is the bare electron-nuclear potential or its
ionic pseudopotential evaluated
at nuclear equilibrium position $\bm{\tau}^0_{rl}$.

To derive $D_{rl\alpha,sl'\beta}(tt')$, we apply Schwinger’s functional
derivative method, but, rather than introducing external scalar fields
that couple to the nuclear charge density, we use a time-dependent vector
field $\mathbf{F}_{rl}(t)$, which directly couples to the
displacements via a perturbative term $\hat{H}_3(t) = \sum_{rl}
\mathbf{F}_{rl}(t) \cdot \Delta\hat{\bm{\tau}}_{rl}(t)$.
Within this framework, the displacement-displacement correlation function
is given by~\cite{baym61}
\be\label{eq:D_def}
D_{rl\alpha,sl'\beta}(tt') =
\frac{\delta\braket{\Delta\hat{\tau}_{r l \alpha}(t)}}{\delta F_{sl'\beta}(t')}\, .
\ee
To evaluate the displacement-displacement correlation function
$D_{rl\alpha,sl'\beta}(tt')$, we use the field-theoretic approach
developed by Baym~\cite{baym61}, later extended in
Refs. \onlinecite{keating68,hedin_lundqvist69,gillis70,maksimov76}.
This method hinges on the time dependence of the displacement
operator $\Delta\hat{\bm{\tau}}_{rl}$ governed by its equation
of motion in the Heisenberg picture~\cite{baym61}.
By applying the time evolution (\ref{eq:psi_time_evolution}) to the displacement
operator, we have
\be\label{eq:equation_motion_displacements}
i\hbar \frac{\partial }{\partial t} \Delta\hat{\bm{\tau}}_{rl}(t) =
\bigl[\Delta\hat{\bm{\tau}}_{rl},
  \hat{H}\bigr](t)  \, ,
\ee
where $\hat{H} = \hat{H}_0 + \hat{H}_3(t)$.
With the aim of describing oscillating nuclei around
their equilibrium positions within the harmonic approximation,
we take an additional time derivative, resulting in the following
nuclear equation of motion
\be\label{eq:equation_motion_displacements2}
 \frac{\partial^2 }{\partial t^2} \Delta\hat{\bm{\tau}}_{rl}(t) =
 -\frac{1}{\hbar^2}\bigl[\bigl[\Delta\hat{\bm{\tau}}_{rl}, \hat{H} \bigr],  \hat{H} \bigr](t).
\ee
By evaluating the expectation value
of Eq. (\ref{eq:equation_motion_displacements2}), taking the
functional derivative with
respect to $\mathbf{F}_{sp}(t)$, and using the definition provided in Eq.
(\ref{eq:D_def}), we derive the equation of motion for the
displacement-displacement correlation function
\begin{multline}
\label{eq:eq_motion_D}
m_r \frac{\partial^2}{\partial t^2} D_{rl\alpha,sl'\beta}(tt')=
-\delta_{r\alpha l,s\beta l'}
\delta(t-t') \\
=- \sum_{t\gamma l''} \int dt''\, \Pi_{r\alpha l, t\gamma l''}(tt'')
D_{tl''\gamma,sl'\beta}(t''t') ,
\end{multline}
where $\Pi_{r\alpha l, t\gamma l''}(tt')$ is the \textit{phonon self-energy}.
By invoking the adiabatic and  harmonic approximations,
the phonon self-energy in the frequency domain becomes~\cite{baym61,giustino17}
\begin{align}\label{eq:phonon_self_energy}
& \Pi_{r\alpha l, s\beta l'}(\omega) = \int_\Omega d\R \int_\Omega d\R'
\biggl[  Z_r\frac{\partial\delta(\R-\bm{\tau}^0_{rl})}{\partial r_\alpha}
  W_e(\R,\R';\omega) + \nonumber \\
  & \,\,\,\,\,\,\, + \delta_{rs}\delta_{ll'}
  \nabla_\alpha \braket{\hat{n}(\R)}
  v(\R-\R')\biggr]
 Z_{s} \frac{\partial \delta(\R'-\bm{\tau}^0_{sl'})}{\partial r'_\beta} \, .
\end{align}
In the present case, we omit the detailed
derivation of Eqs. (\ref{eq:eq_motion_D})
and (\ref{eq:phonon_self_energy}) for brevity. For a comprehensive discussion,
the reader is referred to Ref. \onlinecite{baym61}.

In this work, we focus on the zero-frequency limit of the phonon self-energy,
referred to as the adiabatic approximation, defined as
$\Pi_{r\alpha l, s\beta l'}^{\text{A}}\equiv\Pi_{r\alpha l, s\beta l'}(\omega=0)$.
The final term in Eq. (\ref{eq:phonon_self_energy}) can be decomposed into
two parts due to the separability of the total charge density,
$\braket{\hat{n}(\R)}=\braket{\hat{n}_{e}(\R)}+\braket{\hat{n}_n(\R)}$.
The electronic contribution is recast in terms of
linear response theory~\cite{gillis70} as
\be\label{eq:linear_response_gradn_def}
\nabla_\alpha \braket{\hat{n}_e(\R)} =
\sum_{tn}\int_\Omega d\R' \chi_{e}(\R,\R';0)
\frac{\partial V^{(0)}_{tn}(\R')}{\partial r^{'}_\alpha} \, ,
\ee
where $\chi_e(\R,\R';0)$ denotes the static
reducible electronic density response.
In symbolic notation, it is related to the electronic inverse dielectric
function $\varepsilon^{-1}_e$ through $\varepsilon^{-1}_e = 1 + v\chi_e$,
as derived from the
electronic version of the Dyson equation (\ref{eq:eps_expansion}).
The derivation of Eq. (\ref{eq:linear_response_gradn_def}) follows from
an acoustic sum rule arising from the translational invariance of
the electronic density under rigid crystal translations.
A detailed derivation is provided in \sectionp
S.5 of the \suppinfo .
Using this result, the integral over $\R$
in Eq. (\ref{eq:phonon_self_energy}) can be rewritten as
\begin{align}\label{eq:chi2eps}
& \int_\Omega d\R \, v(\R'-\R)
\nabla_\alpha \braket{\hat{n}(\R)} =
\int_\Omega d\R \, v(\R'-\R)
\nabla_\alpha \braket{\hat{n}_n(\R)} + \nonumber \\
& + \sum_{tn} \biggl( \int_\Omega d\R'' \, \varepsilon^{-1}_e(\R',\R'';0)
\frac{\partial V^{(0)}_{tn}(\R'')}{\partial r^{''}_\alpha} -
\frac{\partial V^{(0)}_{tn}(\R')}{\partial r^{'}_\alpha} \biggr) \, ,
\nonumber \\
\end{align}
where the relationship $v\chi_e = \varepsilon^{-1}_e - 1$ follows
from the definition of the reducible polarizability.
At equilibrium, when the external forces $\mathbf{F}_{rl}(t)$ are set to zero,
the expectation values of the nuclear displacement operators vanish
$\braket{\Delta\hat{\bm{\tau}}_{tl}} = 0$, and the nuclear charge
density reduces to its equilibrium form
$\braket{\hat{n}_n(\R)} = n_n^0(\R) = -\sum_{tn} Z_t\delta(\R-\bm{\tau}^0_{tn})$.
Applying this result to Eq. (\ref{eq:chi2eps}) and using the identity
\be
 Z_r\frac{\partial}{\partial r_\alpha}\delta(\R-\bm{\tau}^0_{rl})
= - \int d\R'v^{-1}(\R-\R')\frac{\partial V^{(0)}_{rl}(\R')}{\partial r^{'}_\alpha}
\ee
as derived from Eq. (\ref{eq:v_vinv}), reveals that the first and last
terms in Eq. (\ref{eq:chi2eps}) cancel \textit{exactly}.
Using the definition of the electronic contribution to the
screened Coulomb interaction
\begin{align}\label{eq:We_def1}
W_e(\R,\R',\omega) &=
\int_\Omega d\R'' \varepsilon_e^{-1}(\R,\R'',\omega) v(\R''-\R') \\
\label{eq:We_def2}
&= \int_\Omega d\R'' v(\R-\R'') \varepsilon_e^{-1}(\R',\R'',\omega),
\end{align}
we rewrite the adiabatic phonon self-energy as
\begin{align}\label{eq:phonon_selfenergy_adiabatic}
& \Pi_{r\alpha l, s\beta l'}^{\text{A}}  = \sum_{tn} \biggl(\delta_{rt}\delta_{ln}-
\delta_{rs}\delta_{ll'}\biggr) \int_\Omega \int_\Omega \int_\Omega
d\R d\R' d\R''   \nonumber \\
& \,\,\,\,\,\,\,\,
\times \frac{\partial V^{(0)}_{sl'}(\R)}{\partial r_\beta}
v^{-1}(\R- \R'')\varepsilon_{e}^{-1}(\R'', \R';0)
\frac{\partial V^{(0)}_{tn}(\R')}{\partial r'_\alpha} \, ,
\end{align}
which corresponds to Eq. (\ref{eq:ifc_manybody}). Additionally,
the symmetric definitions of the screened Coulomb interaction in
Eqs. (\ref{eq:We_def1}) and (\ref{eq:We_def2}) ensure the symmetry of
the \ifc, i.e.,
$\Pi_{r\alpha l, s\beta l'}^{\text{A}} = \Pi_{s\beta l' , r\alpha l}^{\text{A}}$.

The adiabatic approximation naturally leads to a system of
\textit{non-interacting phonons}. By substituting the phonon self-energy
in Eq. (\ref{eq:eq_motion_D}) with the static limit of its spectral
representation expressed in terms of the
adiabatic eigenmodes $\bm{e}_{r,\nu}(\Q)$ and eigenfrequencies $\omega_{\nu\Q}$,
as introduced in \sectionp \ref{sec:phonons},
we obtain an explicit expression for the adiabatic
displacement-displacement correlation
function~\cite{schrieffer83,shafer_wegener02,giustino17}
\be\label{eq:displ_displ_corr_function2}
D^\A_{rl\alpha,sl'\beta}(\omega) = \frac{1}{N_\Q}\sum_{\Q\nu}
 \frac{e^*_{r\alpha,\nu}(\Q)e_{s\beta,\nu}(\Q)}
      {2\omega_{\Q\nu} \sqrt{m_rm_s}} e^{i\Q\cdot(\RR_{l'}-\RR_l)}
   D^\A_{\Q\nu}(\omega) \, ,
\ee
where the \textit{adiabatic phonon propogator} $D^\A_{\Q\nu}(\omega)$
is defined as
\be\label{eq:DAw}
D^\A_{\Q\nu}(\omega) = \frac{1}{\omega-\omega_{\Q\nu} +i\eta} -
\frac{1}{\omega + \omega_{\Q\nu} - i\eta} \, ,
\ee
with $\eta$ being a positive infinitesimal.

Non-adiabatic (NA) corrections to the phonon
self-energy~\cite{giustino17,saitta08,calandra10}, $\bm{\Pi}^{\text{NA}}$,
arise from the differences between the dynamical and static
screened Coulomb interactions.
The real part of this correction shifts the adiabatic phonon frequencies,
while the imaginary part accounts for the spectral broadening of resonances.
The full phonon propagator can be computed via a Dyson-like scheme,
i.e., $\bm{D}(\omega) = \bm{D}^\A(\omega) + \bm{D}^\A(\omega)
\bm{\Pi}^{\text{NA}}(\omega) \bm{D}(\omega)$.
However, NA corrections are typically small compared to adiabatic phonon
frequencies, so we will replace the fully interacting phonon propagator
$\bm{D}(\omega)$ with the adiabatic counterpart given by Eq. (\ref{eq:DAw}).

By combining  Eqs. (\ref{eq:Wph2}), (\ref{eq:displ_displ_corr_function2}),
and (\ref{eq:DAw}), we obtain the \eph contribution
to the screened Coulomb interaction within the adiabatic and harmonic
approximations~\cite{hedin_lundqvist69}
\be\label{eq:Wph_A}
W^\A_{ph}(\R_1,\R_2;\omega) = \frac{1}{N_\Q\Omega_0}\sum_{\Q\nu}
D^\A_{\Q\nu}(\omega) g^*_{\Q\nu}(\R_1;0) g_{\Q\nu}(\R_2;0) \, ,
\ee
where the static electronic inverse dielectric function replaces its
dynamical counterpart in the $\omega=0$ limit.
In Eq. (\ref{eq:Wph_A}), the \eph \textit{coupling function}
$g_{\Q\nu}(\R;0)$ is defined as
\begin{widetext}
\be\label{eq:g_function}
g_{\Q\nu}(\R)=g_{\Q\nu}(\R;0) \equiv \sum_{r\alpha l}
\sqrt{\frac{\Omega_0}{2 m_r \omega_{\Q\nu}}}
e^{i\Q\cdot\RR_l} e_{r\alpha,\nu}(\Q)
\int_\Omega d\R' \varepsilon_e^{-1}(\R,\R';0)
\frac{\partial V_{rl}^{(0)}(\R')}{\partial r'_{\alpha}}  \, ,
\ee
\end{widetext}
which satisfies $g^*_{\Q\nu}(\R)=g_{-\Q\nu}(\R)$.

The primary goal of this paper is to evaluate the matrix element
$\elmat{\psi^\sigma_{i,\bar{\Q}}} {g_{\Q\nu}}
{\psi^\sigma_{n,\K}}_\Omega$, which represents an integral
over the BvK macrocrystal, with the integrand
involving the product of Bloch functions.
Using the translational and rotational invariance of the
electronic inverse dielectric function, it is easy to show that $g_{\Q\nu}(\R)$
transforms as a Bloch function, i.e.,
\be
g_{\Q\nu}(\R+\RR) = e^{i\Q\cdot\RR} g_{\Q\nu}(\R).
\ee
This property allows us to restrict the integral to the Bloch-periodic
part of the unit cell, leading to
\begin{align}
\elmat{\psi^\sigma_{i,\bar{\Q}}} {g_{\Q\nu}}{\psi^\sigma_{n,\K}}_\Omega &=
\delta_{\bar{\Q},\K+\Q} N_\K
\elmat{u^\sigma_{i,\K+\Q}} {g_{\Q\nu}}{u^\sigma_{n,\K}}_{\Omega_0} \nonumber \\
&= \delta_{\bar{\Q},\K+\Q} N_\K \sqrt{\Omega_0/\hbar} \,
g^\sigma_{in,\nu}(\K,\Q) \, ,
\end{align}
where
\begin{equation}\label{eq:gmat}
  g^\sigma_{in,\nu}(\K,\Q) = \sqrt{\hbar/\Omega_0}
  \elmat{u^\sigma_{i,\K+\Q}}{g_{\Q\nu}}{u^\sigma_{n,\K}}_{\Omega_0}
\end{equation}
is the \eph \textit{matrix element}. This quantity
describes, for a given spin-polarization channel,
the scattering amplitude of an electronic state $\ket{\K,n,\sigma}$
with energy $\varepsilon^\sigma_{n\K}$ to a state $\ket{\K+\Q,i,\sigma}$
of energy  $\varepsilon^\sigma_{i\K+\Q}=\varepsilon^\sigma_{n\K}\pm\hbar\omega_{\Q\nu}$
involving the absorption or
emission of a phonon with wave vector $\Q$ and energy $\hbar\omega_{\Q\nu}$.
The prefactor $\sqrt{\hbar/\Omega_0}$ ensures proper energy units,
derived from substituting Eq. (\ref{eq:Wph_A}) into the self-energy definition
in Eq. (\ref{eq:sigma_el_ph}) for a coupled electron-phonon system.
The self-energy $\Sigma^\sigma(\R,\R';\omega)$ is decomposed into purely
electronic and \eph contributions
$\Sigma^\sigma(\R,\R';\omega) = \Sigma^\sigma_e(\R,\R';\omega)+
\Sigma_\sigma^{\text{eph}}(\R,\R';\omega)$, where the \eph self-energy is expressed as
\begin{align}
& \Sigma^{\text{eph}}_\sigma(\R,\R';\omega) = \nonumber \\
& \,\,\,\,\,\,\,\,= \frac{i\hbar}{2\pi}
\int d\omega' W^\A_{ph}(\R,\R';\omega') G^\sigma(\R,\R';\omega-\omega') e^{i\eta\omega'} .
\end{align}
This result holds within the RPA, where
the Dyson equation (\ref{eq:vertex}) for the three-point vertex
simplifies to $\Gamma(12,3)\approx \delta(1-2)\delta(1-3)$.
A more detailed analysis of Hedin's equations
within the RPA is presented in \ssections \ref{sec:qsgw} and
\ref{sec:bse}. In the remainder of this work we limit our
analysis to spin-unpolarized systems, omitting the spin index to
streamline the notation. The developments presented can be readily
extended to spin-polarized cases.

Substituting Eq. (\ref{eq:g_function}) into the \eph matrix element
definition (\ref{eq:gmat}), we obtain
\begin{equation}\label{eq:gmatrix}
g_{in,\nu}(\K,\Q) = \sum_{r\alpha}
\sqrt{\frac{\hbar}{2 m_r \omega_{\Q\nu}}} e_{r\alpha,\nu}(\Q)\,
 \xi_{in}^{r\alpha}(\Q,\K)  \, ,
\end{equation}
where we introduce the {\it reduced \eph matrix elements}
\begin{equation}\label{eq:ximat}
\xi_{in}^{r\alpha}(\Q,\K) = \sum_{l} e^{i\Q\cdot\RR_l}
\elmat{\psi_{i,\K+\Q}}{\xi^{r\alpha}_{\,l}}{\psi_{n,\K}}_{\Omega_0} \, ,
\end{equation}
with $\xi^{r\alpha}_l(\R) = \xi^{r\alpha}(\R-\RR_l)$ the {\it reduced
electron-phonon coupling function} defined as
\begin{equation}\label{eq:xifun}
\xi^{r\alpha}(\R-\RR_l)= \int_\Omega d\R' \varepsilon_e^{-1}(\R,\R';0) \,
\frac{\partial V^{(0)}_{rl}(\R')}{\partial r_{\alpha}^{'}}  \, .
\end{equation}

This section summarizes the foundational elements necessary to
understand the Green’s function formalism for a coupled electron-phonon
system, with a particular emphasis on \eph coupling. In the
subsequent sections and the \suppinfo,
we provide detailed insights into the implementation of this
framework within the \texttt{Questaal} electronic structure suite.

A notable distinction between this approach and the dielectric formalism
outlined in \sectionp \ref{sec:phonons} lies in the
nature of the perturbation used for treating the screening.
In the framework of DFPT, screening is treated via a static
perturbation~\cite{giustino17}, specifically the variation of the
\textit{external potential} $\delta U_{en}$. This perturbation
is implicitly frequency-independent, then leading
to the adiabatic approximation. By contrast,
the dynamical \ifc described in Eq. (\ref{eq:phonon_self_energy})
incorporate retardation effects, recovering the adiabatic approximation
when the dynamical screening $\varepsilon^{-1}_e(\R,\R';\omega)$ is
replaced with its static counterpart, $\varepsilon^{-1}_e(\R,\R';0)$.

As elaborated in \sectionp \ref{sec:bse}, the Hedin-Baym framework
outlined in these sections also allows for the inclusion of
both \textit{excitonic effects and
static exciton-phonon coupling}. These effects are accounted for
by evaluating the electronic inverse dielectric function through
the Bethe-Salpeter Equation (BSE), used to compute the electronic
polarization propagator (\ref{eq:electron_pol}) and, consequently,
the electronic dielectric function (\ref{eq:electron_eps}).

Importantly, as demonstrated in Eqs. (\ref{eq:g_function}) and
(\ref{eq:phonon_self_energy}), the interaction between electrons and
phonons is fundamentally governed by the electronic dielectric response,
represented by $W_e$.
This observation underscores the pivotal
role of the electronic
inverse dielectric function $\varepsilon^{-1}_e(\R,\R';\omega)$ in the
field-theoretic description of the \eph problem.

\subsection{Incomplete-Basis-Set corrections in the field-theoretic framework
for a system of interacting electrons and phonons}
\label{sec:pulay}

The accuracy of electronic structure calculations depends critically
on the choice of the basis set used to describe wavefunctions
and derived quantities. In practical implementations, basis sets
are often incomplete, leading to inaccuracies that manifest differently
depending on the property being calculated and the subspace
of the Hilbert space involved.

Basis functions may be inadequate for capturing
specific subspaces of the Hilbert space
required for calculating response functions. For instance, as it will be
extensively discussed in \sectionp \ref{sec:core}, the finite Hilbert
space spanned by the LMTO basis inadequately
represents electronic core wave functions and their response to
external perturbations. Nonetheless, basis sets
explicitly dependent on nuclear positions, such as those in the
LMTO method, introduce additional complexities. In such cases, the
derivatives of wave function-dependent quantities---like the electronic
density---with respect to
nuclear displacements include extra terms arising from the explicit
dependence of the basis functions on nuclear coordinates.
These additional terms are referred to by some authors
as \textit{Pulay-like corrections},
a terminology derived from Pulay's pioneering work on forces
in KS-DFT~\cite{pulay69}. However, the concept extends more broadly
to \textit{incomplete-basis-set corrections} (\ibcs) that account for basis
dependence in specific subspaces relevant to the property under study.

While the analysis of such corrections is often complex and
strongly dependent on the specific basis set employed, the purpose
of this section is to determine whether these corrections are
necessary when formulating the problem of interacting electrons
and phonons within a field-theoretic framework.

The theoretical derivation of
the \eph coupling function [Eq. (\ref{eq:g_function})] presented
in \sectionp \ref{sec:Wph}, as well as the detailed derivation of the phonon
self-energy [Eq. (\ref{eq:phonon_self_energy})] reported in Ref.
\onlinecite{baym61}, does not involve explicit
nuclear displacements of physical quantities dependent
on the electronic wave functions. Consequently,
Eqs. (\ref{eq:g_function}) and (\ref{eq:phonon_self_energy}) do not require
\textit{Pulay-like} \ibcs~\cite{pulay69,savrasov92,savrasov94,savrasov96}
when the Bloch functions are expanded using a localized basis set that
depends on the equilibrium nuclear positions $\{\bm{\tau}^0_{rl}\}$.
Therefore, in such cases,
formulating the \eph problem within a field-theoretic framework
effectively eliminates the need to evaluate Pulay-like \ibcs.

Additionally, \ibccs can be implicitly incorporated into any
theoretical and computational framework employed to calculate
$\varepsilon^{-1}_e(\R,\R';\omega)$.
In Refs. \onlinecite{betzinger12,betzinger13,friedrich15},
the J\"ulich group formulated a linear response theory
that accounts for \ibcs, albeit in a different context.
While we do not explore this approach further in the present work,
we reserve a detailed analysis for future studies.

Equation (\ref{eq:phonon_self_energy}) also involves the gradient
of the electron density, $\nabla_\alpha \braket{\hat{n}_e(\R)}$.
The treatment of the core density introduces
additional considerations regarding basis set completeness, which
are addressed in \sectionp \ref{sec:core}. Finally,
as elaborated in detail in \sectionp S.5 of the \suppinfo,
reformulating the derivative of the electron density with respect to
nuclear displacements using linear response theory introduces correction terms
to account for the dependence of the basis functions on nuclear
displacements.
Under these conditions, it can be demonstrated that the
translational invariance of the electron density under a rigid
crystal translation leads to the following sum rule for the gradient
$\nabla \braket{\hat{n}_e}$
\bea\label{eq:linear_response_gradn_def_pulay}
&&
\nabla_\alpha \braket{\hat{n}_e(\R)} =
\sum_{rl}\int_\Omega d\R''\chi_{e}(\R,\R'';0)
\frac{\partial V^{(0)}_{rl}(\R'')}{\partial r^{''}_\alpha} - \nonumber \\
&& \,\,\,\,\,\,\,\,\,\,\,\,\,\,\,\,\,\,\,\,\,\,\,\,\,\,\,\,
\,\,\,\,\,\,\,\,\,\,\,\,\,\,\,\,\,\, - \sum_{rl} \left.
\frac{\partial\braket{\hat{n}_e(\R)}}{\partial \tau_{rl \alpha}}
\right|_{\bm{\tau}_{rl}^0,\{\bm{z}^\K\}} \, .
\eea
This replaces Eq. (\ref{eq:linear_response_gradn_def}),
where, in addition to the perturbation responsible for changes
in the wave function, the second term on the right-hand side
also accounts for contributions arising from variations in the basis set.
The derivative
$\left.\partial\braket{\hat{n}_e(\R)}/ \partial \tau_{rl\alpha}
\right|_{\bm{\tau}_{rl}^0,\{\bm{z}^\K\}}$
is evaluated while keeping the expansion coefficients
$\{\bm{z}^\K\}$ (introduced in \sectionp \ref{sec:lmto}) constant.
This derivative captures the parametric dependence of
the basis functions on the equilibrium nuclear positions.
Therefore, by substituting Eq. (\ref{eq:linear_response_gradn_def_pulay}) into
Eq. (\ref{eq:phonon_self_energy}), the adiabatic phonon self-energy,
corrected for the explicit dependence of the basis functions on nuclear displacements,
$\widetilde{\Pi}_{r\alpha l, s\beta l'}^{\text{A}}$, takes the following form
\begin{align}\label{eq:phonon_selfenergy_pulay}
& \widetilde{\Pi}_{r\alpha l, s\beta l'}^{\text{A}} =
\Pi_{r\alpha l, s\beta l'}^{\text{A}} -
\nonumber \\ & \,\,\,\,\,\,\,\,\,\,\,\,\,
- \delta_{rs}\delta_{ll'} \sum_{tn} \int_\Omega d\R
\frac{\partial V^{(0)}_{sl'}(\R)}{\partial r_\beta} \left.
\frac{\partial\braket{\hat{n}_e(\R)}}{\partial \tau_{t n \alpha}}
\right|_{\bm{\tau}_{tn}^0,\{\bm{z}^\K\}}
\end{align}
where $\Pi_{r\alpha l, s\beta l'}^{\text{A}}$ is defined as in Eq.
(\ref{eq:phonon_selfenergy_adiabatic}).
As a result, the acoustic sum rule (\ref{eq:acoustic_sum_rule}), which ensures
the translational invariance of the \ifc,
\begin{align}\label{eq:acoustic_sum_rule_pulay}
\sum_{sl'} \widetilde{\Pi}_{r\alpha l, s\beta l'}^{\text{A}} &=
-\sum_{tn} \int_\Omega d\R
\frac{\partial V^{(0)}_{rl}(\R)}{\partial r_\beta} \left.
\frac{\partial\braket{\hat{n}_e(\R)}}{\partial \tau_{tn \alpha}}
\right|_{\bm{\tau}_{tn}^0,\{\bm{z}^\K\}} \nonumber \\ &= 0
\end{align}
will be satisfied under either of the following conditions: (i)
$\left.\partial\braket{\hat{n}_e(\R)}/ \partial \tau_{tn \alpha}
\right|_{\bm{\tau}_{tn}^0,\{\bm{z}^\K\}} = 0$,
which applies when the basis functions exhibit no
parametric dependence on the equilibrium nuclear positions, or
(ii) when the basis functions form a \textit{complete set}, as discussed
in detail in \sectionp S.5 of the \suppinfo.
In the latter case, the completeness of the basis ensures that any parametric
variation of the nuclear positions is fully captured, allowing for
a precise description of the electron density and preserving the sum rule.
A more detailed analysis of this issue is provided in
\sectionp \ref{sec:dynmat}.

\subsection{Quasiparticle Self-Consistent \gw Approximation}
\label{sec:qsgw}

\texttt{Questaal}~\cite{questaal20} implements a Green's function theory which,
at its lowest level, begins with the
\textit{Quasiparticle Self-Consistent} \gw (QS\gw)
approximation~\cite{mark06qsgw,Faleev04,Kotani07}.  Through
self-consistency, it finds, by construction, an optimal one-body non-interacting
\textit{G}\textsubscript{0}, which enables MBPT to converge as efficiently
as possible with increasing diagram order.  Self-consistency
is one of the primary reasons QS\gw consistently exhibits
higher fidelity than most implementations of MBPT~\cite{Acharya21a}.
The errors are small, systematic, and their origins are
generally well understood. Furthermore, QS$GW$
surmounts the problematic reliance on the
starting point, typically KS-DFT, for
which MBPT methods are often criticized. This allows improvements
to be introduced where needed in a controllable, accurate, and
hierarchical manner, without parameterization
or heuristics.  A particularly notable illustration of this is the
improvement realized when ladder diagrams are incorporated into the
polarizability~\cite{Cunningham2023}.  Since plasmons are the dominant
many-body effect in \gw, it is not
surprising that the RPA used in \gw (also referred to
as the time-dependent Hartree approximation), can be overly simplistic.
In virtual excitations, electron-hole pairs are treated as
independent (bubble diagrams), whereas in reality they should
attract each other. Their attraction is responsible for the formation of
excitons and also enhances screening.  The
omission of electron-hole attraction in the polarizability is the primary
source of errors in QS\gw.  The most significant
discrepancies with experiments---such as a systematic tendency to overestimate
bandgaps~\cite{mark06qsgw}, a blue shift in
the plasmon peaks in the dielectric function~\cite{Kotani07}, and a systematic
underestimation of the static
$\varepsilon_\infty$ by $\sim$20\%---are all related to this omission.
Including this attraction through the ladder approximation
largely mitigates these discrepancies, particularly in weakly or
moderately correlated systems~\cite{Cunningham2023}.  We
denote QS\gw with ladder diagrams included in the
polarizability (\ref{eq:electron_pol}) as QS$G\widehat{W}$.
The high fidelity with which the inverse electronic dielectric
function (\ref{eq:electron_eps}) is described is particularly important
in this context, as the \eph coupling
function (\ref{eq:g_function}) is primarily determined by it.

Even though self-consistency is important, it has long been known that
full self-consistency in \textit{Self-Consistent}
\gw (sc\gw)
can perform poorly in solids~\cite{Shirley96,holm98}.  A recent
re-examination of some semiconductors~\cite{Grumet18} confirms that
the dielectric function (and the concomitant quasi-particle levels)
indeed worsen when $G$ is fully self-consistent, as explained
in Appendix A of Ref. \onlinecite{Kotani07}.  Fully sc\gww
becomes even more problematic in transition metals~\cite{BelashchenkoLocalGW}.
Moreover, while sc\gww is a
conserving approximation in $G$ with respect to physical quantities such
as charge density and current density, it violates conservation
laws in the screened Coulomb interaction $W_e$. Specifically, sc\gww fails
to satisfy the $f$-sum rule for the inverse dielectric function
within the RPA~\cite{Tamme99}, leading to a loss of its usual physical
interpretation as a response function.  As a result, sc\gw tends to
smear out spectral functions in transition
metals~\cite{BelashchenkoLocalGW}, often yielding less accurate results
compared to KS-DFT within the LDA.

For these reasons, QS\gw is generally considered superior to
fully sc\gw, unless additional vertex corrections are incorporated into
sc\gw. Nevertheless, QS\gww has its own limitations: first,
it cannot be constructed from standard diagrammatic expansion, and second, it
relies on a single-reference starting point. Although the self-consistent
solution does not correspond to a stationary point in the Luttinger-Ward or
Klein functional, Ismail-Beigi demonstrated that it is stationary
with respect to the gradient of the Klein functional~\cite{Beigi17}.

An optimal one-body $G_0$ is determined by minimizing a norm
(within a prescribed level of approximation for the
many-body part), which serves as a measure of the difference between $G_0$
and the interacting $G$ generated by it. While there is no unique
prescription for this norm~\cite{Kotani07}, a well regarded choice for the
static non-local \xc potential within the QS\gww framework
is~\cite{mark06qsgw,Kotani07}
\begin{align}\label{eq:mapping_vxc}
\hat{V}_{xc} = \tfrac{1}{2}\sum_{ij}
        |\psi_i\rangle
        \left\{ {{\rm Re}[\Sigma_e({\varepsilon_i})]_{ij}+
          {\rm Re}[\Sigma_e({\varepsilon_j})]_{ij}} \right\}
        \langle\psi_j| \, ,
\end{align}
where $\text{Re}[\Sigma_e]$ denotes the Hermitian part of the electron
self-energy.
This choice, as demonstrated by Ismail-Beigi~\cite{Beigi17}, minimizes
the gradient of the Klein functional. Furthermore, at self-consistency, the
poles of $G$ coincide with the poles of $G_0$, allowing the energy bands of
QS\gww to be properly interpreted as excitation energies, unlike
KS and generalized KS-DFT~\cite{seidl96}.

\section{The implementation}
\label{sec:implementation}

\subsection{Questaal's LMTO Basis Set}
\label{sec:lmto}

\texttt{Questaal} utilizes an \textit{all-electron}
augmented-wave basis set, which represents an optimized variant of the LMTO
method originally proposed by O. K. Andersen~\cite{Andersen75}, with
a generalization of the Hankel functions of the Andersen's
envelope functions. For extensive details on this method,
the reader is referred to Ref. \onlinecite{questaal20}.
In \texttt{Questaal}, envelope functions used to
represent one-particle Bloch functions
consist of convolutions of gaussian and Hankel functions, namely the
\textit{smooth Hankel functions}. These functions offer some key
advantages. First, they are nonsingular, making
full-potential implementations tractable.
They have more flexibilty than Hankel functions,  and can be
better tailored to the potential of real solids, whereas ordinary Hankel functions
are exact only for a muffin-tin potential.
This basis attains an accuracy approaching the linearized augmented plane-wave
(LAPW) method while offering more compactness and reduced computational cost~\cite{questaal20}.
The envelope functions $\mathcal{H}_{L}(\varepsilon,r_s;\mathbf{r})$
are centered at each nucleus and characterized by angular
momentum $L=\{l m\}$, an energy parameter
$\varepsilon=-\kappa^2$ that controls the
exponential decay at large distances,
and a smoothing radius $r_s$ that governs the
degree of smoothing near the nucleus.
The Fourier representation of these functions has a closed form
\begin{align}
\label{eq:defhlq}
\widehat{\mathcal{H}}_{ L}(\varepsilon,r_s;{{\mathbf{q}}})
&= \mathcal{Y}_L(-i{{\mathbf{q}}})
\widehat{h}_{0}(\varepsilon,r_s;{q}) \\
\widehat{h}_{0}(\varepsilon,r_s;{q}) &=
-\frac{4\pi}{\varepsilon-q^2}e^{r_s^2(\varepsilon-q^2)/4} \, , \nonumber
\end{align}
where $\mathcal{Y}_L(\R) = r^lY_{lm}(\widehat{\R})$
are spherical harmonics polynomials,
and $\widehat{h}_{0}$ is the Fourier transform of the $l = 0$ function.
In real space, ${h}_{0}(\varepsilon,r_s;r)$ is obtained as a
convolution of the ordinary Hankel function $h_0(\kappa;r) = e^{-\kappa r}/r$
with a Gaussian function, the
latter smoothing the $1/r$ singularity of the Hankel
function near the nucleus.
For small $r$, $h_0$ behaves as a Gaussian, while for $r \gg r_s$,
it asymptotically approaches the ordinary Hankel form.
In real space,
\begin{align}
  \mathcal{H}_{L}(\varepsilon,r_s;{\mathbf{r}}) =
  \mathcal{Y}_L(-\nabla) h_0(\varepsilon,r_s;r)
\label{eq:defhl}
\end{align}
can be recursively constructed from $h_0(\varepsilon,r_s;r)$.
For a more detailed exposition on the envelope functions, the reader is
referred to Ref. \onlinecite{bott98}, and to Ref. \onlinecite{questaal20} for the envelope and basis set
functions.

In a periodic system, the electronic eigenfunctions are expanded as
linear combinations of Bloch-summed basis functions
\begin{eqnarray}
\label{eq:lmtopsi}
      {\psi_{n,\K}}(\R) = \sum_{{\bm{\tau}L}{j}}
      z^{\mathbf{k}}_{{\bm{\tau}L}{j},n}
      \chi^{\mathbf{k}}_{{\bm{\tau}L}{j}}(\mathbf{r}),
\end{eqnarray}
where $n$ is the band index,
$z^{\mathbf{k}}_{\bm{\tau}L{j},n}$ are the eigenvectors, and
$\chi^{\mathbf{k}}_{{\bm{\tau}L}{j}}(\mathbf{r})$ are the lattice-summed
envelope functions, augmented by
partial waves. Here, $\bm{\tau}$ identifies the atomic
site where the envelope function
is centered, and $j$ distinguishes different shapes of the envelope
function (as defined by $\varepsilon$ and $r_{s}$) within the primitive cell.
In the interstitial region, the envelope functions are represented by a
discrete Fourier series of plane waves~\cite{questaal20}
\begin{align}
{\mathcal{H}^{\mathbf{k}}_{L}(\varepsilon,r_s;\mathbf{r})}
=\frac{1}{\Omega}\sum\limits_{\mathbf{G}}
{
{\widehat{\mathcal{H}}_{L}}(\varepsilon,r_s;{\mathbf{k+G}})
{e^{i({\mathbf{k+G}}) \cdot \mathbf{r}}}
}
\label{eq:blochsmh}
\end{align}
where its plane wave representation is given
by substituting Eq. (\ref{eq:defhlq})
for $\widehat{\mathcal{H}}_{L}(\varepsilon,r_s;{\mathbf{k+G}})$.
Within the augmentation sphere centered at $\bm{\tau}$, the envelope functions
are smoothly matched in a differentiable manner to a linear
combination of radial functions
($\varphi_{\bm{\tau}l}$, $\dot\varphi_{\bm{\tau}l}$,
$\varphi^z_{\bm{\tau}l}$) at that site.
Here, $\varphi_{\bm{\tau}l}(\varepsilon_\nu;r)$ is the solution to the
radial Schr\"odinger (Dirac) equation at some specific energy $\varepsilon_\nu$.
$\dot\varphi_{\bm{\tau}l}(\varepsilon_\nu;r)$ denotes the energy-derivative
of $\varphi_{\bm{\tau}l}(\varepsilon_\nu;r)$, necessary for the
linearization $\varphi_{\bm{\tau}l}(\varepsilon;r) =
\varphi_{\bm{\tau}l}(\varepsilon_\nu;r) + (\varepsilon-\varepsilon_\nu)
\dot\varphi_{\bm{\tau}l}(\varepsilon_\nu;r) + \dots $.
To enhance accuracy, additional local orbitals
$\varphi^z_{\bm{\tau}l}(\varepsilon;r)$ may be introduced, corresponding
to radial solutions at energies well above or below $\varepsilon_\nu$.  These
radial functions can be succinctly labeled in a compact notation as
$\{\varphi_{\bm{\tau}u}\}$, where $u$ is a composite index incorporating $L$ and
one of the functions $\varphi_{\bm{\tau}l}$, $\dot\varphi_{\bm{\tau}l}$,
$\varphi^z_{\bm{\tau}l}$.

Thus, the total wave function (\ref{eq:lmtopsi}) can be rewritten as the sum of
interstitial and augmentation parts
\begin{eqnarray}
{\psi_{n,\K}}(\mathbf{r})
= \sum_{\bm{\tau}u}      \alpha^{{\mathbf{k}}n}_{\bm{\tau}u}
\varphi^\mathbf{k}_{\bm{\tau}u}(\mathbf{r})
+ \sum_\mathbf{G}  \beta^{{\mathbf{k}}n}_\mathbf{G}
P^\mathbf{k}_\mathbf{G}(\mathbf{r}),
\label{def:psiexp}
\end{eqnarray}
where the interstitial plane wave (IPW) is defined as
\begin{eqnarray}\label{eq:ipw_basis}
P^\mathbf{k}_\mathbf{G}(\mathbf{r}) =
\begin{cases}
 0                           & \text{if \textbf{r}} \in \text{any MT} \\
 e^{i (\mathbf{k+G})\cdot\mathbf{r}}/\sqrt{\Omega_0} & \text{otherwise}
\end{cases}
\end{eqnarray}
and $\varphi^\mathbf{k}_{\bm{\tau} u}$ represents the
Bloch sums of the augmented radial functions $\varphi_{\bm{\tau} u}$
\begin{eqnarray}\label{eq:aug_basis}
  \varphi^\mathbf{k}_{\bm{\tau} u}(\mathbf{r}) &\equiv&
  \frac{1}{\sqrt{\Omega_0}}\sum_\RR
  \varphi_{\bm{\tau} u}(\R-\bm{\tau}-\RR) e^{i \K\cdot\RR} \, ,
\end{eqnarray}
where $\RR$ is a lattice translation vector and
$\varphi_{\bm{\tau} u}(\R)$ is nonzero only within the augmentation
spheres centered at $\bm{\tau}$.
This formalism is applicable to both
LMTO and LAPW frameworks,
and eigenfunctions from both types of methods have been used~\cite{Miyake02}.

\subsection{The mixed product basis formalism}
\label{sec:mpb}

Two-particle quantities, such as Coulomb integrals and polarizability,
entail matrix elements of operators involving
four Bloch functions, a pair at $\mathbf{r}$ and another at $\mathbf{r}'$.
To efficiently reduce the number of required wave function indices from
4 to 2, a \textit{mixed product basis} (MPB) $\{M_I^\K\}$
can be introduced, consisting of products of
Bloch functions from Eqs. (\ref{eq:ipw_basis}) and (\ref{eq:aug_basis}).
This approach leverages the completeness of the MPB set,
which contains both plane wave and augmentation components,
making it an efficient choice for expanding products of ${\psi_{n,\K}}$.

\subsubsection{MPB in the interstitial region}
\label{sec:mpb_ipw}

The products involving the plane wave components of the set $\{M^\K_I\}$
naturally result in other plane waves. However,
by construction, the set of plane waves $\{P^\K_\G(\R)\}$ does not span the
entire space within the unit cell, leading to non-orthogonality, where the
overlap matrix deviates from the identity matrix.
A straightforward computational strategy to evaluate overlaps
and integrals involves the following definition
\begin{align}\label{eq:pw}
& \sqrt{\Omega_0}P^\K_\G(\R) = e^{i(\K+\G)\cdot\R} - \nonumber \\
& \,\,\,\,\,\,\,\,\,\,\,\, -
\sum_\RR \sum_r\sum_{L} P^{\K+\G}_{\bm{\tau}_r+\RR L}(\R)
\theta(s_r-\abs{\R-\bm{\tau}_{r,\RR}})
\end{align}
where $\R-\bm{\tau}_{r,\RR}$, with $\bm{\tau}_{r,\RR} = \bm{\tau}_r+\RR$,
accounts for the Bloch-sum of the non-periodic
function $P^{\K+\G}_{\bm{\tau}_r L}$ and for
the vector position $\R-\bm{\tau}_{r,\RR}$
inside the $r$-th augmentation sphere in the primitive unit cell.
The Heaviside step function $\theta(s_r-\abs{\R - \bm{\tau}_{r,\RR}})$
limits the integration domain
to the interstitial region. The plane wave component inside the
augmentation spheres is represented by
\begin{equation}\label{eq:pw_sphere}
 P^{\K+\G}_{\bm{\tau}_r L}(\R) = 4\pi i^l
e^{i(\K+\G)\cdot\bm{\tau}_r} j_l(\abs{\K+\G}r)
 Y_{lm}^*(\widehat{\K+\G}) Y_{lm}(\widehat{\R}) \, ,
\end{equation}
where $j_l(\abs{\K+\G}r)$ is the Bessel
function of the first kind of order $l$ within the standard convention,
$Y_{lm}(\widehat{\R})$ denotes the real spherical harmonic, and
$\widehat{\R}=\R/\abs{\R}$ is the unit vector in the direction of $\R$.
The quantity $\sum_{L} P^{\K+\G}_{\bm{\tau}_r L}$
corresponds to the expansion coefficients of the
plane wave $\exp[i(\K+\G)\cdot\R]$ within the $r$-th augmentation sphere.
Consequently, in the $l\to\infty$ limit, the contributions from
all the augmentation spheres vanish in Eq. (\ref{eq:pw}), thereby recovering
the definition given in Eq. (\ref{eq:ipw_basis}).

It is worth noting that when integrating $P^\K_\G(\R)$ over the primitive
unit cell, the Bloch-sum transformation of $P^{\K+\G}_{\bm{\tau}_r L}$ can be
bypassed by considering only the $\RR=\mathbf{0}$ contribution, which is
multiplied by the phase factor $\exp[i(\K+\G)\cdot\RR]$ when summing over
$\RR$. The Heaviside step function $\theta(s_r-\abs{\R-\bm{\tau}_{r,\RR}})$
also restricts the integration
to the volume of the $r$-th augmentation sphere, denoted as $\Omega_r$,

\subsubsection{MPB in the augmentation region}
\label{sec:mpb_aug}

Inside the $r$-th augmentation sphere,
the functions $\{B_{\bm{\tau}_r \mu L}^{\K}(\R)\}$ form the
radial components of the set $\{M_I^\K\}$. These are
Bloch sums of the product
basis set $\{B_{\bm{\tau}_r \mu L}(\R)\}$,
expressed as
\begin{equation}\label{eq:aug}
  B_{\bm{\tau}_r \mu L}^{\K}(\R) =
  \sum_\RR   \frac{e^{i\K\cdot\RR}}{\sqrt{\Omega_0}}
  B_{\bm{\tau}_r \mu L}(\R-\bm{\tau}_{r,\RR})
  \theta(s_r - \abs{\R-\bm{\tau}_{r,\RR}})  \, ,
\end{equation}
where $\mu$ indicates the index corresponding
to the $\mu$-th product basis function.
The radial component of the product functions, $B_{\bm{\tau}_r \mu l}(r)$,
satisfies orthonormality conditions and is computed from the
products $b_{\bm{\tau}_r \mu l}(r)= \varphi_{\bm{\tau}_r p l'}(r)
\varphi_{\bm{\tau}_r q l''}(r)$,
where the index $l$ ranges within $\abs{l'-l''}\le l \le l'+l''$ and
$\mu$ labels the combination $(p,q)$. The set of radial product
functions $\{ b_{\bm{\tau}_r \mu l}\}$ generally lacks orthonormality.
As shown in Ref. [\onlinecite{Kotani07}], we establish an orthonormal
set of basis functions via the linear combination
$B_{\bm{\tau}_r \mu l}= \frac{1}{\sqrt{p_\mu}} \sum_\nu z_{\mu\nu}
b_{\bm{\tau}_r \nu l}$, such that
\begin{align}
\int_0^{s_r} dr r^2 B_{\bm{\tau}_r \mu l} B_{\bm{\tau}_r \mu' l} &=
\sum_{\nu \nu'} \frac{z_{\mu\nu}z_{\mu'\nu'}}{\sqrt{p_\mu p_{\mu'}}}
\int_0^{s_r} dr r^2 b_{\bm{\tau}_r \nu l} b_{\bm{\tau}_r \nu' l} \nonumber \\
&= \frac{1}{\sqrt{p_\mu p_{\mu'}}} [\mathbf{z} \mathbf{O}_b
\mathbf{z}^{T}]_{\mu\mu'} = \delta_{\mu\mu'} \, ,
\end{align}
or, in matrix notation,
$\mathbf{O}_b \mathbf{z}^T = \mathbf{z}^T \mathbf{p}$.
Solving this eigenvalue problem provides the coefficients and prefactor
for expanding $B_{\bm{\tau}_r \mu L}$.
In the LMTO formalism, $B_{\bm{\tau}_r \mu l}(r)$
vanishes in the interstitial region, but may assume non-zero values
at the boundary of the augmentation spheres.

\subsubsection{The orthonormal MPB set $\{E_\mu^\K\}$}
\label{sec:Ebasis}

To describe the products of wave functions, one can define the MPB set
$\{M_I^\K\}\equiv\{P_{\G}^{\K},B^\K_{\bm{\tau} \mu L}\}$, where
the index $I\equiv\{\G,\bm{\tau} \mu L\}$ classifies the basis functions
associated with the interstitial and augmentation regions, respectively.
Due to the nature of the IPW basis functions, the MPB functions
are not orthogonal. To enforce orthonormality,
we introduce a second basis set, the {\it biorthogonal set}~\cite{friedrich09},
defined as
\begin{equation}\label{eq:mixbasis}
  \ket{\widetilde{M}_I^\K} = \sum_{I'}\ket{M_{I'}^\K} O^{\K,-1}_{I'I}
\end{equation}
where $O^\K_{II'}=\bra{M_{I}^\K}M_{I'}^\K\rangle_{\Omega_0}$ is the overlap
matrix, and the integration is performed over the primitive unit cell.
This ensures orthonormality between the two bases, expressed as
\be
\bra{\widetilde{M}_I^\K} M_J^\K \rangle_{\Omega_0} =
  \delta_{IJ} \, .
\ee

A significant simplification of the
mathematical formalism required to compute the main physical
quantities within Hedin's framework can be achieved by performing
a basis transformation $\{M_I^\K\}\to\{E_\mu^\K\}$ that
diagonalizes the Coulomb interaction matrix.
The \textit{orthonormal product basis
functions} can then be determined by evaluating the
eigenfunctions of the generalized eigenvalue problem
\begin{equation}\label{eq:eigen_prob}
\sum_J\bigl( v_{IJ}^\K - v_\mu^\K O^\K_{IJ}\bigr) w_{\mu J}^\K =0 \, ,
\end{equation}
where $v_{IJ}^\K = \bra{M_I^\K} \hat{v}^\K\ket{M_J^\K}$ is the Coulomb
interaction matrix and $v_\mu^\K$ its $\mu$-th eigenvalue.
Consequently, the Coulomb interaction can
be expanded in terms of the eigenvectors $w_{\mu J}^\K$, allowing us to express
the operator $\hat{v}$
\begin{equation}\label{eq:v_E_expansion}
\hat{v}^\K = \sum_\mu \ket{E_\mu^\K} v_\mu^\K \bra{E_\mu^\K} \, ,
\end{equation}
in terms of a new set $\{E_\mu^\K\}$
of orthonormal product basis functions defined as
\begin{align}\label{eq:Ebasis}
E_\mu^\K&=\sum_I w_{\mu I}^\K M^\K_I \nonumber \\
&= \sum_\G w_{\mu \G}^\K P^\K_\G + \sum_{r \nu L} w_{\mu, \bm{\tau}_r \nu L}^\K
 B_{\bm{\tau}_r \nu L}^{\K} \, ,
\end{align}
which diagonalize the Coulomb interaction
kernel $\hat{v}^\K$. In matrix form, Eq. (\ref{eq:eigen_prob}) can be rewritten as
$\widetilde{\mathbf{v}}^\K [\mathbf{W}^\K]^T = \mathbf{O}^\K [\mathbf{W}^\K]^T
\mathbf{v}^\K$, representing a generalized eigenvalue problem,
where $\widetilde{\mathbf{v}}_{IJ}^\K = v^\K_{IJ}$,
$\mathbf{W}^\K_{\mu J} = w_{\mu J}^\K = \bra{\widetilde{M}^\K_J}
E^\K_\mu\rangle_{\Omega_0}$
are the expansion coefficients,
and $\mathbf{v}^\K_{\mu\nu}=v_\mu^\K\delta_{\mu\nu}$ is the diagonal
Coulomb matrix.
Since $\mathbf{W}^\K$ is not a unitary matrix, its inverse can be
found by enforcing the orthonormality condition
for the set of basis functions $\{E_\mu^\K\}$
\begin{align}
  \bra{E_\mu^\K} E_\nu^\K \rangle_{\Omega_0} &= \sum_{IJ} w_{\mu I}^{\K\,*}
w_{\nu J}^{\K} \bra{M_I^\K} M_J^\K \rangle_{\Omega_0} \nonumber \\
&= \sum_{IJ} w_{\mu I}^{\K\,*} O^\K_{IJ} w_{\nu J}^{\K} =
\delta_{\mu\nu} \, ,
\end{align}
or, equivalently, $\mathbf{W}^{\K} \mathbf{O}^{\K\,*}
\mathbf{W}^{\K\,\dagger}=\mathbf{1}$, from
which it follows that
$[\mathbf{W}^{\K}]^{-1}= \mathbf{O}^{\K\,*} \mathbf{W}^{\dagger}
= \mathbf{O}^{\K\, T} \mathbf{W}^{\K \,\dagger}$,
given that the overlap matrix
$\mathbf{O}^\K$ is Hermitian, i.e. $\mathbf{O}^{\K\,*}=
[\mathbf{O}^{\K\,\dagger}]^*=\mathbf{O}^{\K\, T} $.

In the following, we will focus on expanding the electronic
inverse dielectric function
using both the biorthogonal set $\{M_I^\K\}$ and the
orthogonal set $\{E_\mu^\K\}$. In terms of the biorthogonal set,
the inverse dielectric function can be expressed as
\begin{equation}\label{eq:eps_inv_Mexpansion}
\varepsilon^{-1}_e(\R,\R';\omega) =
\frac{1}{N_\K} \sum_{\K\in\BZ} \sum_{IJ}
\widetilde{M}_I^\K(\R) \varepsilon^{-1}_{e,IJ}(\K;\omega) M_J^{\K\, *}(\R') \, ,
\end{equation}
where the matrix $\bm{\varepsilon}^{-1}_{e,\text{M}}(\K;\omega)$ is defined as
\be\label{eq:eps_inv_Mmatrix}
\varepsilon_{e,IJ}^{-1}(\K;\omega) =
\frac{1}{\Omega} \int_\Omega\int_\Omega d\R d\R'
M_I^{\K\,*}(\R) \varepsilon_e^{-1}(\R,\R';\omega) \widetilde{M}_J^\K(\R') \, .
\ee
The transformation of the inverse dielectric matrix
from the biorthogonal to the orthogonal basis,
$\bm{\varepsilon}^{-1}_{e,\text{M}}(\K;\omega) \to
\bm{\varepsilon}^{-1}_{e,\text{E}}(\K;\omega)$,
can be obtained using the completeness relation $\sum_I\ket{\widetilde{M}_I^\K}
\bra{M_I^\K} = \sum_I\ket{M_I^\K}\bra{\widetilde{M}_I^\K}=1$.
Thus, the matrix element of the inverse
dielectric operator in the new basis set reads
\begin{align}\label{eq:M_to_E_eps-inv}
\varepsilon^{-1}_{e,\mu\nu}(\K,\omega) &=
\elmat{E_\mu^\K}{\hat{\varepsilon}^{-1}_e(\omega)}{E_\nu^\K}_\Omega \nonumber \\
&= \sum_{IJ} \bra{E_\mu^\K} \widetilde{M}_I^\K\rangle_\Omega
\elmat{M_I^\K}{\hat{\varepsilon}_e^{-1}}{\widetilde{M}_J^\K}_\Omega
\bra{M_J^\K} E^\K_\nu \rangle_\Omega \nonumber \\
&= \sum_{IJK}  \varepsilon^{-1}_{e,IJ}(\K,\omega) O^\K_{JK} w_{\nu K}^{\K}
 w_{\mu I}^{\K\,*} \, ,
\end{align}
or in matrix notation
\be\label{eq:M_to_E_eps-inv_matrix}
\bm{\varepsilon}^{-1}_{e,\text{E}}(\K;\omega) = \mathbf{W}^{\K\,*}
\bm{\varepsilon}^{-1}_{e,\text{M}}(\K;\omega) \mathbf{O}^\K \mathbf{W}^{\K\,T}\, .
\ee
The subscripts M and E are used to distinguish the inverse dielectric matrix
expressed in terms of the biorthogonal basis set $\{M_I^\K\}$ and the orthogonal
basis set $\{E_\mu^\K\}$, respectively.
The inverse transformation is similarly straightforward and can be written as
\be\label{eq:E_to_M_eps-inv_matrix}
\bm{\varepsilon}^{-1}_{e,\text{M}}(\K;\omega) =
\mathbf{O}^\K \mathbf{W}^{\K\,T}
\bm{\varepsilon}^{-1}_{e,\text{E}}(\K;\omega) \mathbf{W}^{\K\,*}\, .
\ee

\subsection{Ladder diagrams in the Polarizability}
\label{sec:bse}

At the RPA level, the irreducible electron polarizability $P_e$ is
approximated by a bubble diagram, symbolically expressed as
$P_e\approx  \chi_e^0=-i\hbar \sum_\sigma G^\sigma_0 G^\sigma_0$.
The inverse electronic dielectric function $\varepsilon_e^{-1}$
is then computed from $P_e$
by inverting the electronic dielectric function (\ref{eq:electron_eps}).
The screened Coulomb interaction is evaluated in the standard manner as
$W_e = \varepsilon_e^{-1} v$,
where $v$ represents the bare Coulomb interaction.
In \texttt{Questaal}, $W_e$, $v$ and $\chi^0_e$
are represented in the biorthogonal mixed
basis $\{M^\K_{I}\}$ introduced in \sectionp \ref{sec:mpb}.

Furthermore, the electron polarizability $P_e(12)$ can be viewed as
a contraction of a more general four-point polarizability, expressed as
$P_e(12)=P_e(1122)=P_e(1324)\delta(1-3)\delta(2-4)$. Within the RPA,
this simplifies to $P_e(12)\approx
\chi^0_e(1324)\delta(1-3)\delta(2-4)$, where
\be\label{eq:chi0e}
\chi_e^0(1324)=-i\hbar \sum_\sigma G_0^\sigma(13)G_0^\sigma(42) \, .
\ee

To go beyond the RPA, it becomes necessary to evaluate
the vertex function $\Gamma$ provided in
Eq. (\ref{eq:vertex}), which involves the functional derivative
$\delta\Sigma_e/\delta G$, under the assumption of a negligible
\eph self-energy contribution.  However, computing
$\delta\Sigma_e/\delta G$ is highly challenging. Following
a common approximation, we neglect
$\delta W_e/\delta G$~\cite{onida_electronic_2002}, resulting in the
following expression for the electron polarizability
\begin{equation}
P_e(12)= \chi_e^0(12)-\int{\rm d}(34)\chi_e^0(1134)W_e(34)P_e(3422) \, ,
\label{eq:BSE_PW}
\end{equation}
exhibiting the structure of a Dyson equation.
When $W_e$ is assumed to be static, as is often the case,
Eq.~(\ref{eq:BSE_PW}), can be simplified to diagonalizing an effective
two-particle Hamiltonian rather than a computationally expensive inversion
in the geometric series of Eq. (\ref{eq:BSE_PW}). This is achieved by
introducing a basis of single-particle eigenfunctions that diagonalize
the RPA polarization.
While we omit the details here, specific information on its
implementation in \texttt{Questaal} can be found
in Ref. \onlinecite{Cunningham2023}.

Equation (\ref{eq:BSE_PW}) replaces the RPA electron polarizability
in the evaluation of the electronically screened Coulomb interaction.
Vertex corrections, approximated by ladder diagrams with a
static kernel $W_e^{\text{RPA}}(\omega = 0)$, enable the inclusion
of excitonic contributions to the electron self-energy in a
self-consistent manner. Incorporating these ladder
corrections within the QS\gw framework leads to the QS$G\widehat{W}$
approximation, as discussed in \sectionp \ref{sec:qsgw}.
Within this framework, the electronic dielectric function in
Eq. (\ref{eq:electron_eps}) is modified by vertex corrections,
which consequently influence the \eph matrix elements, as
these explicitly depend on $\varepsilon_e$.
This approach inherently integrates excitonic effects into
the evaluation of the \eph coupling.
The inclusion of vertex corrections
naturally enhances the accuracy of the \eph interaction by
improving the electronic screening.
The practical impact of this method
is illustrated in the case of graphene, as
discussed in \sectionp \ref{sec:results-ladders}.

We conclude this section by emphasizing that the BSE framework
has the potential to account for exciton-phonon coupling,
a topic of considerable importance and ongoing
research~\cite{cudazzo20,louie22,schebek24,adamska21,chen20,park22,paleari22}.
Most existing studies approach this problem by combining the \eph matrix
elements for electrons and holes with the eigenvectors of the two-particle
BSE Hamiltonian to evaluate the exciton-phonon coupling
without utilizing a field-theoretic framework.
In contrast, incorporating phonon vertex corrections through a
field-theoretic approach---such as including the static \eph contribution
($W^{\text{A}}_{ph}$) in the kernel of the two-particle Hamiltonian---could
offer a more comprehensive description of the exciton-phonon coupling.
This method inherently accounts for an infinite series of \eph vertex
diagrams within the BSE-corrected inverse dielectric matrix,
providing deeper insights into these interactions.

\section{The electron-phonon matrix elements in the mixed product
basis formalism}
\label{sec:eph_mpb}

Calculating the \eph matrix elements in Eqs. (\ref{eq:gmatrix})-(\ref{eq:xifun})
can be reformulated by expanding them in the
MPB and applying the identity in Eq. (\ref{eq:M_to_E_eps-inv_matrix}).
This leads to an expression for the reduced \eph matrix
elements (\ref{eq:ximat}) in terms of the inverse dielectric
matrix $\bm{\varepsilon}^{-1}_{e,\text{E}}(\K;\omega)$,
as given in Eq. (\ref{eq:M_to_E_eps-inv_matrix}),
\begin{align}\label{eq:ximat1}
& \xi_{in}^{r\alpha}(\Q,\K) =
\frac{1}{N_\K}\sum_{\K'\in\BZ}\sum_{\mu\nu}\sum_l
e^{i\Q\cdot\RR_l} \varepsilon_{e,\mu\nu}^{-1}(\K',0)  \times \nonumber \\
&\,\, \times \bra{\psi_{i,\K+\Q}} \psi_{n,\K} E^{\K'}_\mu\rangle_{\Omega_0}
\int_{\Omega} d\R E_\nu^{\K'\,*}(\R)
\frac{\partial V^{(0)}_{rl}(\R) }{\partial r_\alpha} \,,
\end{align}
where, due to Bloch's theorem, the conservation of crystal momentum
imposes the condition
$\bra{\psi_{i,\K+\Q}} \psi_{n,\K} E^{\K'}_\mu\rangle_{\Omega_0} =
\delta_{\K'\Q} \bra{\psi_{i,\K+\Q}} \psi_{n,\K} E^{\Q}_\mu\rangle_{\Omega_0}$.
The bare nuclear potential $V^{(0)}_{rl} (\R)$ can be expressed as an
inverse Bloch sum
\be\label{eq:bloch_sum_nuc_pot}
V_{rl}^{(0)}(\R) = \frac{1}{N_\K} \sum_{\K\in\BZ} e^{-i\K\cdot\RR_l}
V_r^{(0)\,\K}(\R) \, ,
\ee
enabling the integral over the BvK macrocrystal to be cast as
an integral over the unit cell
\be\label{eq:applying_bloch_sum_to_pot}
\int_{\Omega} d\R E_\nu^{\Q\,*}(\R)
\frac{\partial V^{(0)}_{rl}(\R)}{\partial r_\alpha}  =
e^{-i\Q\cdot\RR_l} \int_{\Omega_0} d\R E_\nu^{\Q\,*}(\R)
 \frac{\partial V^{(0)\,\Q}_{r}(\R)}{\partial r_\alpha} \, .
\ee
As a result of this periodic symmetry, the reduced \eph matrix elements
simplify to
\begin{eqnarray}\label{eq:ximat_final}
\xi_{in}^{r\alpha}(\Q,\K) &=&
\sum_{\mu\gamma} \varepsilon_{e,\mu\gamma}^{-1}(\Q)
\bra{\psi_{i,\K+\Q}}\psi_{n,\K} E_\mu^{\Q}\rangle_{\Omega_0} \times \nonumber \\
&&\,\, \times \int_{\Omega_0} d\R E_\gamma^{\Q\,*}(\R)
\frac{\partial V^{(0)\,\Q}_r(\R)}{\partial r_\alpha}
\, ,
\end{eqnarray}
which serves as the foundation for further analyses, including i)
separating the short- and long-range contributions to the \eph matrix
elements in the long-wavelength limit and ii) deriving an alternative
mathematical framework for the short-range \eph matrix elements, aimed at
simplifying their evaluation under rotations of phonon wave vectors.

\subsection{Scattering at long-wavelengths:
  short- and long-range contributions to the electron-phonon
  matrix elements}
\label{sec:long-range}

A critical challenge in modeling the \eph matrix elements for polar
materials---those where two or more atoms possess non-zero
Born effective charges~\cite{born_huang54}---arises from the non-analytic
behavior of the \eph matrix elements.
Within a field-theoretic formalism, starting from
Eq. (\ref{eq:ximat_final}), the \eph matrix elements can be naturally
decomposed into short- and long-range contributions in
the long-wavelength limit.
This separation, however, is not straightforward in DFPT,
where the KS potential does not inherently distinguish
between these two contributions.
In polar materials, longitudinal optical (LO) phonon modes at long-wavelengths
induce macroscopic electric fields that strongly couple with electrons
and holes. This coupling results in the Fr\"ohlich
interaction~\cite{frohlich54}, a key feature in polar materials.
Incorporating this interaction into
\abinitio calculations of \eph matrix elements is a relatively
recent achievement. This has been realized by introducing
an electrostatic long-range potential, which generalizes
Fr\"ohlich’s model to account for anisotropic lattices
and multiple phonon modes~\cite{verdi15}.

Recent advancements in analyzing the non-analytic behavior of \eph
matrix elements include contributions from long-range quadrupolar fields,
applicable to both polar and nonpolar materials~\cite{brunin20,jhalani20}.
Vogl’s~\cite{vogl76} mean-field dielectric approach also accounts
for dipole and quadrupole potentials, providing a more formal treatment
of long-range effects in \eph interactions.

This section focuses on deriving a Fr\"ohlich-like term as a long-range
correction to describe electrons coupling with dipolar electrostatic fields
within a field-theoretic framework.
Within the MPB formalism used in the LMTO framework,
addressing electronic polarization and dielectric response at long-wavelengths
is crucial. The basis transformation $\{M_I^\K\} \to \{E_\mu^\K\}$,
introduced in \sectionp \ref{sec:mpb}, simplifies this task by isolating the
divergence of the Coulomb matrix into a single eigenvalue\cite{friedrich10},
$v_1(\Q) = 4\pi e^2/q^2$. The corresponding eigenfunction, analytically
expressed as $E_1^\Q(\R) = \exp(i\Q\cdot\R)/\sqrt{\Omega_0}$, enables efficient
computational handling of terms such as $\bm{v}^{-1}\bm{\varepsilon}_e^{-1}$,
with the irreducible polarization function $\chi_e^0$ replacing the
standard polarization in Eq. (\ref{eq:electron_pol}), i.e.,
$P_e \approx \chi_e^0$.

The symmetrized inverse dielectric matrix
$\tilde{\bm{\varepsilon}}_{e,\text{E}}^{-1}$ within the RPA is given by
\be\label{eq:invseps}
 \mathbf{v}^{-1} \bm{\varepsilon}_{e,\text{E}}^{-1} =
 \mathbf{v}^{-1/2} \tilde{\bm{\varepsilon}}_{e,\text{E}}^{-1} \mathbf{v}^{-1/2}
     \, ,
\ee
where $\tilde{\bm{\varepsilon}}_{e,\text{E}}^{-1} =
\bigl[ \mathbf{1} -
  \mathbf{v}^{1/2} \bm{\chi}_{e,\text{E}}^0\mathbf{v}^{1/2}\bigr]^{-1}$
is Hermitian in the zero-frequency limit.
Here, $\bm{\chi}_{e,\text{E}}^0$ represents the irreducible polarizability
in the orthogonal basis set $\{E_\mu^\K\}$.
Incorporating the symmetrization (\ref{eq:invseps}) into
Eq. (\ref{eq:ximat_final}), the reduced \eph matrix elements
can be reformulated as
\be\label{eq:xi2}
\xi_{in}^{r\alpha}(\Q,\K) =
 \sum_{\mu\gamma} \tilde{\varepsilon}_{e,\mu\gamma}^{-1}(\Q)
\biggl(\frac{v_\mu^\Q}{v_\gamma^\Q}\biggr)^{\frac{1}{2}}
C^{\mu}_{in}(\K,\Q) \mathcal{I}^{r\alpha}_\gamma(\Q) \, ,
\ee
where we define
\begin{equation}\label{eq:c}
C^{\mu}_{in}(\K,\Q) \equiv
\bra{\psi_{i,\K+\Q}}\psi_{n,\K} E_\mu^{\Q}\rangle_{\Omega_0}
\end{equation}
and
\begin{equation}\label{eq:Imat}
\mathcal{I}^{r\alpha}_\gamma(\Q) \equiv \int_{\Omega_0} d\R E_\gamma^{\Q\,*}(\R)
\frac{\partial V^{(0)\,\Q}_r(\R)}{\partial r_\alpha} .
\end{equation}
In the long-wavelength limit, the symmetrized inverse dielectric matrix
$\tilde{\bm{\varepsilon}}_{e,\text{E}}^{-1}$ can be expressed using a
block-matrix formalism
\be\label{eq:eps_block}
\tilde{\bm{\varepsilon}}_{e,\text{E}}
= \begin{pmatrix}
1-\frac{4\pi e^2}{q^2}\chi_{e,11}^0      &&  \ub{b}^\dagger \\
\ub{b}                              &&  \uub{\tilde{\varepsilon}} \, ,
\end{pmatrix}
\ee
where non-bold quantities, such as $\chi_{e,11}^0(\Q)$, denote
the head element of the matrix, while the wings are indicated by bold
single-underlined quantities,
such as $\ub{b}\equiv -\frac{\sqrt{4 \pi e^2}}{q}\uub{v}^{1/2}
\ub{\chi}_e^0$, and bold double-underlined quantities refer to the body
of the matrix.

Inverting the matrix in Eq. (\ref{eq:eps_block}) using a block-matrix formalism,
the inverse of the symmetrized electronic dielectric matrix in the
orthonormal basis $\{E_\mu^\Q\}$ can be expressed in the
long-wavelength limit as
\begin{equation}\label{eq:inveps_q0}
\tilde{\bm{\varepsilon}}_{e,\text{E}}^{-1} \xrightarrow[q \to 0]{}
 \begin{pmatrix}
0             && \ub{0}^T   \\
\ub{0}        && \uub{\tilde{\varepsilon}}^{-1}
\end{pmatrix} +
\frac{1}{\tilde{\varepsilon}_{11}}
\begin{pmatrix}
1 && - \ub{b}^\dagger\,\uub{\tilde{\varepsilon}}^{-1} \\
- \uub{\tilde{\varepsilon}}^{-1} \ub{b}  &&
\uub{\tilde{\varepsilon}}^{-1} \ub{b} \ub{b}^\dagger\,
\uub{\tilde{\varepsilon}}^{-1}
\end{pmatrix} \, ,
\end{equation}
where
\begin{align}\label{eq:rho_q0}
\tilde{\varepsilon}_{11} &= 1-\tfrac{4\pi e^2}{q^2} \chi_{e,11}^0 -
\ub{b}^\dagger \uub{\tilde{\varepsilon}}^{-1} \ub{b} \nonumber \\
&= 1-\tfrac{4\pi e^2}{q^2} \bigl[\chi_{e,11}^0 +
  \ub{\chi}_e^{0\,\dagger} \uub{v}^{1/2}
  \uub{\tilde{\varepsilon}}^{-1} \uub{v}^{1/2}
\ub{\chi}_e^{0}\bigr]
\end{align}
is the \textit{electronic macroscopic dielectric constant} incorporating
\textit{crystal local field effects}
via $\ub{b}^\dagger \uub{\tilde{\varepsilon}}^{-1} \ub{b}$.
For insulators in the $\{E_\mu^\Q\}$ basis in the long-wavelength limit,
the head and wings of the electron polarizability
behave as $\sim q^2$ and $\sim q$, respectively,
while the body remains constant.
More specifically, in the frequency-dependent case, the head behaves as
$\chi_{e,11}^0(\Q,\omega) \sim
 q^2 \widehat{\Q}^T \mathbf{H}(\omega) \widehat{\Q}$, where $\mathbf{H}(\omega)$
is a $3\times 3$ matrix and $\widehat{\Q}=\Q/\abs{\Q}$.
The wings scale as $\ub{\chi}^{0\,\dagger}_{e,j}(\omega) \sim
 q \,\widehat{\Q}^T \mathbf{s}_j(\omega)$ and
$\ub{\chi}^0_{e,j}(\omega) \sim
 q \,\widehat{\Q}^T \mathbf{s}^*_j(\omega)$
for $j=2,3,\dots,n$, where
$\mathbf{s}_j(\omega)$ is a $3$-dimensional vector corresponding to the
$j$-th row wing element. In matrix notation, we can write
$\ub{\chi}^{0\,\dagger}_e \sim q \widehat{\Q}^T \mathbf{S}$ and
$\ub{\chi}_e^{0} \sim q\mathbf{S}^\dagger \widehat{\Q}$.
Equation (\ref{eq:inveps_q0}) can then be rewritten in a more compact form as
\begin{equation}\label{eq:inveps_q0_2}
\tilde{\bm{\varepsilon}}_{e,\text{E}}^{-1} \xrightarrow[q \to 0]{}
 \begin{pmatrix}
0             && \ub{0}^T   \\
\ub{0}        && \uub{\tilde{\varepsilon}}^{-1}
\end{pmatrix} +
\begin{pmatrix}
\tilde{\varepsilon}_{11}^{-1} &&  \tilde{\ub{\varepsilon}}^{-1\,\dagger} \\
\tilde{\ub{\varepsilon}}^{-1}  && \tilde{\varepsilon}_{11}
\tilde{\ub{\varepsilon}}^{-1} \tilde{\ub{\varepsilon}}^{-1\,\dagger}
\end{pmatrix} \, ,
\end{equation}
where Eq. (\ref{eq:rho_q0}) has the form
\begin{align}\label{eq:macr_eps}
\tilde{\varepsilon}_{11} &=
1-4\pi e^2\bigl[\widehat{\Q}^T \mathbf{H} \widehat{\Q} +
\widehat{\Q}^T \mathbf{S} \uub{v}^{1/2} \uub{\tilde{\varepsilon}}^{-1}
\uub{v}^{1/2} \mathbf{S}^\dagger \widehat{\Q}\bigr] \nonumber \\
&= \widehat{\Q}^T \bm{\varepsilon}_\infty \widehat{\Q}
\nonumber \\
\end{align}
with $\widehat{\Q}^T \bm{\varepsilon}_\infty \widehat{\Q} = \sum_{\alpha\beta}
\widehat{q}_\alpha \varepsilon_\infty^{\alpha\beta} \widehat{q}_\beta$ and the
\textit{high-frequency dielectric permittivity tensor}
\begin{equation}\label{eq:high-freq_eps}
\bm{\varepsilon}_\infty = 1-4\pi e^2 \mathbf{H} - 4\pi e^2 \mathbf{S}
\uub{v}^{1/2} \uub{\tilde{\varepsilon}}^{-1} \uub{v}^{1/2} \mathbf{S}^\dagger.
\end{equation}
Using this notation, the wings become
\begin{align}\label{eq:roweps}
\tilde{\ub{\varepsilon}}^{-1} &= - \tilde{\varepsilon}^{-1}_{11}
\uub{\tilde{\varepsilon}}^{-1} \ub{b} \nonumber \\
&= \sqrt{4\pi e^2}\tilde{\varepsilon}^{-1}_{11}
\uub{\tilde{\varepsilon}}^{-1} \uub{v}^{1/2} \mathbf{S}^\dagger \widehat{\Q}\, ,
\end{align}
and
\begin{equation}\label{eq:coleps}
\tilde{\ub{\varepsilon}}^{-1\,\dagger} =
\sqrt{4\pi e^2}\tilde{\varepsilon}^{-1}_{11} \widehat{\Q}^T \mathbf{S}
\uub{v}^{1/2} \uub{\tilde{\varepsilon}}^{-1} \, .
\end{equation}

The block-matrix formulation (\ref{eq:inveps_q0_2}) allows for a
reformulation of the reduced \eph coupling matrix from Eq. (\ref{eq:xi2}) in the
long-wavelength limit as
\begin{widetext}
\be\label{eq:xi3}
\xi^{r\alpha}_{in}(\K,\Q) \xrightarrow[q \to 0]{}
\sum_{\mu\ne1}\sum_{\gamma\ne 1}
\tilde{\uub{\varepsilon}}_{\mu\gamma}^{-1}(\Q)
\biggl(\frac{v_\mu^\Q}{v_\gamma^\Q}\biggr)^{\frac{1}{2}}
C^{\mu}_{in}(\K,\Q) \mathcal{I}_\gamma^{r\alpha}(\Q) +
\tilde{\varepsilon}_{11}(\widehat{\Q}) \biggl(\sum_{\mu}
\tilde{\varepsilon}_{e,\mu 1}^{-1}(\widehat{\Q}) \sqrt{v_\mu^\Q}
C^{\mu}_{in}(\K,\Q)\biggr)\biggl(
\sum_{\gamma} \tilde{\varepsilon}_{e,1\gamma}^{-1}(\widehat{\Q})
\frac{\mathcal{I}_\gamma^{r\alpha}(\Q)}{\sqrt{v_\gamma^\Q}}
\biggr) \, .
\ee
The second term on the right-hand side is
non-analytic, and its behavior depends on the direction in which
the limit $\Q\to 0$ is approached. This term reflects the presence of
a long-range interaction with a macroscopic field associated with the
LO phonon modes, contributing to the reduced \eph matrix
with diverging behavior. As a result, the reduced \eph coupling matrix
can be decomposed into short- and long-range components, i.e.,
$\xi^{r\alpha}_{in} =\xi^{r\alpha,\mathcal{S}}_{in} +
\xi^{r\alpha,\mathcal{L}}_{in}$, with the
\textit{analytic} short-range term given by
\begin{equation}\label{eq:sxi}
\xi^{r\alpha,\mathcal{S}}_{in}(\K,\Q) \xrightarrow[q \to 0]{}
 \sum_{\mu\ne1}\sum_{\gamma\ne 1}
\tilde{\uub{\varepsilon}}_{\mu\gamma}^{-1}(\Q)
\biggl(\frac{v_\mu^\Q}{v_\gamma^\Q}\biggr)^{\frac{1}{2}}
C^{\mu}_{in}(\K,\Q) \mathcal{I}_\gamma^{r\alpha}(\Q)
\end{equation}
and the \textit{non-analytic} long-range term defined as
\begin{equation}\label{eq:lxi}
\xi^{r\alpha,\mathcal{L}}_{in}(\K,\Q) \xrightarrow[q \to 0]{}
\tilde{\varepsilon}_{11}(\widehat{\Q}) \biggl(\sum_{\mu}
\tilde{\varepsilon}_{e,\mu 1}^{-1}(\widehat{\Q}) \sqrt{v_\mu^\Q}
C^{mu}_{in}(\K,\Q)\biggr)\biggl(
\sum_{\gamma} \tilde{\varepsilon}_{e,1\gamma}^{-1}(\widehat{\Q})
\frac{1}{\sqrt{v_\gamma^\Q}}
\mathcal{I}_\gamma^{r\alpha}(\Q)\biggr) \, .
\end{equation}

We now turn out our attention to the \textit{long-range reduced \eph
matrix element} (\ref{eq:lxi}), and in particular to the first term
in the summation over the basis index $\gamma$
\be\label{eq:z1}
\tilde{\varepsilon}_{11}(\widehat{\Q})
\sum_{\gamma}
\frac{\tilde{\varepsilon}_{e,1\gamma}^{-1}(\widehat{\Q})}{\sqrt{v_\gamma^\Q}}
\mathcal{I}_\gamma^{r\alpha}(\Q) 
= \frac{q}{\sqrt{4\pi}} \int_{\Omega_0}d\R E_1^{\Q\,*}(\R)
\frac{\partial V^{(0)\,\Q}_r(\R)}{\partial r_\alpha}   + 
\tilde{\varepsilon}_{11}(\widehat{\Q})
\sum_{\gamma\ne 1} \frac{ \tilde{\varepsilon}_{e,1\gamma}^{-1}(\widehat{\Q})}
{\sqrt{v_\gamma^\Q}} \int_{\Omega_0}d\R E_\gamma^{\Q\,*}(\R)
\frac{\partial V^{(0)\,\Q}_r(\R)}{\partial r_\alpha}  \, .
\ee
\end{widetext}
In the long-wavelength limit, the first term on the right-hand side
can be simplified by using the explicit definition
of the product basis function
$E_1^\Q = \exp[-i\Q\cdot\R]/\sqrt{\Omega_0}$
\begin{align}\label{eq:non_analytic_first_term}
& \int_{\Omega_0}d\R E_1^{\Q\,*}(\R)
\frac{\partial V^{(0)\,\Q}_r(\R) }{\partial r_\alpha} =
\int_{\Omega_0}d\R \frac{e^{-i\Q\cdot\R}}{\sqrt{\Omega_0}}
\frac{\partial V^{(0)\,\Q}_r(\R)}{\partial r_\alpha}  \nonumber \\
& \,\,\,\,\,\,\,\,\,\,\,\,\,\,\,\,\,\,\,
= -\frac{i 4\pi Z_r e \widehat{q}_\alpha}{\sqrt{\Omega_0} q}
e^{-i\Q\cdot\bm{\tau}_r} \xrightarrow[q \to 0]{} -
\frac{i 4\pi Z_r e \widehat{q}_\alpha}
{\sqrt{\Omega_0} q} \, ,
\end{align}
where we have combined the inverse Bloch sum
$V^{(0)\,\Q}_r(\R) = \sum_l V^{(0)}_{rl}(\R) \exp(i\Q\cdot\RR_l)$ with the
inverse Fourier transform of the nuclear potential $V^{(0)}_{rl}$, as
outlined in Eq. (S.6) of the \suppinfo.
Combining this result with Eqs. (\ref{eq:coleps}), (\ref{eq:z1}),
and (\ref{eq:invseps}), and using
$\tilde{\uub{\varepsilon}} = \mathbf{1} - \uub{v}^{1/2}
\uub{\chi}_e^0 \uub{v}^{1/2}$, we find
\be\label{eq:z3}
\tilde{\varepsilon}_{11}(\widehat{\Q})
\sum_{\gamma}
\frac{\tilde{\varepsilon}_{e,1\gamma}^{-1}(\widehat{\Q})}{\sqrt{v_\gamma^\Q}}
\mathcal{I}_\gamma^{r\alpha}(\Q) =
-i \sqrt{\frac{4\pi}{\Omega_0}} e
\sum_\beta \widehat{q}_\beta Z_{r}^{\beta\alpha} \, ,
\ee
where $\bm{Z}_{r}$ is the {\it Born effective charge tensor}, with
components defined as
\be\label{eq:born-charge}
Z_{r}^{\beta\alpha} = Z_r \delta_{\alpha\beta} + Z_{r}^{el,\beta\alpha} \, .
\ee
Here, the second term accounts for the contribution to the
classical nuclear charge $Z_r$
arising from the polarization of the electron density in the $\Q\to 0$ limit
\be\label{eq:born-charge-el}
Z_{r}^{el,\beta\alpha} = i \frac{\sqrt{\Omega_0}}{e}
\sum_{\substack{\eta\ne 1 \\ \gamma\ne 1}}
\mathbf{S}_{\beta\eta} \, \uub{\varepsilon}_{\eta\gamma}^{-1}(\Q)
\int_{\Omega_0}d\R E_\gamma^{\Q\,*}(\R)
 \frac{\partial V^{(0)\,\Q}_r(\R)}{\partial r_\alpha}  \, .
\ee
This expression mirrors the typical structure
of a linear response problem, closely resembling the
short-range \eph coupling matrix element in Eq. (\ref{eq:sxi}), with
$i \mathbf{S}_{\beta\eta}$ substituting $C^{\eta}_{in}(\K,\Q)$.
It aligns with Eq. (3.6) of Ref. \onlinecite{vogl76},
which employs a dielectric approach.

In non-polar materials or certain
polar diatomic materials with
specific symmetries, the electronic screening cancels out the
classical nuclear charge, yielding
$Z_{r}^{el,\beta\alpha} = -Z_r\delta_{\beta\alpha}$,
and the Born effective charge tensor components $Z_{r}^{el,\beta\alpha}$
become zero. Conversely, in polar materials, external electric fields
associated with LO phonon modes induce
electronic polarization, leading to over- or under-screening
in certain regions of the primitive unit cell.
Consequently, the Born effective charge tensor components deviate from
zero, although they remain constrained by the charge neutrality condition,
i.e., summing to zero over the unit cell
\be\label{eq:neutrality_condition}
\sum_r Z_{r}^{\beta\alpha} = 0 \, .
\ee

Finally, by combining Eqs. (\ref{eq:z3}) and (\ref{eq:lxi}),
with the definitions in Eq. (\ref{eq:roweps}) for
$\tilde{\varepsilon}_{e,\mu 1}^{-1}$, Eq. (\ref{eq:c}) for the coefficients
$C^{\mu}_{in}$, and Eq. (\ref{eq:macr_eps}) for the high-frequency
dielectric permittivity tensor, we obtain the expression for the
long-range reduced \eph matrix element as follows
\begin{widetext}
\begin{equation}\label{eq:xi-long-range}
\xi^{r\alpha,\mathcal{L}}_{in}(\K,\Q) \xrightarrow[q \to 0]{}
- i\, \frac{4\pi e}{\Omega_0} \biggl[\frac{1}{q} 
\elmat{\psi_{i,\K}}{e^{i\Q\cdot\R}}{\psi_{n,\K}}_{\Omega_0}
  + \sqrt{\Omega_0}\sum_{\mu\ne 1}\sum_{\eta\ne 1}\sum_{\gamma}
\tilde{\uub{\varepsilon}}_{\mu \eta}^{-1}(\Q) \sqrt{v_\eta^\Q}\,
\mathbf{S}^{*}_{\gamma\eta} \widehat{\Q}_\gamma
\bra{\psi_{i,\K}}\psi_{n,\K} E_\mu^{\Q}\rangle_{\Omega_0}\biggr]
\frac{ \sum_\beta \widehat{q}_\beta Z_{r}^{\beta\alpha}}
{\widehat{\Q}^T \bm{\varepsilon}_\infty \widehat{\Q}} \, ,
\end{equation}
\end{widetext}
where the projection coefficient
in the leading diverging term behaves as
$\elmat{\psi_{i,\K}}{e^{i\Q\cdot\R}}{\psi_{n,\K}}_{\Omega_0} \sim \delta_{in} +
\mathcal{O}(\Q)$ in the $\Q\to 0$ limit, owing to the orthonormality
of the Bloch states---here normalized to the unit cell.
While the first term in Eq. (\ref{eq:xi-long-range}) diverges as
$\sim 1/q$, the second term yields a finite value. Despite this,
the second term should not be disregarded, as the square modulus of
the \eph matrix elements, which typically exhibit a divergence
of $\sim 1/q^2$, is often the quantity of interest.
Additional contributions may arise from mixing terms scaling
as $\sim 1/q$. Although the second term can be
neglected when $\Q=0$, it may provide significant contributions
within the $\Gamma$ cell of the BZ.

Consequently, the \eph coupling matrix in Eq. (\ref{eq:gmatrix}) decomposes
into the sum of short-range and long-range components
\begin{align}\label{eq:gmatrix_lr_sr}
g_{in,\nu}(\K,\Q) &\xrightarrow[q \to 0]{} \sum_{r,\alpha}
\sqrt{\frac{\hbar}{2 m_r \omega_{\Q\nu}}}
e_{r\alpha,\nu}(\Q)\times \nonumber \\
& \,\,\,\,\,\,\,\,\,\,\,\,\,
\times \, \biggl[\xi_{in}^{r\alpha,\mathcal{S}}(\Q,\K) +
 \xi_{in}^{r\alpha,\mathcal{L}}(\Q,\K) \biggr] \nonumber \\
&= g_{in,\nu}^{\mathcal{S}}(\K,\Q) + g^{\mathcal{L}}_{in,\nu}(\K,\Q) \, .
\end{align}
Note that the leading term in $g^{\mathcal{L}}_{in,\nu}(\K,\Q)$
\be\label{eq:long_range_g}
-i \sum_{r}\frac{4\pi e}{q \Omega_0}
\sqrt{\frac{\hbar}{2 m_r \omega_{\Q\nu}}}
\frac{\widehat{\Q}\cdot\bm{Z}_r\cdot\bm{e}_{r,\nu}(\Q)}
{\widehat{\Q}^T \bm{\varepsilon}_\infty \widehat{\Q}}
\ee
exhibits the same structure as the leading dipolar term described in
Ref. \onlinecite{vogl76} using a dielectric approach and in Ref.
\onlinecite{verdi15} using an electrostatic formalism. In the latter,
the anisotropic Poisson's equation is solved for the dipole-induced
electrostatic potential, screened by the
high-frequency electronic permittivity (\ref{eq:high-freq_eps}).
The non-diverging term in Eq. (\ref{eq:xi-long-range}) arises from
crystal local-field corrections, accounting for the influence
of local electrostatic fields generated by the electron density's response
to the macroscopic perturbing field associated with LO phonon modes.

It is important to highlight that transverse optical (TO) phonon modes
do not contribute to the long-range term. This is because the scalar product
$\widehat{\Q}\cdot\bm{Z}_r \cdot \bm{e}_{r,\nu}(\Q)$ vanishes for these modes,
due to the orthogonality between the TO polarization vector and
the wave vector $\Q$. This cancellation is particularly evident when the tensor
$\bm{Z}_r$ is isotropic or aligned with the propagation direction of the
phonon modes. In contrast, LO modes induce electronic polarization
and contribute significantly to long-range electrostatic interactions.

In practice, the long-range contribution $g^{\mathcal{L}}_{in,\nu}$
should be combined with the short-range term
$g^{\mathcal{S}}_{in,\nu}$ only after interpolating from a coarse
to a fine BZ mesh, which is commonly done to reduce computational cost.
Currently, $g_{in,\nu}^{\mathcal{S}}$ is computed using the reduced \eph matrix
elements derived from Eq. (\ref{eq:sxi}). In contrast, the \eph
matrix elements calculated within the framework of DFPT inherently incorporate
the long-range contribution in the long-wavevelength
limit\cite{verdi15}, as the formalism evaluates changes
in the effective KS potential
\begin{align}
\partial_{r\alpha} v^{\text{KS}}(\R) &=
\partial_{r\alpha} V_{n}(\R) + \nonumber \\
&\,\,\,\,\,\,\,+ \int_\Omega d\R' \biggl[v(\R-\R') + f_{xc}(\R,\R')\biggr]
\partial_{r\alpha} n(\R')  \nonumber \\
&= \int_\Omega d\R' \varepsilon^{-1}_{e,\text{TDDFT}}(\R,\R')
\partial_{r\alpha} V_{n}(\R') \, .
\end{align}
As the head of the Coulomb kernel in reciprocal space scales
as $\sim 1/q^2$ in the long-wavelength limit, it is necessary
within DFPT (i) to subtract the long-range term in the $\Q \to 0$ limit
before applying the interpolation scheme, and then (ii) to reintroduce
$g^{\mathcal{L}}_{in,\nu}$ at the end of the interpolation.
Notably, the \xc kernel $f_{xc}$ in standard LDA/GGA
approximations does not exhibit the correct
$\sim 1/q^2$ behavior for insulators in the long-wavelength limit,
leading to a mismatch. Consequently, the \xc contribution, which is removed
when subtracting the long-range term (\ref{eq:long_range_g}) and implicitly
included into the Born effective charge tensor
$Z_r^{\beta\alpha}$ and high-frequency dielectric permittivity tensor
$\varepsilon_\infty^{\alpha\beta}$, lacks an equivalent counterpart
in the non-analytical component of $\varepsilon^{-1}_{e,\text{TDDFT}}$.
This results in a \textit{spurious removal} of the non-analytic \xc contribution
within the short-range \eph matrix elements.

In the following sections, we focus exclusively on the short-range \eph
matrix elements $g^{\mathcal{S}}_{in,\nu}$.
A thorough discussion of the long-range correction will be
addressed in future works.

\subsection{Irreducible wedge of the Brillouin zone}
\label{sec:bz}

The evaluation of the short-range \eph matrix elements,
as defined in Eqs. (\ref{eq:sxi}) and (\ref{eq:gmatrix_lr_sr}),
is computationally intensive, with the most demanding step being the
computation of the electronic inverse dielectric matrix
$\bm{\varepsilon}^{-1}_{e,\text{E}}(\Q)$ for each phonon wave vector
$\Q$ in the coarse BZ mesh. Therefore, it is crucial to limit the calculations
to the irreducible wedge of the Brillouin zone (IBZ),
thereby reducing redundancies. The irreducible $\Q$-points are determined
by identifying vectors that are inequivalent under the symmetry
operations of the crystal point group, denoted $\{\mathcal{S}\}$ and labeled
following the Seitz convention~\cite{litvin11} as
\begin{align}\label{eq:def_vbar}
\bm{\tau}_r = \mathcal{S}\bm{\tau}_s &\equiv
\{\rot|\bm{v}\} \bm{\tau}_s + \bar{\bm{v}}_{\bm{\tau}_s}(\mathcal{S})
\nonumber \\
&= \rot\bm{\tau}_s + \bm{v}(\mathcal{S}) +
\bar{\bm{v}}_{\bm{\tau}_s}(\mathcal{S}),
\end{align}
where $\rot$ represents a proper or improper rotation,
$\bm{v}(\mathcal{S})$ a fractional translation, and
$\bar{\bm{v}}_{\bm{\tau}_s}(\mathcal{S})$ a lattice vector
translation required to map the rotated
vector position $\{\rot|\bm{v}\} \bm{\tau}_s$
onto an equivalent vector $\bm{\tau}_r$ within the same sublattice.
Once a $\Q$-point in the IBZ is identified, its entire \textit{star} is
generated by applying all crystal symmetry operations.

The \eph matrix elements for phonon momentum transfers $\rot\Q$
within the star of $\Q\in\text{IBZ}$ can then be determined
by leveraging the transformation properties of vibrational eigenmodes under
symmetry operations. This is governed by the established transformation
rule~\cite{giustino07}
\be\label{eq:original_rotation}
g^{\mathcal{S}}_{in,\nu}(\K,\rot\Q) = g^{\mathcal{S}}_{in,\nu}(\rot^{-1}\K,\Q) \, .
\ee
A proof of this relation within a field-theoretic framework is provided
in \sectionp S.2 of the \suppinfo. In practice, applying
Eq. (\ref{eq:original_rotation}) requires knowledge of
$\bm{g}^{\mathcal{S}}_\nu(\K,\Q)$ across the full set of electronic
wave vectors $\K$ in the BZ, even when only a subset of $\K$-points
in the IBZ is of primary interest. To address this challenge,
the subsequent section introduces an alternative mathematical formulation
for the short-range \eph matrix elements, designed to simplify their
evaluation under rotations of the phonon wave vectors.

\subsubsection{Expansion of the \eph matrix elements
using projection coefficients expressed in terms of biorthogonal basis
functions}
\label{sec:Giv_q}

In this section, we reformulate Eqs. (\ref{eq:gmatrix})-(\ref{eq:xifun}) of
\sectionp \ref{sec:Wph} by projecting onto the biorthogonal MPB functions,
$\{M_I^\Q\}$, leveraging the analytical properties of
rotations in the BZ. In contrast,
when using the orthonormal set $\{E_\mu^\Q\}$, a more intricate algorithm
is required to account for the
rotation rules for the eigenvectors $\{w_{\mu I}^\Q\}$.

By combining the definition of the reduced \eph matrix elements
(\ref{eq:ximat})-(\ref{eq:xifun}), the expansion of the electronic
inverse dielectric function in terms of the biorthogonal MPB functions
as given by Eq. (\ref{eq:eps_inv_Mmatrix}), and the identity
in Eq. (\ref{eq:mixbasis}), we obtain
\be\label{eq:xi_reduced_new}
\xi^{r\alpha}_{in}(\K,\Q)
= \sum_I \zeta^{r\alpha}_{\Q,I}
\bra{\psi_{i,\K+\Q}} \psi_{n,\K} M_I^{\Q}\rangle_{\Omega_0} \, ,
\ee
where the coefficients $\zeta^{r\alpha}_{\Q,I}$ are defined as
\begin{eqnarray}\label{eq:zeta}
  \zeta^{r\alpha}_{\Q,I} &=&
  \sum_{J}\biggl(\sum_{I'} O_{I I'}^{\Q\,-1} \varepsilon_{e,I'J}^{-1}(\Q)\biggr)
  \times \nonumber \\ && \,\,\,\,\,\,\,\,\,\,\,\,\,\,\,\,\,\,\,\,\,\, \times
  \int_{\Omega_0} d\R\, M_J^{\Q\,*}(\R)
\frac{\partial V_r^{(0)\,\Q}(\R)}{\partial r_{\alpha}}  \, .
\end{eqnarray}
Here, $\bigl(\mathbf{O}^{\Q\,-1} \bm{\varepsilon}_{e,\text{M}}^{-1}\bigr)_{IJ}$
can be rewritten using the identity in Eq. (\ref{eq:E_to_M_eps-inv_matrix}) as
\be\label{eq:invO_invEM}
\bigl(\mathbf{O}^{\Q\,-1} \bm{\varepsilon}_{e,\text{M}}^{-1}\bigr)_{IJ} =
\bigl(\mathbf{W}^{\Q,T} \bm{\varepsilon}^{-1}_{e,\text{E}}(\Q) \mathbf{W}^{\Q,*}
\bigr)_{IJ} \, .
\ee
Equations (\ref{eq:xi_reduced_new}) and (\ref{eq:zeta}) give rise to the
\eph coupling matrix elements given in Eq. (\ref{eq:gmatrix_lr_sr}).
However, it can be straightforwardly demonstrated
that the short-range contribution
to the reduced \eph matrix elements, $\xi^{r\alpha,\mathcal{S}}_{in}(\K,\Q)$, can
be derived within this formalism by replacing
Eq. (\ref{eq:invO_invEM}) with
\be\label{eq:invO_invEM_sr}
\bigl(\mathbf{O}^{\Q\,-1} \bm{\varepsilon}_{e,\text{M}}^{-1}\bigr)_{IJ} \to
\sum_{\mu\ne 1}\sum_{\gamma\ne 1} w^{\Q}_{\mu I} \varepsilon^{-1}_{e,\mu\gamma}(\Q)
w^{\Q\,*}_{\gamma J} \, .
\ee
To determine the transformation rules for the augmentation and
interstitial region contributions to $\xi^{r\alpha}_{in}(\K,\rot\Q)$,
and consequently to $\xi^{r\alpha,\mathcal{S}}_{in}(\K,\rot\Q)$,
we introduce the following quantity
\begin{align}\label{eq:eps_inv_Mexpansion2}
\bar{\varepsilon}_{e,IJ}^{-1}(\Q) &=
\bigl(\mathbf{O}^{\Q\,-1} \bm{\varepsilon}_{e,\text{M}}^{-1}\bigr)_{IJ}
\nonumber \\
&= \frac{1}{\Omega} \int_\Omega \int_\Omega d\R\,d\R'
\widetilde{M}_I^{\Q\,*}(\R) \varepsilon^{-1}_{e}(\R,\R';0)
\widetilde{M}_J^{\Q}(\R') \, , \nonumber \\
\end{align}
instead of using Eq. (\ref{eq:invO_invEM_sr}).
Substituting Eq. (\ref{eq:eps_inv_Mexpansion2}) into the definition
of the coefficients $\zeta^{r\alpha}_{\Q,I}$, we arrive at
\be\label{eq:zeta_new}
\zeta^{r\alpha}_{\Q,I} =
  \sum_{J} \bar{\varepsilon}_{e,IJ}^{-1}(\Q)
  \int_{\Omega_0} d\R\, M_J^{\Q\,*}(\R)
\frac{\partial V_r^{(0)\,\Q}(\R)}{\partial r_{\alpha}}  \, .
\ee
Finally, substituting this result into the definition of
reduced \eph matrix element in Eq. (\ref{eq:xi_reduced_new}) and of
\eph matrix element in Eq. (\ref{eq:gmatrix}), we derive
\be\label{eq:eph2}
g_{in,\nu}(\K,\Q) =  \sum_I G_{I\nu}^\Q \bra{\psi_{i,\K+\Q}} \psi_{n,\K} M_I^{\Q}
\rangle_{\Omega_0}
\ee
where the coefficients $G_{I\nu}^\Q$ are defined as
\be\label{eq:Gdef}
G_{I\nu}^\Q = \sum_{r\alpha}
\sqrt{\frac{\hbar}{2 m_r \omega_{\Q\nu}}}
e_{r\alpha,\nu}(\Q) \, \zeta^{r\alpha}_{\Q,I} \, .
\ee
Equation (\ref{eq:eph2}) provides an expansion of the \eph matrix elements,
$g_{in,\nu}(\K,\Q)$, in terms of projection coefficients involving the
biorthogonal MPB functions $\{M_I^\Q\}$. The coefficients in
Eq. (\ref{eq:Gdef}) also enable the expansion of the \eph coupling
function in terms of the biortogonal MPB functions as
$g_{\Q\nu}(\R) = \sum_I G_{I\nu}^\Q M_I^\Q(\R)$.

In \sectionp S.3 of the \suppinfo, we demonstrate that the rotated \eph
matrix $\bm{g}_\nu(\K,\rot\Q)$, corresponding to a phonon momentum transfer
$\rot\Q$, can be computed directly from the unrotated coefficients
$G_{I\nu}^\Q$. This allows for an efficient computation of the \eph coupling
matrix elements $g_{in,\nu}(\K,\rot\Q)$ using only rotations of the projection
coefficients $\bra{\psi_{i,\K+\Q}} \psi_{n,\K} M_I^{\Q}\rangle_{\Omega_0}$.

Accordingly, Eq. (\ref{eq:Gdef}) can be alternately rewritten as
\be\label{eq:GIVmat}
G_{I\nu}^\Q = \sum_J \bar{\varepsilon}_{e,IJ}^{-1}(\Q) \Pi_{J\nu}^\Q \, ,
\ee
with
\begin{align}\label{eq:PIdef}
\Pi_{J\nu}^\Q &=
\sum_{r\alpha} \sqrt{\frac{\hbar}{2 m_r \omega_{\Q\nu}}}
e_{r\alpha,\nu}(\Q) \times \nonumber \\ & \,\,\,\,\,\,\,\,\,\,\,\,\,\,\,\times
\int_{\Omega_0} d\R\, M_J^{\Q\,*}(\R)
\frac{\partial V_r^{(0)\,\Q}(\R)}{\partial r_\alpha}  \, .
\end{align}
The \suppinfo\ further analyzes the transformation rules for
$\bar{\varepsilon}_{e,IJ}^{-1}(\rot\Q)$ and $\Pi_{J\nu}^{\rot\Q}$ when evaluating
$G_{I\nu}^\Q$ under symmetry operations.
Additionally, Eqs. (\ref{eq:eph2}) and (\ref{eq:GIVmat}) sum over
contributions from IPWs, $\{P_\G^\Q\}$, and atomic
sphere augmentation functions, $\{B^\Q_{\bm{\tau} a l m}\}$.
The \eph matrix elements can thus be decomposed into interstitial (\textit{ipw})
and augmentation (\textit{aug}) contributions
\begin{equation}\label{eq:gmat_decomposition}
g_{in,\nu}(\K,\rot\Q) = g^{ipw}_{in,\nu}(\K,\rot\Q) +
g^{aug}_{in,\nu}(\K,\rot\Q),
\end{equation}
with the interstitial term defined as
\begin{align}\label{eq:eph_ipw}
&g^{ipw}_{in,\nu}(\K,\rot\Q) =
\sum_{\G} G_{\G,\nu}^{\Q}\times \nonumber \\&\,\,\,\,\,\,\times
\biggl( e^{-i\rot(\Q+\G)\cdot \bm{v}(\mathcal{S})}
\bra{\psi_{i,\K+\rot\Q}} \psi_{n,\K}
P_{\rot\G}^{\rot\Q} \rangle_{\Omega_0} \biggr)  \, ,
\end{align}
and the augmentation contribution as
\begin{align}\label{eq:eph_aug_final}
& g^{aug}_{in,\nu}(\K,\rot\Q) =
\sum_{\bm{\tau} a l \eta}
G_{\bm{\tau} a l \eta, \nu}^{\Q} \times
\nonumber \\ &  \times
\biggl( e^{i\rot\Q\cdot \bar{\bm{v}}_{\bm{\tau}}(\mathcal{S})}
\sum_{m} \widetilde{D}^l_{\eta m}(\mathcal{S})\,
\bra{\psi_{i,\K+\rot\Q}} \psi_{n,\K}
B_{\mathcal{S}\bm{\tau} a l m}^{\rot\Q}
\rangle_{\Omega_0} \biggr)  \, . \nonumber \\
\end{align}
In Eq. (\ref{eq:eph_aug_final}),
$\widetilde{D}^l_{\mu m}(\mathcal{S})$ is the matrix element of the orthogonal
Wigner $\widetilde{D}$-matrix $\widetilde{\mathbf{D}}^l(\mathcal{S})$ used
to rotate real spherical harmonics as
$Y_{lm}(\widehat{\rot\R}) = \sum_{\mu=-l}^l \widetilde{D}^l_{\mu m}(\mathcal{S})
Y_{l\mu}(\widehat{\R})$.
The quantities in brackets in Eqs. (\ref{eq:eph_ipw}) and
(\ref{eq:eph_aug_final}) represent the transformation rules for
the projection coefficients as implemented in the \texttt{Questaal}
code. \textit{Our algorithm does not require evaluating the
projection coefficients for rotated electronic wave vectors
$\invrot \K$ and can be easily extended to electronic wave vectors
that do not belong to the BZ mesh}.
Expressions (\ref{eq:gmat_decomposition})-(\ref{eq:eph_aug_final})
can be directly extended to the short-range \eph matrix elements.
These formulations, along with Eqs. (\ref{eq:zeta_new})-(\ref{eq:Gdef}),
have been fully implemented in the \texttt{Questaal} electronic structure suite.

\section{Treatment of the core}
\label{sec:core}

A precise description of the \eph scattering requires an accurate
evaluation of the inverse electronic dielectric function at the
equilibrium positions of the nuclei.
To illustrate this assertion, we begin by examining the
static screening of the electron-nuclear
potential $V_{rl}^{(0)}$, as defined in Eq. (\ref{eq:nuclear_potential}).
Using the definition in Eq. (\ref{eq:nuclear_density_operator}) for
the nuclear charge density operator and assuming nuclei at
their equilibrium positions, we express the electronic screening as
\begin{align}
& \int_\Omega d\R' \varepsilon_e^{-1}(\R,\R';0) V_{rl}^{(0)}(\R') =
-Z_r e\int_\Omega d\R_1 \delta(\R_1-\bm{\tau}_{rl}^0)
\times \nonumber \\
& \,\,\,\,\,\,\,\,\,\,\,\,\,\,\,\,\,\, \times
\biggl[   \int_\Omega d\R' \varepsilon_e^{-1}(\R,\R';0)
  v(\R_1 - \R') \biggr] \nonumber \\
& \,\,\,\,\,\,\,\,\,\,\,\,\,\,\,\,\,\,= -Z_r e W_e(\R,\bm{\tau}_{rl}^0;0) \, .
\end{align}
By differentiating this expression with respect to nuclear
displacements, employing Eq. (\ref{eq:xifun}), and solving for the reduced
\eph coupling function, we derive
\bea\label{eq:reduced_eph_function}
\xi^{r\alpha}_{\,l}(\R) = \int_\Omega d\R'
\left.\frac{\partial\varepsilon_e^{-1}(\R,\R';0)}{\partial\tau_{r l \alpha}}
\right|_{\bm{\tau}^0_{rl}}
V_{rl}^{(0)}(\R') + \nonumber \\
+ \, Z_r e \left.
\frac{\partial W_e(\R,\bm{\tau}^0_{rl};0)}{\partial\tau_{r l \alpha}}
\right|_{\bm{\tau}^0_{rl}} \, .
\eea
Equation (\ref{eq:reduced_eph_function}) shows that $\xi^{r\alpha}_{\,l}(\R)$
relies on two components: the derivative of the
inverse dielectric function with respect to the nuclear coordinates and
the derivative of the screened electron-nuclear Coulomb interaction,
$W_e(\R,\bm{\tau}^0_{rl};0)$. Furthermore, leveraging
the symmetric property of the screened Coulomb interactions, i.e.,
$W_e(\R,\bm{\tau}^0_{rl};0) = W_e(\bm{\tau}^0_{rl},\R;0)$, we conclude that
the evaluation of the reduced \eph coupling function depends on the nuclear
gradient of $W_e(\bm{\tau}^0_{rl},\R;0) = \int_\Omega d\R'
\varepsilon_e^{-1}(\bm{\tau}^0_{rl},\R';0) v(\R'-\R)$,
with the inverse dielectric function evaluated
at the nuclear position $\bm{\tau}^0_{rl}$.

However, assessing the inverse dielectric function at nuclear
positions is hindered
by the incompleteness of the LMTO basis set in all-electron methods.
The finite Hilbert space spanned by the LMTO basis inadequately
represents electronic wave functions (and thus response functions) in
the high-density regions near nuclear sites. Additionally, the
treatment of core electrons in all-electron methods may not fully
capture their response to external perturbations around nuclear sites.
The treatment of core electrons is crucial in an all-electron method,
such as those implemented in \texttt{Questaal}, which explicitly includes core
electrons in its calculations. It is essential for accurately
capturing the physical interactions and electronic responses
in the vicinity of nuclear regions, where core electrons are tightly
localized around their respective nuclei.
To address these issues, \ibccs are necessary,
with the response of basis functions involving their dependence on an effective
potential~\cite{betzinger12,betzinger13,friedrich15}. Nonetheless,
we do not pursue this approach here, leaving it for future works.

Here, we mitigate the impact of the basis set incompleteness by
introducing a \textit{screened nuclear potential} $\widetilde{V}_{ls}(\R)$,
defined as the bare nuclear potential $V_{ls}(\R)$ screened by the
potential generated by a \textit{spherically symmetric core electron
density}. This is done by decomposing the total electron density
$n_e$ into valence ($n_e^v$) and core ($\sum_{rl} n_{e,rl}^c$) contributions, with
\be\label{eq:ncore_def}
n_{e,rl}^c(\R) = n_{e,r}^c(\R-\bm{\tau}^0_r-\RR_l)
\theta(s_r - \abs{\R-\bm{\tau}^0_r-\RR_l}) \, ,
\ee
where $n_{e,rl}^c(\R)$ represents the core electron density around
the $r$-th nucleus inside the $l$-th unit cell within the BvK macrocrystal, and where
$\theta(s_r - \abs{\R-\bm{\tau}^0_r-\RR_l})$ is the
Heaviside step function, which limits the integration
domain solely to the augmentation region located at the
nuclear position $\bm{\tau}^0_{rl}$.
Under this decomposition, the core
contribution to the Hellmann-Feynman force vanishes when
applied to an equilibrium nuclear configuration
\be
\int_\Omega d\R\, n_e^c(\R)\frac{\partial V_n(\R)}{\partial\tau_{rl \alpha}}=0 \,,
\ee
for materials with an inversion symmetry---satisfying the condition
$\bm{\tau}_r = - \bm{\tau}_s \,\,\,\forall r\ne s$---and assuming a
spherically symmetric core density,
$n_{e,r}^c(\R) = n_{e,r}^c(r) Y_{00}(\widehat{\R})$.
This result can be demonstrated using a method similar to that outlined
in \sectionp S.1 of the \suppinfo. For a spherically symmetric core density and
in presence of an inversion symmetry, the Bloch-summed structure constant
$S^{\bm{\tau}_i\bm{\tau}_r}_{lm,00}(\kappa;\Q)$---defined as in Eqs. (S.15)-(S.17) and
(S.112) in the \suppinfo---is always zero in the long-wavelength limit $\Q\to 0$
for $l=1$ and $m$ in the range $[-1,0,1]$. Physically, this indicates that a
spherically symmetric core density, in materials with an
inversion symmetry, remains rigidly bound to the nucleus during
nuclear vibrations, remaining unperturbed. As a result, only perturbations
in electron densities lacking spherical symmetry contribute to electronic
forces.

Screening a bare nuclear potential effectively reduces the nuclear charge
$Z_r e$ by an amount corresponding to the number of core electrons.
This relationship can be expressed through the Poisson's equation
\begin{align}
& \nabla_\R^2 \widetilde{V}_n(\R) = \sum_{rl} \nabla_\R^2 \widetilde{V}_{lr}(\R)
\nonumber \\
& = -4\pi \sum_{lr}\bigl[-Z_r e \delta(\R-\bm{\tau}^0_r-\RR_l) +   n_{e,rl}^c(\R)
  \bigr] \, ,
\end{align}
where $-Z_r e\delta(\R-\bm{\tau}^0_r-\RR_l)$ represents the
nuclear charge localized at $\bm{\tau}^0_{rl}$
(according to the definition of Eq. (\ref{eq:nuclear_density_operator})),
and $n_{e,rl}^c(\R)$
denotes the core electron density in the augmentation sphere centered at
$\bm{\tau}^0_{rl}$, expressed as in Eq. (\ref{eq:ncore_def}).
Using the linearity of Poisson's equation, the potential contributions
can be separated as
\be\label{eq:V_plus_Vc}
\widetilde{V}_{rl}^{(0)}(\R) =
V_{rl}^{(0)}(\R) + V^c_{rl}(\R) \,,
\ee
where $V^{(0)}_{rl}(\R)$ is defined by Eq. (\ref{eq:nuclear_potential}),
and where the core electronic potential $V^c_{rl}(\R)$ is
\begin{subequations}
\begin{align}
& V^c_{rl}(\R) = \int_\Omega d\R'
\frac{n_{e,r}^c(\R'-\bm{\tau}^0_r-\RR_l)
  \theta(s_r-\abs{\R'-\bm{\tau}^0_r-\RR_l})}
     {\abs{\R-\R'}} \nonumber \\
     \label{eq:def_Vcore_1}  \\
& \,\,\,\,\,\,\,\,\,\,\,\,\,\,\,\,
= \int_{\Omega_r} d\R' \,\frac{n^c_{e,r}(\R')}
     {\abs{\R'-\R- \bm{\tau}^0_r-\RR_l}}
     \label{eq:def_Vcore_2} \, .
\end{align}
\end{subequations}
Therefore, the reduced \eph coupling function can be expressed as
\be\label{eq:xifun_val_core}
\xi^{r\alpha}_l(\R) = \int_\Omega d\R' \varepsilon_{e,v}^{-1}(\R,\R';0) \,
\frac{\partial \widetilde{V}^{(0)}_{rl}(\R')}{\partial r_{\alpha}^{'}}
\ee
where $\varepsilon_{e,v}^{-1}(\R,\R';0)$ is the static inverse
dielectric function that accounts only for valence electron contributions.
When introducing in Eq. (\ref{eq:xifun_val_core}) the expansion
of $\varepsilon_{e,v}^{-1}(\R,\R';0)$
expressed in terms of biortogonal MPB functions,
as given by Eq. (\ref{eq:eps_inv_Mexpansion}), we can rewrite
the reduced \eph coupling matrix elements $\xi^{r\alpha}_{in}(\K,\Q)$ as
\begin{align}\label{eq:xifun_val_core_2}
\xi^{r\alpha}_{in}(\Q,\K) &=  \sum_{IJ} \widetilde{C}^I_{in}(\K,\Q)
\,\varepsilon^{-1}_{e,v,IJ}(\Q;0) \times \nonumber \\
& \,\,\,\,\,\,\,\,\,\,\, \times
\int_{\Omega_0} d\R' M_J^{\Q\, *}(\R') \frac{\partial
\widetilde{V}^{(0)\,\Q}_{r}(\R')}
{\partial r_{\alpha}^{'}} \, , \nonumber \\
\end{align}
where Eq. (\ref{eq:ximat}) has been used and where
$\widetilde{C}^I_{in}(\K,\Q) = \bra{\psi_{i,\K+\Q}}\psi_{n,\K}
\widetilde{M}_I^{\Q}\rangle_{\Omega_0}$.
Further details on the computation of the unit-cell integrals
\be\label{eq:starting_point_no_core}
\int_{\Omega_0} d\R M_I^{\Q\,*}(\R)
\frac{\partial\widetilde{V}^{(0)\,\Q}_r(\R)}{\partial r_\alpha}
\, ,
\ee
are provided in \ssections S.1 A-D of the \suppinfo.
As noted in \sectionp S.5 of the \suppinfo, this integral does not
involve Pulay-like \ibcs, as the spherically symmetric core
density $n^c_{e,r}(r)$ remains unaffected by nuclear displacements.
Equation (\ref{eq:xifun_val_core_2}) is then straightforwardly expressed
in terms of the orthogonal MPB set $\{E_\mu^\Q\}$ by using
the transformation (\ref{eq:E_to_M_eps-inv_matrix}), which results in
\begin{align}\label{eq:xifun_val_core_2_Ebasis}
\xi^{r\alpha}_{in}(\Q,\K) &= \sum_{\mu\gamma} C^\mu_{in}(\K,\Q)
\,\varepsilon^{-1}_{e,v,\mu\gamma}(\Q;0) \times \nonumber \\
& \,\,\,\,\,\,\,\,\,\,\, \times
\int_{\Omega_0} d\R' E_\gamma^{\Q\, *}(\R') \frac{\partial
\widetilde{V}^{(0)\,\Q}_{r}(\R')}
{\partial r_{\alpha}^{'}} \, . \nonumber \\
\end{align}

\subsection{Justification of the scheme: downfolding to
a subspace of high-energy transitions from core states}
\label{sec:downfolding}

To justify Eq. (\ref{eq:xifun_val_core}),
we begin by combining Eq. (\ref{eq:def_Vcore_2}) with the derivative of
$\widetilde{V}_{rl}$ with respect to nuclear displacements,
yielding
\begin{align}
&\left.\frac{\partial\widetilde{V}^{(0)}_{rl}(\R)}
{\partial \tau_{rl\alpha}}\right|_{\bm{\tau}^0_{rl}} =
\int_{\Omega_r} d\R' \,
\biggl[ \delta(\R-\R') + \nonumber \\ & \,\,\,\,\,\,\,\,\,
\,\,\,\,\,\,\,\,\,\,\,\,\,\,\,\,\,\,\,\,\,\,\,\,\,\,\,\,\,\,
\,\,\,\,\,\,\,\,+
\frac{n^c_{e,r}(\R-\R')}{Z_r} \biggr]
\left.\frac{\partial V^{(0)}_{rl}(\R')}{\partial \tau_{rl\alpha}}
\right|_{\bm{\tau}^0_{rl}} .
\end{align}
This equation is reminiscent of a linear response expression for
an inverse dielectric function that \textit{partially} screens
the bare electron-nuclear potential generated by the nucleus
located at $\bm{\tau}^0_{rl}$
\be\label{eq:screened_Vder}
\left.\frac{\partial\widetilde{V}^{(0)}_{rl}(\R)}
{\partial \tau_{rl\alpha}}\right|_{\bm{\tau}^0_{rl}} = \int_{\Omega_r} d\R' \,
\varepsilon_{c,r}^{-1}(\R,\R';0)
\left.\frac{\partial V^{(0)}_{rl}(\R')}{\partial \tau_{rl\alpha}}
\right|_{\bm{\tau}^0_{rl}} \, ,
\ee
where
\be\label{eq:inv_eps_core}
\varepsilon_{c,r}^{-1}(\R,\R';0) =
\delta(\R-\R') + \frac{n^c_{e,r}(\R-\R')}{Z_r}
\ee
denotes the static inverse dielectric function with contributions from
core states, confined to the $r$-th augmentation sphere.
To prove Eq. (\ref{eq:inv_eps_core}) we use its formal definition
\be\label{eq:inv_eps_core_def}
\varepsilon_{c,rl}^{-1}(\R,\R';0) =
\delta(\R-\R') + \int_\Omega
d\R_1 v(\R-\R_1) \chi_{e}^{c,rl}(\R_1,\R';0) \, ,
\ee
where $\chi_{e}^{c,rl}(\R_1,\R';0)$ is the static reducible
polarizability from core states within the augmentation sphere
located at $\bm{\tau}^0_{rl}$,
\be\label{chi_core_def}
\chi_{e}^{c,rl}(\R_1,\R';0) =
\frac{\delta n^c_{e,rl}(\R_1)}{\delta V^{(0)}_{rl}(\R')} \, .
\ee
The assumption of spherically symmetric core densities implies
that radial core charge distributions are isotropic and insensitive
to perturbations in the nuclear potentials for materials possessing
an inversion symmetry.
Additionally, the screening effects of valence electrons ensure
that core electron densities are negligibly affected by external
potential perturbations, even in systems lacking an inversion center.
As a result, the core density $n^c_{e,rl}$ remains invariant under
displacements of the other surrounding nuclei located at
$\bm{\tau}^0_{tm}$, where $r,l \ne \{t,m\}$. However, it is
affected by displacements $\Delta\bm{\tau}_{rl}$ of its own
nucleus due to variations in the Heaviside step function.
This observation justifies the replacement of $V_n^{(0)}$ with
$V^{(0)}_{rl}$ in Eq. (\ref{chi_core_def}).
This assumption also simplifies the functional dependence between
the core density and the nuclear potential.
This allows us to approximate the polarizability as
\be
\chi_{e}^{c,rl}(\R_1,\R';0) \approx
\frac{n^c_{e,rl}(\R_1)}{V^{(0)}_{rl}(\R')}
\ee
Substituting into the dielectric function
(\ref{eq:inv_eps_core_def}) yields
\bea\label{eq:inv_eps_core_def_1}
&& \varepsilon_{c,rl}^{-1}(\R,\R';0) \approx
\delta(\R-\R') + \int_\Omega
d\R_1 v(\R-\R_1)  \frac{n^c_{e,rl}(\R_1)}{V^{(0)}_{rl}(\R')}
 \nonumber \\
 && \,\,\,\,\,\,\,\,\,\,\,\,\,\,\,\,
 = \delta(\R-\R') + \frac{1}{V^{(0)}_{rl}(\R')}
\int_\Omega d\R_1 v(\R-\R_1) \times \nonumber \\
&& \,\,\,\,\,\,\,\,\,\,\,\,\,\,\,\,  \,\,\,\,\,\,
\times \, n_{e,r}^c(\R_1-\bm{\tau}^0_r-\RR_l)
\theta(s_r-\abs{\R_1-\bm{\tau}^0_r -\RR_l}) \,. \nonumber \\
\eea
and with a change of variables $\R_2 = \R_1-\bm{\tau}^0_r -\RR_l$,
\begin{align}
\varepsilon_{c,rl}^{-1}(\R,\R';0) &\approx
\delta(\R-\R') + \nonumber \\
& \,\,\,\,\,\,\,\,\,+\frac{e^2}{V^{(0)}_{rl}(\R')}
\int_{\Omega_r} d\R_2
\frac{n_{e,r}^c(\R_2)}{\abs{\R -\R_2-\bm{\tau}^0_{r}-\RR_l}} \, .
\end{align}
With a further substitution of variables $\R_1 = \R -\R_2$, we
obtain
\be
\varepsilon_{c,rl}^{-1}(\R,\R';0) \approx
\delta(\R-\R') + \int_{\Omega_r} d\R_1
\frac{n_{e,r}^c(\R-\R_1)}{Z_r}
\frac{V^{(0)}_{rl}(\R_1)}{V^{(0)}_{rl}(\R')} \, .
\ee
Finally, we approximate the ratio of the electron-nuclear
potentials with a Dirac delta, and drop the index $l$ in the
notation for $\varepsilon_{c,rl}^{-1}$
\begin{align}
\varepsilon_{c,r}^{-1}(\R,\R';0) &\approx
\delta(\R-\R') + \int_{\Omega_r} d\R_1
\frac{n_{e,r}^c(\R-\R_1)}{Z_r}
\delta(\R_1 - \R') \nonumber \\
&= \delta(\R-\R') + \frac{n_{e,r}^c(\R-\R')}{Z_r} \, ,
\end{align}
thus proving Eq. (\ref{eq:inv_eps_core}).

If we define the \textit{bare} \eph \textit{coupling function} as
\be\label{eq:gb_function}
g^b_{\Q\nu}(\R) \equiv \sum_{r\alpha l}
\sqrt{\frac{\Omega_0}{2 m_r \omega_{\Q\nu}}} e^{i\Q\cdot\RR_l}
e_{r\alpha,\nu}(\Q)
\frac{\partial V_{rl}^{(0)}(\R)}{\partial r_{\alpha}}  \, ,
\ee
then we introduce within this framework the \textit{partially screened} \eph
\textit{coupling function}
\begin{align}\label{eq:gp_function}
g^p_{\Q\nu}(\R) &= \sum_{r\alpha l}
\sqrt{\frac{\Omega_0}{2 m_r \omega_{\Q\nu}}} e^{i\Q\cdot\RR_l}
e_{r\alpha,\nu}(\Q)
\frac{\partial \widetilde{V}_{rl}^{(0)}(\R)}{\partial r_{\alpha}}
\nonumber \\
&= \int_{\Omega} d\R' \,
\varepsilon_{e,c}^{-1}(\R,\R';0)\,  g^b_{\Q\nu}(\R') \, .
\end{align}
Here, $\varepsilon_{e,c}^{-1}$ is integrated solely within the $l$-th unit cell
and the $r$-th augmentation sphere
\begin{widetext}
\begin{align}\label{eq:inveps_core_nonlocal}
\int_{\Omega} d\R' \, \varepsilon_{e,c}^{-1}(\R,\R';0)
\frac{\partial V_{rl}^{(0)}(\R')}{\partial r'_{\alpha}} &=
\int_{\Omega} d\R'  \biggl\{\sum_{tn} \delta_{tr}\delta_{nl}
\varepsilon_{c,tn}^{-1}(\R,\R';0)
\theta(s_t-\abs{\R'-\bm{\tau}^0_t -\RR_n})
\frac{\partial V_{tn}^{(0)}(\R')}{\partial r'_{\alpha}}\biggr\}
\nonumber \\
&= \int_{\Omega_r} d\R' \,
\varepsilon_{c,r}^{-1}(\R,\R';0)
\frac{\partial V^{(0)}_{rl}(\R')}{\partial r'_{\alpha}} \, ,
\end{align}
i.e., it acts as a non-local operator with respect to the indices
$tn$ and $rl$.
Notably, Eq. (\ref{eq:gp_function}) is equivalent to the
diagrammatic equation (15) in Ref. \onlinecite{berges23}, which
splits the screening into two parts: (i) \textit{downfolding}
to a subspace of high-energy transitions from core states
to an infinite number of empty bands
contributing to the \textit{irreducible core polarizability}
$\chi_{c,rl}^{0}$, and (ii) a \textit{renormalization} step
to account for the most relevant low-energy transitions from
valence states, which contribute to the screening through the
\textit{irreducible valence polarizability} defined as $\chi_v^{0} \equiv
\chi^0_e - \sum_{rl}\chi_{c,rl}^{0}$. We then express the
full \eph coupling function as~\cite{berges23}
\begin{align}\label{eq:g_full_def}
g_{\Q\nu}(\R) &= \int_{\Omega} d\R'\biggl\{\delta(\R-\R') +
\int_\Omega d\R_1\chi^{0}_{v}(\R,\R_1;0)W_e(\R_1,\R';0)\biggr\}
g^p_{\Q\nu}(\R') \nonumber \\
&= \int_{\Omega} d\R' \mathcal{E}_e^{-1}(\R,\R';0)
\, g^p_{\Q\nu}(\R') \, ,
\end{align}
\end{widetext}
with $\mathcal{E}_e^{-1} = 1 + \chi^{0}_{v} W_e \ne \varepsilon_{e}^{-1}$ and
where $W_e = \varepsilon_{e}^{-1} v$ includes screening from
both core and valence states.
It is straightforward to show that
Eq. (\ref{eq:g_full_def}) is equivalent to the more familiar
definition (\ref{eq:g_function}) for the \eph coupling function
\be\label{eq:g_from_gbare}
 g_{\Q\nu}(\R) = \int_{\Omega} d\R'
\varepsilon^{-1}_e(\R,\R';0) \, g^b_{\Q\nu}(\R') \, .
\ee

Within the framework of this section, the accuracy of
Eq. (\ref{eq:g_full_def}) is constrained by the incomplete nature
of the LMTO basis functions, as the \textit{fully screened}
Coulomb matrix $W_e$ is still required.
Additionally, \textit{core} and \textit{valence} eigenfunctions
are treated differently in the \texttt{Questaal} framework~\cite{Kotani07}.
\textit{Valence} states are determined by diagonalizing
a secular matrix for the LMTOs, ensuring complete orthogonality
between them.
High-energy shallow core states (within $\sim$2 Ry below the Fermi
level) can only be reliably treated if they are incorporated
into the valence manifold and computed using local orbitals specifically
tailored to these states.

In contrast, deep \textit{core} states, which are well-localized
eigenfunctions within the augmentation spheres, exhibit
minimal screening effects. Consequently, they are treated as
exchange-only contributions in the electron self-energy.
\textit{Core} eigenfunctions are obtained by solving the radial
Schr\"odinger equation using a DFT-KS potential, with the
density provided by the effective static
non-local QS\gww Hamiltonian. However, due to the imperfect
orthogonality between \textit{core} and \textit{valence} states,
a small but uncontrollable error is introduced.
To mitigate these limitations, we adopt the approximation
$\mathcal{E}_e^{-1} \approx
\varepsilon_{e,v}^{-1}$ and calculate the \textit{screened} \eph
\textit{coupling function} as
\be\label{eq:g_full_def_approx}
g_{\Q\nu}(\R) = \int_{\Omega} d\R'
\varepsilon_{e,v}^{-1}(\R,\R';0) \, g^p_{\Q\nu}(\R') \, ,
\ee
which can be derived starting from the screened reduced \eph coupling function
(\ref{eq:xifun_val_core}).

Symbolically, the error $\Delta$ introduced by this approximation can be
derived by combining Eqs. (\ref{eq:gp_function}), (\ref{eq:g_from_gbare}), and
(\ref{eq:g_full_def_approx})
\begin{align}\label{eq:error_g_full_approx}
\Delta &= \varepsilon_{e}^{-1} g^b_{\Q\nu} -
\varepsilon_{e,v}^{-1} \, g^p_{\Q\nu} \nonumber \\
&= (\varepsilon_{e}^{-1}
- \varepsilon_{e,v}^{-1} \varepsilon_{e,c}^{-1}) \, g^b_{\Q\nu} \, .
\end{align}
The inverse dielectric function can be formally decomposed
by means of a geometric series as
$\varepsilon_{e}^{-1} = \varepsilon_{e,v}^{-1} +
\varepsilon_{e,c}^{-1} - 1 + \Delta^{2,\infty}_{c,v}$, where
\begin{align}\label{eq:Delta_mixed_terms}
\Delta^{2,\infty}_{c,v} &= \sum_{n=2}^\infty \sum_{k=1}^{n-1}
\sum_{\sigma\in \mathcal{S}_k^n} \prod_{i=1}^n T_i^\sigma \nonumber \\
&= v\chi_c^0 v \chi_v^0 + v\chi_v^0 v
\chi_c^0 + v\chi_c^0 v \chi_v^0 v \chi_c^0 + \dots
\end{align}
represents an infinite sum of products involving both
$v\chi_c^0$ and $v \chi_v^0$ and
accounting for the interplay between core and valence screening.
In Eq. (\ref{eq:Delta_mixed_terms}), $\mathcal{S}_k^n$ denotes
the set of all permutations of $k$ core ($v\chi_c^0$) and $n-k$
valence ($v\chi_v^0$) contributions, and $T_i^\sigma$ specifies the
order of these terms in the product.
Consequently, the error becomes
\begin{align}\label{eq:error_g_full_approx__2}
\Delta &= \bigl[\Delta^{2,\infty}_{c,v}  - (\varepsilon_{e,v}^{-1} - 1)
  (\varepsilon_{e,c}^{-1} - 1) \bigr] g^b_{\Q\nu} \nonumber \\
&= \biggl[\sum_{n=2}^\infty \sum_{k=1}^{n-1}
\sum_{\sigma\in \mathcal{S}_k^n} \prod_{i=1}^n T_i^\sigma -
\sum_{n=1}^\infty\sum_{k=1}^\infty \bigl(v\chi_v^0\bigr)^n \bigl(v \chi_c^0\bigr)^k
\biggr] g^b_{\Q\nu} \nonumber \\
&= \bigl( v\chi_c^0 v \chi_v^0 + v\chi_c^0 v \chi_v^0 v \chi_c^0 + \dots
\bigr) g^b_{\Q\nu}
\end{align}
i.e., at least of second order in the irreducible polarizabilities.
The leading contribution to the error $\Delta$ is $v\chi_c^0 v \chi_v^0$,
which can be considered small under the assumption that
$v\chi_c^0 \ll v\chi_v^0$. This assumption is well-justified because core
states are highly localized around the nuclei, resulting in a negligible
contribution to the screening. The error $\Delta$ can then be considered
negligible for practical purposes.

\section{The Dynamical matrix in the mixed product basis formalism}
\label{sec:dynmat}

In this section, we present a brief analysis of the implementation of
the dynamical matrix within a field-theoretic framework, alongside
a discussion of its accuracy within a LMTO-MPB formalism.
The \ifc, initially defined via Eq. (\ref{eq:ifc_manybody}), where the
static {\it test-electron} dielectric function is typically assessed within
the KS-DFT framework, are now replaced by their
many-body counterpart derived within a
field-theoretic approach under the adiabatic approximation, as expressed in
Eq. (\ref{eq:phonon_selfenergy_adiabatic}).
By invoking the definition of the inverse dielectric function in terms of
the reducible polarizability $\chi_e$, expressed symbolically as
$\varepsilon_e^{-1} = 1 + v\chi_e$, we reframe
Eq. (\ref{eq:phonon_selfenergy_adiabatic}) as the sum of electronic and ionic
contributions to the phonon self-energy
\be
\Pi^{\text{A}}_{r \alpha l,s \beta l'} = \Pi^{\text{A},e}_{r \alpha l,s \beta l'} +
\Pi^{\text{A},i}_{r \alpha l,s \beta l'}
\ee
where $\Pi^{\text{A},e}_{r \alpha l,s \beta l'}$ indicates
the electronic contribution
\begin{align}\label{eq:ifc_manybody_el}
& \Pi^{\text{A},e}_{r \alpha l,s \beta l'} =
\sum_{tn}\biggl( \delta_{ln}\delta_{rt} - \delta_{ll'}\delta_{rs}\biggr)
\times \nonumber \\ & \times
\int_\Omega\int_\Omega
\frac{\partial V^{(0)}_{sl'}(\R)}{\partial r_\beta}  \chi_e(\R,\R';0)
\frac{\partial V^{(0)}_{tn}(\R')}{\partial r'_{\alpha}}  \,d\R\,d\R'
\end{align}
and $\Pi^{\text{A},i}_{r \alpha l,s \beta l'}$ a purely electrostatic term
\begin{align}\label{eq:ifc_manybody_ion}
& \Pi^{\text{A},i}_{r \alpha l,s \beta l'} = \sum_{tn}\biggl( \delta_{ln}\delta_{rt}-
\delta_{ll'}\delta_{rs}\biggr)
\times \nonumber \\ & \times
\int_\Omega\int_\Omega
 \frac{\partial V^{(0)}_{sl'}(\R)}{\partial r_\beta} v^{-1}(\R-\R')
 \frac{\partial V^{(0)}_{tn}(\R')}{\partial r'_{\alpha}}
 \,d\R\,d\R' \, .
\end{align}

It can be demonstrated that
Eq. (\ref{eq:ifc_manybody_ion}) corresponds to
the second derivative of the nuclear-nuclear interaction potential
energy\cite{sahni75},
as given in Eq. (\ref{eq:Unn})
\be\label{eq:ifc_manybody_ion2}
\Pi^{\text{A},i}_{r \alpha l,s \beta l'} =
\left.\frac{\partial^2 \braket{\hat{U}_{nn}(\{\bm{\tau}_{rl}\})}}
      {\partial \tau_{r l \alpha} \partial \tau_{sl' \beta}}
      \right|_{\{\bm{\tau}^0_{rl}\}} \, .
\ee
A proof of this relation is presented in \sectionp S.7 of the \suppinfo.
The long-range nature of the Coulomb contribution to
the dynamical matrix requires a specific approach. The Bloch transform of
Eq. (\ref{eq:ifc_manybody_ion2}) can be efficiently computed using the
Ewald-Kellerman summation method~\cite{ewald21,kellerman40}. This method
involves splitting the sum over lattice vectors into two separate parts.
The first part, of short-range nature, requires a
summation over a limited region in real space. The second part,
of long-range nature, entails a summation in
reciprocal space and is
designed to explicitly avoid singularities in the long-wavelength
limit when evaluating the \textit{short-range phonon
self-energy} (see \sectionp \ref{sec:long-range}).

We then shift our focus on the electronic contribution.
By invoking the definition of the reducible polarizability, expressed
symbolically as $\chi_e = \chi_e^0\varepsilon_e^{-1}$, we reformulate
the electronic contribution in Eq. (\ref{eq:ifc_manybody_el}) as
\begin{align}\label{eq:ifc_manybody_el__2}
& \Pi^{\text{A},e}_{r \alpha l,s \beta l'} =
\sum_{tn}\biggl( \delta_{ln}\delta_{rt} - \delta_{ll'}\delta_{rs}\biggr)
\times \nonumber \\ &
\,\,\,\,\,\,\,\,\,\,\,\,\,\,\, \times
\int_\Omega\int_\Omega
\xi^{s\beta,b}_{l'}(\R)  \chi_e^0(\R,\R';0) \,\xi^{t\alpha}_{n}(\R') \,d\R\,d\R' \,,
\end{align}
where $\xi^{s\beta,b}_{l'}(\R) = \partial V^{(0)}_{sl'}(\R)/\partial r_\beta$ is
the \textit{bare reduced \eph coupling function}.
In Eq. (\ref{eq:ifc_manybody_el__2}) we express
$\Pi^{\text{A},e}_{r \alpha l,s \beta l'}$ as an electron-hole bubble
connected to a bare and a screened electron-phonon vertex~\cite{berges23}.
Following the downfolding approach
for the full phonon propagator, $\bm{D}(\omega)$,
outlined in Eqs. (21)-(29) of Ref. \onlinecite{berges23},
the electronic response part of the phonon self-energy can be symbolically
rewritten within the adiabatic approximation as
\bea\label{eq:downfolding_phonon_selfenergy}
\Pi^{\text{A},e} &=& \xi^b \chi_e^0 \xi \nonumber \\
&=&  \xi^p \chi_v^0 \xi +
\xi^b \chi_c^0  \xi^p \, ,
\eea
where $\xi^p = \varepsilon_{e,c}^{-1} \xi^b$ denotes the \textit{partially
screened reduced \eph coupling function}, equivalent to
Eqs. (\ref{eq:screened_Vder}) and (\ref{eq:inveps_core_nonlocal}).
Under the assumption of highly localized core states
with negligible contributions to screening
(see \sectionp \ref{sec:core}), the core contribution $\xi^b \chi_c^0  \xi^p$
in Eq. (\ref{eq:downfolding_phonon_selfenergy}) can be neglected.
This leads to the simplified expression
\begin{align}\label{eq:ifc_manybody_el__3}
& \Pi^{\text{A},e}_{r \alpha l,s \beta l'} =
\sum_{tn}\biggl( \delta_{ln}\delta_{rt} - \delta_{ll'}\delta_{rs}\biggr)
\times \nonumber \\ & \,\,\, \times \int_\Omega\int_\Omega
\frac{\partial\widetilde{V}^{(0)}_{sl'}(\R)}{\partial r_{\beta}}
\chi_v^0(\R,\R';0) \,\xi^{t\alpha}_{n}(\R') d\R\,d\R' \, .
\end{align}
Equation (\ref{eq:ifc_manybody_el__3}) is particularly useful, as
$\xi^{t\alpha}_{n}(\R')$ can be replaced using Eq. (\ref{eq:xifun_val_core}),
yielding
\begin{align}\label{eq:ifc_manybody_el__4}
& \Pi^{\text{A},e}_{r \alpha l,s \beta l'} =
\sum_{tn}\biggl( \delta_{ln}\delta_{rt} - \delta_{ll'}\delta_{rs}\biggr)
\times \nonumber \\ & \,\,\, \times \int_\Omega\int_\Omega
\frac{\partial\widetilde{V}^{(0)}_{sl'}(\R)}{\partial r_{\beta}}
\chi_e^v(\R,\R';0)
\frac{\partial\widetilde{V}^{(0)}_{tn}(\R')}{\partial r'_{\alpha}} d\R\,d\R' \, .
\end{align}
In Eq. (\ref{eq:ifc_manybody_el__4}), the electronic contribution to the phonon
self-energy incorporates the electron-nuclear
potential screened by core electrons and the static reducible
electronic polarizability $\chi^v_e(\R,\R';0)$, which includes
contributions from valence states only~\cite{sahni75}. As detailed in
\sectionp \ref{sec:core}, Eq. (\ref{eq:ifc_manybody_el__4}) provides a
framework to address the incompleteness inherent in the LMTO basis set
in the evaluation of the phonon self-energy within a field-theoretic
formalism.

By introducing the expansion of the reducible polarizability
in terms of the orthonormal MPB set functions
$\{E^\Q_\mu\}$, we reformulate Eq. (\ref{eq:ifc_manybody_el__4}) as
\begin{align}\label{eq:ifc_manybody_el2}
& \Pi^{\text{A},e}_{r \alpha l,s \beta l'} = \frac{1}{N_\K} \sum_{\Q\in\BZ}
\sum_{\mu\nu} \sum_{tn}\biggl( \delta_{ln}\delta_{rt} -
\delta_{ll'}\delta_{rs}\biggr)
\times \nonumber \\ & \,\,\,\,\,\,\,\,\,\,\,\,\,\,\,\,\,\,\, \times
e^{-i\Q\cdot(\RR_n - \RR_{l'})}
\chi^v_{e,\mu\nu}(\Q;0) \widetilde{\mathcal{I}}^{s\beta}_\mu(\Q)
\widetilde{\mathcal{I}}^{t\alpha\,*}_\nu(\Q) \, ,
\end{align}
where Eqs. (\ref{eq:applying_bloch_sum_to_pot}) and (\ref{eq:Imat}) have
been used. Here, we introduce the notation $\widetilde{\mathcal{I}}^{s\beta}_\mu$
to emphasize the use of the screened electron-nuclear
potential $\widetilde{V}^{(0)}_{sl}$.
The Bloch transforms of Eq. (\ref{eq:ifc_manybody_el2}) and
(\ref{eq:ifc_manybody_ion2}), with the bare nuclear potential
$V^{(0)}_{rl}(\R)$ replaced by its screened counterpart
$\widetilde{V}^{(0)}_{rl}(\R)$, yield
\be\label{eq:dynmat2}
D^{\alpha\beta}_{rs}(\Q) = D^{\alpha\beta}_{e,rs}(\Q) + D^{\alpha\beta}_{i,rs}(\Q) \, ,
\ee
with the {\it short-range} electronic contribution to the dynamical matrix
defined as
\be
D^{\alpha\beta}_{e,rs}(\Q) = \frac{1}{\sqrt{m_rm_s}} \Pi^{\alpha\beta}_{rs}(\Q) -
\frac{\delta_{rs}}{m_r} \sum_t \Pi^{\alpha\beta}_{tr}(\mathbf{0})
\ee
and
\be\label{eq:dynmat_el}
\Pi^{\alpha\beta}_{rs}(\Q) =  \sum_{\mu\ne 1 \nu\ne 1}
\chi^v_{e,\mu\nu}(\Q;0) \widetilde{\mathcal{I}}^{s\beta}_\mu(\Q)
\widetilde{\mathcal{I}}^{r\alpha\,*}_\nu(\Q)
\ee
Notably, in Eq. (\ref{eq:dynmat_el}) only the body of the inverse dielectric
matrix is considered, following a
block-matrix derivation (not reported here)
akin to the approach employed in \sectionp \ref{sec:long-range}.

The formalism described thus far
is appropriate for a set of basis functions forming a complete set.
However, as previously discussed in \sectionp \ref{sec:pulay},
a Pulay-like \ibcc term must be considered
to account for contributions arising
from variations in the LMTO-MPB formalism
due to nuclear displacements. This leads to
a dynamical matrix corrected for the explicit dependence of
the basis functions on nuclear displacements
\be\label{eq:dynmat2_pulay}
\widetilde{D}^{\alpha\beta}_{rs}(\Q) =
D^{\alpha\beta}_{e,rs}(\Q) + D^{\alpha\beta}_{i,rs}(\Q) -
 \frac{\delta_{rs}}{m_r} \sum_t P^{\alpha\beta}_{tr}(\mathbf{0})  \, ,
\ee
where $P^{\alpha\beta}_{tr}(\mathbf{0})$ denotes the Bloch transform of the second
term on the right-hand side of Eq. (\ref{eq:phonon_selfenergy_pulay}),
evaluated in the long-wavelength limit.
As detailed in \sectionp \ref{sec:pulay} and in \sectionp S.5 of
the \suppinfo, the acoustic sum rule (\ref{eq:acoustic_sum_rule})
is no longer satisfied unless the basis functions exhibit no
parametric dependence on the equilibrium nuclear positions or
form a \textit{complete set}. In the absence of these conditions, both
dynamical matrices (\ref{eq:dynmat2}) and (\ref{eq:dynmat2_pulay}) become
inaccurate and unphysical.
An alternative formulation for the phonon self-energy,
free from Pulay-like \ibccs and
unaffected by the incompleteness inherent in the LMTO-MPB set, is given
by Eq. (\ref{eq:phonon_self_energy}). Within the adiabatic regime, this
expression can be reformulated as
\be
\Pi^{\text{A}}_{r \alpha l,s \beta l'} =
\Lambda^{\text{A},e}_{r \alpha l,s \beta l'} +
\Pi^{\text{A},i}_{r \alpha l,s \beta l'} \, ,
\ee
with the electronic contribution defined as
\begin{align}
& \Lambda^{\text{A},e}_{r \alpha l,s \beta l'} =
\int_\Omega\int_\Omega
\frac{\partial\widetilde{V}^{(0)}_{sl'}(\R)}{\partial r_\beta}  \chi^v_e(\R,\R';0)
\frac{\partial\widetilde{V}^{(0)}_{rl}(\R')}{\partial r'_{\alpha}}  \,d\R\,d\R' -
\nonumber \\ & \,\,\,\,\,\,\,\,\,\,\,\,\,\,\,\,\,\,\,\,\,\,\,\,
- \delta_{rs}\delta_{ll'} \int_\Omega d\R
\nabla_\alpha n^v_e(\R)
\frac{\partial\widetilde{V}^{(0)}_{sl'}(\R)}{\partial r_\beta}
\, .
\end{align}
In the case of a complete set of basis functions, the condition
$\bm{\Lambda}^{\text{A},e} = \bm{\Pi}^{\text{A},e}$ will be fulfilled.

However, this approach may yield to inaccuracies due to
inconsistencies in the definition of the electron density within the
QS$GW$ formalism~\cite{Kotani07}.
The electron density is connected to the total energy of the system
via the functional derivative of
the latter with respect to the nuclear potential $V_n$
\be\label{eq:dEdVext}
\frac{\delta E[n_e]}{\delta V_n(\R)} = \int_\Omega d\R'
\frac{\delta E[n_e]}{\delta n_e(\R')} \frac{\delta n_e(\R')}{\delta V_n(\R)} +
\left.\frac{\delta E[n_e]}{\delta V_n(\R)}\right|_{n_e} \, .
\ee
For ground state densities satisfying the stationary principle, the
\textit{Euler-Lagrange equation}
$\delta E[n]/\delta n_e(\R') = \mu$, with $\mu$ as
the \textit{chemical potential},
simplifies Eq. (\ref{eq:dEdVext}) to
\begin{align}
\frac{\delta E[n_e]}{\delta V_n(\R)}
&= \mu \int_\Omega d\R' \chi_e(\R',\R) + n_e(\R) \nonumber \\
&= n_e(\R) \, .
\end{align}
Here, the vanishing of $\int_\Omega d\R' \chi_e(\R',\R)$, consistent with
$\int_\Omega d\R' \chi_e^0(\R',\R)=0$, eliminates the first term. The second term
arises from the functional derivative of the
electron-nuclear interaction energy (\ref{eq:Uen}), expressed as
$\int_\Omega d\R \, n_e(\R) V_n(\R)$.
The total energy within a field-theoretic framework corresponds to the
\textit{Galitskii-Migdal energy}\cite{galitskii_migdal_1958,fetter_walecka03},
$E^{\text{GM}}$, with the associated ground state density,
$n_e^{\text{GM}}$, defined as
\begin{align}\label{eq:n_gm}
n_e^{\text{GM}}(\R) &= \frac{\delta E^{\text{GM}}[n_e^{\text{GM}}]}{\delta V_n(\R)}
\nonumber \\
&= \int d(12) \frac{\delta E^{\text{GM}}[n_e^{\text{GM}}]}{\delta G_0(12)}
\frac{\delta G_0(12)}{\delta V_n(\R)} \nonumber \\
&= -\frac{i}{2}\int d(12) \, \Sigma_e(12)
\frac{\delta G_0(12)}{\delta V_n(\R)} \, .
\end{align}
However, the eigenenergies and eigenfunctions used to construct the optimal
one-body non-interacting Green's function $G_0$ within QS$GW$
are solutions of a one-particle non-interacting effective Hamiltonian,
$\hat{H}^{\text{eff}}$. The effective non-local and static potential
for this non-interacting reference system
can be determined self-consistently within Hedin's formalism using
the mapping procedure (\ref{eq:mapping_vxc})\cite{Kotani07}.
Under these conditions, the
ground state density $n_e = (1/N_\K)\sum_n\sum_{\K\in\text{BZ}} f_{n,\K}
\abs{\psi_{n,\K}}^2$ is defined as
\begin{align}\label{eq:n_eff}
n_e(\R) &= \frac{\delta E^{\text{eff}}[n_e]}{\delta V_n(\R)}
\nonumber \\
&= -\frac{i}{2}\int d(12) \, V_{xc}(12)
\frac{\delta G_0(12)}{\delta V_n(\R)} \, .
\end{align}
The discrepancy between the definitions (\ref{eq:n_eff}) and (\ref{eq:n_gm})
for the electron density
\be
n_e^{\text{GM}}(\R) - n_e(\R) = - \frac{i}{2} \int d(12) \biggl\{
\Sigma_e(12) - V_{xc}(12)  \biggr\}
\frac{\delta G_0(12)}{\delta V_n(\R)}
\ee
highlights the inconsistency in our treatment, justifying the decision not
to pursue this direction in the present work.

To assess the impact of the incompleteness of the LMTO-MPB formalism
in computing Eq. (\ref{eq:dynmat_el}), we compared
the phonon dispersions obtained using
our formalism for diamond with those derived from
DFPT within a plane wave framework (not reported
here\footnote{These results have been provided by Nicola Bonini through personal communication})
and then free from Pulay-like \ibcs.
Diamond serves as an ideal test case due to the absence of dipolar long-range
contributions to Eq. (\ref{eq:dynmat_el}),
as the Born effective charge tensor is zero.
DFPT-based phonon dispersion calculations were conducted using our
developmental version of \texttt{QUANTUM ESPRESSO}
\cite{giannozzi09,giannozzi17,giannozzi20}, which is the state of the art
for the characterization of the vibrational properties of materials.
In DFPT calculations, we enforce the acoustic sum rule as in
Eq. (\ref{eq:ifc_manybody}) and set the \xc kernel $f_{xc}$ to zero,
effectively evaluating phonon dispersions within the RPA.
Further computational details are provided in \sectionp S.6 C of the
\suppinfo.
It is important to note the distinction from the approach
in Ref. \onlinecite{ramberger17}, where forces and phonon dispersions were
accurately determined using the adiabatic connection dissipation-fluctuation
theorem, providing a description of the correlation energy within the RPA.
Phonon dispersions obtained via our modified DFPT formalism exhibited imaginary
acoustic phonon modes and overestimated optical phonon frequencies,
corroborating findings from Ref. \onlinecite{martin69}.
Phonon dispersions computed by using
Eqs. (\ref{eq:dynmat2})-(\ref{eq:dynmat_el}) within the MPB formalism
and the RPA demonstrate similar trends but larger deviations compared to
DFPT results.

Restoring the \xc kernel in the DFPT framework yields typical
phonon dispersions evaluated with local \xc functionals when using the
original algorithm. In contrast,
within a field-theoretic approach we incorporate ladder diagram corrections
to the inverse dielectric matrix by solving the BSE equation, as outlined
in \sectionp \ref{sec:bse} and Ref. \onlinecite{Cunningham2023}.
However, dispersions computed using
Eqs. (\ref{eq:dynmat2})-(\ref{eq:dynmat_el}) still exhibit similar
deficiencies to RPA dispersions, albeit to a lesser extent.
In this context, the limitations arising from the
incompleteness of the LMTO-MPB set
and its dependence on nuclear positions can introduce significant
inaccuracies, making it essential to incorporate
\ibcs\cite{betzinger12,betzinger13,friedrich15}. These corrections
aim to enhance the accuracy of the phonon properties by extending
the Hilbert space spanned by the LMTO-MPB functions, thereby accounting
for the perturbative effects associated with nuclear vibrations and
compensating for
the missing contributions arising from the basis set's incompleteness.
While these corrections provide a pathway to mitigate the impact
of the incomplete basis set, they have not been implemented in the present
work, as detailed in \sectionp \ref{sec:core}, leaving this as a potential area
for future research.
Consequently, to accurately model the \eph interaction in this study, we
compute the \eph matrix elements following field-theoretic approach guidelines,
while employing phonon frequencies and polarization vectors provided
by the DFPT implementation in \texttt{QUANTUM ESPRESSO}.

\section{Test calculations: \eph matrix elements from Kohn anomalies
in graphene phonon dispersions}
\label{sec:results}

In this section we examine the implementation of the
\eph matrix elements within a
field-theoretic framework, with details provided
in \sectionp \ref{sec:core} and \sectionp S.1 of the \suppinfo.
While \eph matrix elements represent the coupling strengths between
electronic states and lattice vibrations in materials, they do not directly
correspond to observable physical quantities.
However, Piscanec et al.~\cite{piscanec04} proposed a method linking
certain observable properties to the \eph matrix elements.
Specifically, in graphene, it has been shown, according to perturbation theory,
that the slope of the highest in-plane optical phonon branch in proximity
of Kohn anomalies can be directly linked to the Fermi surface-averaged
square modulus of the \eph matrix elements via the relation
\be\label{eq:slope_to_gmat}
\alpha^\nu_\Q =
\frac{\sqrt{3}\pi^2}{v_F} \braket{g^2_{\Q,\nu}}_F \quad \Q=\mathbf{\Gamma},
\mathbf{K}.
\ee
For momentum transfer $\Q=\bm{\Gamma}$ and LO $E_{2g}$ phonon mode,
\be
\braket{g^2_{\GG,\text{LO}}}_F =  \tfrac{1}{8}
\sum_{\nu}^{\text{LO,TO}}\sum_{i,j}^{\pi,\pi^*}
\abs{g^{\mathcal{S}}_{ij,\nu}(\KK,\GG)}^2 \, ,
\ee
and for $\Q=\KK$ and TO $A_1'$ phonon branch,
\be
\braket{g^2_{\KK,\text{TO}}}_F = \tfrac{1}{4}\sum_{i,j}^{\pi,\pi^*}
\abs{g^{\mathcal{S}}_{ij,\text{TO}}(\KK,\KK)}^2 \, ,
\ee
with the summations running over the two degenerate
$\pi$ bands at the Fermi energy.
This elegant connection provides a means to extract meaningful information
about the \eph interaction from experimental observations.
In Eq. (\ref{eq:slope_to_gmat}), $v_F$ is the Fermi velocity,
which corresponds to the slope of
the $\pi$-bands at the Dirac cone near the electron wave vector $\KK$.
In Ref. \onlinecite{piscanec04} a Fermi velocity $v_F = 14.1 \,
\text{eV}\,\text{Bohr}$ was computed at GGA level of theory and employed
to derive $\braket{g^2_{\Q,\nu}}_F$ from interpolated experimental
measurements.
However, it is well-known that local/semilocal \xc functionals typically
underestimate $v_F$ by approximately 30\%, while QS\gw overestimates
it by $\sim$20\% within the RPA\cite{mark11}.
Using a scaled-$\Sigma$ potential, which approximates the electron self-energy
from a QS$G\widehat{W}$ scheme, provides a more accurate estimation,
resulting in an overestimation of the Fermi velocity
by $\sim$10\%\cite{mark11}.
In this study, we adopt the value $v_F = 12.44 \,
\text{eV}\,\text{Bohr}$ (or $10^6m\cdot s^{-1})$, based on cyclotron
mass measurements of electron and hole estimated as function of their
concentrations~\cite{novoselov05}. This ensures that reference values for
$\braket{g^2_{\Q,\nu}}_F$ are experimentally consistent.

Using an interpolated experimental value of
$\alpha^{\text{LO}}_\GG = 340\,\text{cm}^{-1}$, we determine
$\braket{g^2_{\GG,\text{LO}}}_F = 0.031\,\text{eV}^2 $.
However, data near the symmetry point $\KK$ exhibit significant
scattering and were excluded from this analysis.
Additionally, the  Raman \textit{D}-peak dispersion reflects the slope
of the Kohn anomaly at the symmetry point $\KK$\cite{pocsik98},
albeit providing only a lower limit,
$\braket{g^2_{\KK,\text{TO}}}_F = 0.072 \,\text{eV}^2$.
Nevertheless, from a first-neighbors tight-binding approximation,
it is evident that the \eph matrix elements at the $\GG$ and $\KK$
symmetry points are not independent but related by the expression
\be\label{eq:tb_relation}
\frac{\braket{g^2_{\KK,\text{TO}}}_F \omega_{\KK,\text{TO}}}
     {\braket{g^2_{\GG,\text{LO}}}_F \omega_{\GG,\text{LO}}} = 2 \, ,
\ee
where $\braket{g^2_{\KK,\text{TO}}}_F = 0.076 \,\text{eV}^2$ is derived,
demonstrating reasonable agreement with the lower limit extracted from
Raman $D$-peak dispersions. Equation (\ref{eq:tb_relation}) was solved
using $\omega_{\GG,\text{LO}} \approx 1540\, \text{cm}^{-1}$ and
$\omega_{\KK,\text{TO}} \approx 1250\, \text{cm}^{-1}$, values from
Ref. \onlinecite{piscanec04}, where a 64×64×1 BZ mesh was employed along with
a Hermite-Gauss smearing
of order 1 equivalent to $\sigma = 0.01\,\text{Ry}$.
Alternatively, when $\omega_{\KK,\text{TO}} \approx 1192\, \text{cm}^{-1}$,
as extrapolated by Lazzeri et al. in Ref. \onlinecite{lazzeri08} using a
frozen-phonon approach within the $G_0W_0$ formalism, a value of
$\braket{g^2_{\KK,\text{TO}}}_F = 0.080 \,\text{eV}^2$ is obtained. This value
is adopted in the present work
as our reference for the symmerty point $\KK$. Indeed,
Ref. \onlinecite{lazzeri08} demonstrates that Raman $D$-line frequencies,
calculated using a dynamical matrix model based on a frozen-phonon approach
at the $G_0W_0$ level of theory, align well with experimental results.

It is important to acknowledge that Eq. (\ref{eq:tb_relation}) does not
incorporate \eph vertex diagrams, which describe electron coupling
to multiphonon excitations. The significance of these diagrams has been
explicitly demonstrated in the Renormalization Group (RG) analysis presented
in Ref. \onlinecite{basko08}, where they were shown to play a crucial role
in reproducing the ratio of the integrated intensities $I_{D}/I_{G}$ of the
$D$- and $G$-peaks in two-phonon Raman spectra.
By defining
\be\label{eq:tb_relation_deviation}
\lambda = \frac{\braket{g^2_{\KK,\text{TO}}}_F \omega_{\KK,\text{TO}}}
     {\braket{g^2_{\GG,\text{LO}}}_F \omega_{\GG,\text{LO}}}  \, ,
\ee
the inclusion of \eph vertex diagrams within a RG framework yields
$\lambda=5.19$.
This result highlights the deviation of methodologies
incorporating these diagrams from the condition $\lambda=2$, which is strictly
satisfied in theoretical frameworks that neglect \eph vertex diagrams.

In this study, phonon frequencies and polarization vectors were computed
using \texttt{QUANTUM ESPRESSO}~\cite{QE1,QE2,QE3} with the same computational
parameters outlined in Ref. \onlinecite{piscanec04}. The decision to use DFPT
phonon dispersions instead of a field-theoretic approach is explained in
\sectionp \ref{sec:dynmat}. Notably, in the analysis of
Eq. (\ref{eq:tb_relation_deviation}), the specific phonon frequencies
at $\GG$ and $\KK$ are not of primary concern, as the quantities
$\braket{g^2_{\Q,\nu}}_F \omega_{\Q,\nu}$ do not depend on them.
Consequently, the ratio in Eq. (\ref{eq:tb_relation_deviation}) remains
unaffected by the accuracy of the phonon dispersions obtained through DFPT.
Conversely, polarization vectors---predominantly determined by the material's
symmetry---play a critical role. These vectors are employed to evaluate
the Fermi surface-averaged square modulus of the \eph matrix elements
within a field-theoretic framework.

It is important to highlight that in our investigation,
the \eph contribution in Eq. (\ref{eq:Wph_A}) to the screened Coulomb
interaction was not included in the self-consistency loop
of the QS$\gw$ scheme. Instead, our focus lies in assessing the
accuracy and reliability of the implementation detailed
in \sectionp \ref{sec:core} and \sectionp S.1 of the \suppinfo.
Specifically, we present and analyze \eph matrix elements computed
based on pre-existing QS\gw calculations within the RPA, as well as
those obtained from QS$G\widehat{W}$ without self-consistency.
This choice is motivated by observations of minimal numerical
fluctuations during the computation of the electron self-energy
within the QS$G\widehat{W}$ scheme. Such fluctuations break the degeneracy
of the $\pi$-bands at $\KK$ during the self-consistency process.
Prior analyses have shown that the initial iterative step within
the QS$G\widehat{W}$ scheme captures the majority of the band structure
renormalization~\cite{Cunningham2023}, with subsequent iterations
negligibly affecting the electronic structure and, consequently,
the \eph matrix elements.

\subsection{Convergence analysis of \eph matrix elements}
\label{sec:results_convergence}

Table \ref{tab:k_mesh_convergence} illustrates the convergence of
$\braket{g^2_{\Q,\nu}}_F$ with respect to BZ sampling.
These values were calculated using an inverse dielectric matrix obtained
from pre-existing QS\gw calculations within the RPA and incorporating ladder
diagrams through a QS$G\widehat{W}$ scheme \footnote{Note that readers
interested to the computational details of these
calculations are referred to Ref. \protect\onlinecite{mark11}.}.
In the QS$G\widehat{W}$ calculations, the two-particle Hamiltonian was
constructed within the BSE framework using only four unoccupied states.
This choice was made to enable computationally feasible calculations
on dense wave vector meshes.
\begin{table}[htbp]
    \centering
    \caption{Convergence of $\braket{g^2_{\Q,\nu}}_F$
      (expressed in eV$^2$) with the sampling of the BZ
      for the symmetry points
      $\Q =\GG$,$\KK$ and for the highest optical phonon branch.
      $\braket{g^2_{\Q,\nu}}_F$ values
      are computed using the inverse dielectrix matrix evaluated within the
      RPA or adding ladder diagrams by solving the BSE. In the latter case,
      only 4 unoccupied states were utilized when building the two-particle
      Hamiltonian.}
    \label{tab:k_mesh_convergence}
    \begin{tabular}{ccccc}
        \hline\hline
        $\K$ mesh & $\braket{g^2_{\GG,\text{LO}}}_F^{\text{RPA}} $ & $\braket{g^2_{\KK,\text{TO}}}_F^{\text{RPA}} $ & $\braket{g^2_{\GG,\text{LO}}}_F^{\text{BSE}} $ & $\braket{g^2_{\KK,\text{TO}}}_F^{\text{BSE}} $ \\
        \hline
          6x6x1 & 0.03360 & 0.09165 & 0.03337 & 0.08743 \\
        12x12x1 & 0.03384 & 0.09040 & 0.03321 & 0.08463 \\
        18x18x1 & 0.03388 & 0.08996 &         &         \\
        \hline\hline
    \end{tabular}
\end{table}

A couple of observations emerge from Table I.
The first is that denser wave vectors BZ grids do not significantly alter
the \eph matrix elements for both the QS$GW$ and
QS$G\widehat{W}$ level of theory.
For instance, $\braket{g^2_{\KK,\text{TO}}}_F$ changes by approximately 2\%
at the RPA level when increasing the grid from 6x6x1 to 18x18x1,
and by $\sim$3\% at the BSE level when increasing the grid from
6x6x1 to 12x12x1. In contrast, $\braket{g^2_{\GG,\text{LO}}}_F$ exhibits
negligible variations across these grid changes.

The second is the comparison of DFPT and field-theoretic approaches.
Figure \ref{fig:kmesh_convergence} contrasts the results from DFPT
with those obtained from the QS\gww formalism within the RPA.
At the high-symmetry points $\Q =\GG,\KK$, and for the highest optical phonon
branches, QS\gww results show minimal deviations as the BZ grid increases
from 6x6x1 to 12x12x1. However, DFPT calculations require a denser
grid (at least 18x18x1) to yield results within $\sim$1.4\% and $\sim$2.1\%
of the extrapolated values for $\braket{g^2_{\GG,\text{LO}}}_F$
and $\braket{g^2_{\KK,\text{TO}}}_F$, respectively.

Guandalini et al.~\cite{guandalini24} recently introduced a theoretical
framework combining the multipole approximation (MPA)~\cite{dario_leon_mpa21}
and the W-av method~\cite{guandalini_rimw} to achieve accurate convergence
of quasi-particle (QP) band structures in graphene. The MPA efficiently and
accurately approximates full-frequency response functions using a
limited number of poles, while the W-av method
enhances convergence with respect to BZ sampling in
two-dimensional (2D) materials. This synergistic approach, along with
the incorporation of vanishing intra-band transitions near the Dirac point,
significantly accelerates the convergence of both the QP gap at
$\K = \mathbf{M}$ and the Fermi velocity with respect to the number of
$\K$ points. Nevertheless, a 60x60x1 mesh
is required to reduce the deviation from the converged QP gap to $\sim$30 meV.

Incorporating vanishing intra-band transitions at the Dirac point as a
long-wavelength contribution to the irreducible polarizability is essential for
accurately describing the static dielectric function in this regime,
which is relevant to the present study.
Ref. \onlinecite{guandalini24} demonstrates that the inverse dielectric
function, $\varepsilon_{e,11}^{-1}$, asymptotically approaches a positive
constant in the $\Q\to 0$ limit when the intra-band correction,
$\chi^0_{e,11,\mathcal{D}}(\omega = 0) = -q/4\gamma$, is included in
the head of the irreducible polarizability, $\chi_e^0$.
Here, to zero-th order, $\gamma$ represents the Fermi velocity $v_F$
within a small circular region $\mathcal{D}_\KK$ around the Dirac cone at $\KK$.
The intra-band correction, $\chi^0_{e,11,\mathcal{D}}$,
derived from a Dirac Hamiltonian model, is often omitted in
conventional \gw-based implementations. When this contribution is excluded,
the screening function in graphene erroneously behaves similarly to that
of a 2D semiconductor, necessitating extremely dense $\K$-point meshes
to recover the correct semi-metallic screening behavior.

The \textit{tetrahedron integration method}, as implemented in \texttt{Questaal},
effectively reproduces such dense BZ samplings by dividing the BZ
into tetrahedra and linearly interpolating
eigenenergies at their vertices\cite{Kotani07}.
This approach, akin to the W-av method used in Ref. \onlinecite{guandalini24},
ensures the rapid convergence
of $\omega_{\Q,\nu}\braket{g^2_{\Q,\nu}}_F$ by accelerating the convergence
of the static inverse dielectric matrix, which enters the
definition of the \eph matrix elements.
Rapid convergence of QP band structures with respect to BZ sampling
has similarly been reported by van Schilfgaarde and Katsnelson~\cite{mark11}.

\begin{figure}[htbp]
    \centering
    \includegraphics[width=0.8\columnwidth]{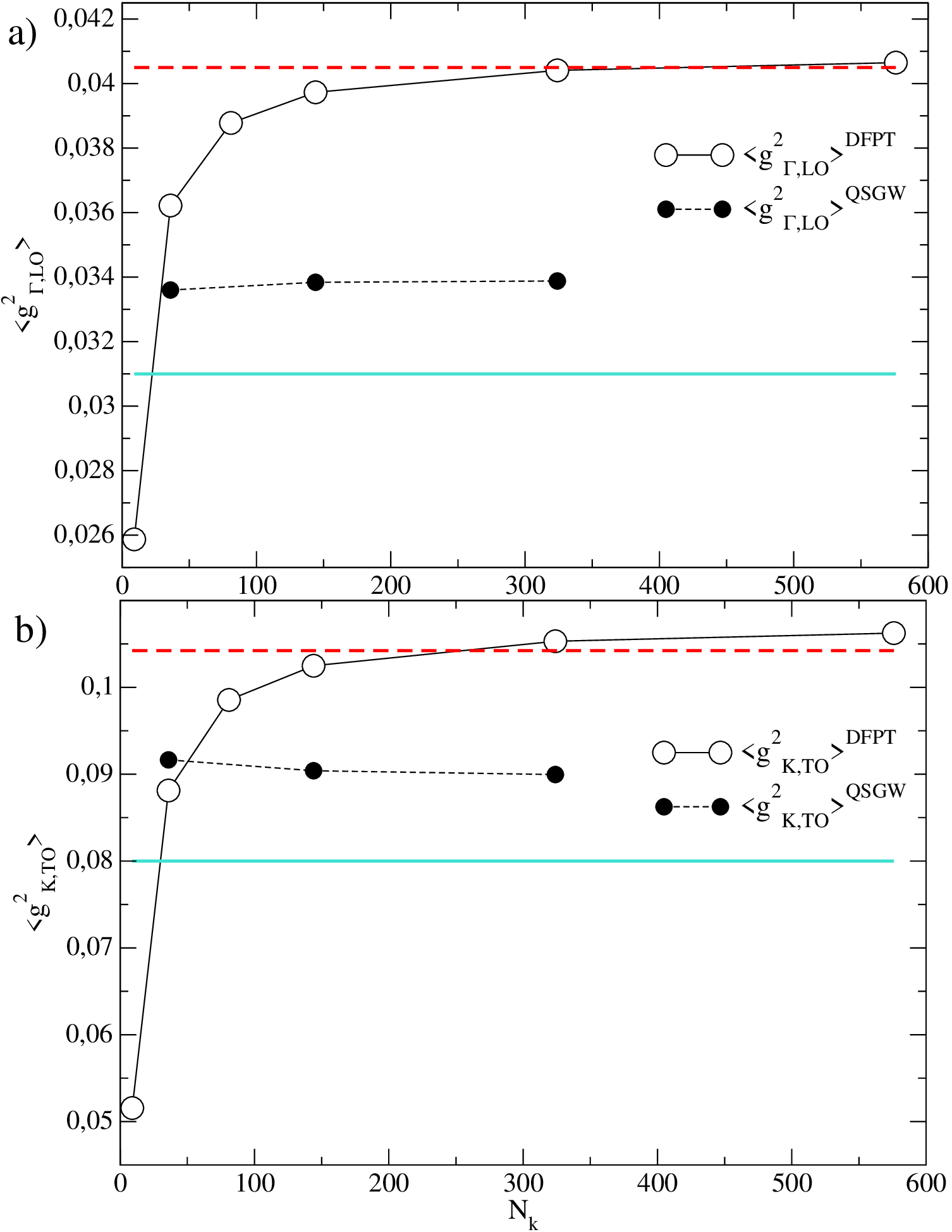}
    \caption{Convergence of $\braket{g^2_{\Q,\nu}}_F$ (expressed in
    eV$^2$) with the sample of the BZ. $\omega_{\Q,\nu}\braket{g^2_{\Q,\nu}}_F$
    values are calculated within DFPT or using the inverse dielectric matrix
    evaluated within the RPA. Results are presented for the high-symmetry points
    $\Q =\GG$,$\KK$ and for the highest optical phonon branch.
    Phonon frequencies of $\omega_{\GG,\text{LO}} =
    1540\,\text{cm}^{-1}$ and $\omega_{\KK,\text{TO}} = 1192\,\text{cm}^{-1}$ are
    employed in the evaluation of both DFPT and QS\textit{GW} Fermi surface-averaged
    square modulus of the \eph matrix elements. The dashed red lines indicate
    the DFPT results from Ref. \onlinecite{piscanec04}, rescaled to reflect
    the phonon frequencies employed in this work. The solid
    turquoise lines represent the reference values used in this study:
    $\braket{g^2_{\GG,\text{LO}}}_F = 0.031\,\text{eV}^2$ and
    $\braket{g^2_{\KK,\text{TO}}}_F = 0.080\,\text{eV}^2$.}
    \label{fig:kmesh_convergence}
\end{figure}

\subsection{Logarithmic divergence of the Fermi velocity near the Dirac point}
\label{sec:results_log}

Close to the Dirac point, strong electronic correlations in intrinsic graphene
induce an ultraviolet logarithmic divergence in the Fermi velocity as
$\K\to\KK$. This divergence, present in both Hartree-Fock and
higher-order theories, is independent of whether the Coulomb interaction
is bare or screened. The QP energy renormalization can be expressed as
$\varepsilon_{\mathcal{D}}(\bar{k}) = \varepsilon(\bar{k}) +
\Sigma_{\mathcal{D}}(\bar{k})$, where $\varepsilon(\bar{k})=\gamma \bar{k}$
and $\bar{k}=\abs{\K - \KK}$. A hyperbolic model derived from a
Dirac Hamiltonian accurately reproduces this renormalization
as~\cite{guandalini24}
\be\label{eq:ren_en}
\varepsilon_{\mathcal{D}}(\bar{k}) = \gamma \bar{k} \biggl[
1 + \frac{f}{2}\biggl(\text{cosh}^{-1}(2/\bar{k}) + \frac{1}{2}\biggr) \biggr]
\ee
where $f$ scales the Hartree-Fock self-energy to account for screening effects.
For small $\bar{\K}$, a Taylor expansion yields
\be\label{eq:ren_en_2}
\varepsilon_{\mathcal{D}}(\bar{k})
\approx \gamma \biggl( \bar{k} + f c_1 \bar{k} -
\frac{f \bar{k}\ln \bar{k}}{2 \ln 10} -
\frac{\bar{k}^3}{32} - \frac{3 \bar{k}^5}{512} -
\mathcal{O}(\bar{k}^7) \biggr) \, ,
\ee
with $c_1 = 1/4 + \log 2$. In the limit $\bar{\K}\to 0$,
it follows that $\bar{k}\ln \bar{k} = - \bar{k}$, leading to the
simplified expression $\varepsilon_{\mathcal{D}}(\bar{k}) =
\gamma_{\mathcal{D}} \bar{k}$ within a small circular region $\mathcal{D}_\KK$
around the Dirac cone, where the renormalized Fermi velocity is given by
$\gamma_{\mathcal{D}} = \gamma\bigl[1+ f\bigl( c_1 +
\frac{1}{2 \ln 10} \bigr)\bigr]$. Consequently, the intra-band correction
to the irreducible polarizability becomes
$\widetilde{\chi}^0_{e,11,\mathcal{D}}(\omega = 0) = -q/4\gamma_{\mathcal{D}}$.
Neglecting the logarithmic divergence, as achieved through the
tetrahedra integration method, introduces a deviation
$(\widetilde{\chi}^0_{e,11,\mathcal{D}} -
\chi^0_{e,11,\mathcal{D}})$, which can be expressed as
$\Delta\chi^0_{e,11,\mathcal{D}}=- \chi^0_{e,11,\mathcal{D}}
(\alpha-1)/\alpha$, where $\alpha = 1+ f\bigl( c_1 + \frac{1}{2 \ln 10} \bigr)=
1.07682$ for $f=0.1$, based on experimental data~\cite{siegel11}.
The deviation $\Delta\chi^0_{e,11,\mathcal{D}}= - 0.07134 \,
\chi^0_{e,11,\mathcal{D}}$ is two orders of magnitude smaller than
$\chi^0_{e,11,\mathcal{D}}$, demonstrating that the ultraviolet logarithmic
divergence minimally affects the accuracy and convergence of the results
presented in this study.

\subsection{Role of ladder diagrams}
\label{sec:results-ladders}

The discrepancies observed between the QS\gw results and reference values in
Fig. \ref{fig:kmesh_convergence} can primarily be attributed to the omission of
ladder diagrams in the calculation of the \eph matrix elements.
Incorporating ladder diagrams in the inverse dielectric matrix introduces
effects in the \eph matrix elements that resemble the band-gap
renormalizations discussed in previous studies~\cite{Cunningham2023}.
As summarized in Table \ref{tab:k_mesh_convergence}, the RPA tends to
overestimate $\braket{g^2_{\KK,\text{TO}}}_F$ while producing only minor
deviations for $\braket{g^2_{\GG,\text{LO}}}_F$ compared to
QS$G\widehat{W}$ results. This RPA overestimation may also arise from
differences in the treatment of \xc effects during the computation of
the inverse dielectric matrix and phonon frequencies.
The Fermi surface-averaged \eph matrix elements can be
expressed as $\braket{g^2_{\Q,\nu}}_F =
\braket{\eta^2_{\Q,\nu}}_F / \omega_{\Q,\nu}$ where
$\braket{\eta^2_{\Q,\nu}}_F$ is computed either at the RPA
or BSE level, while $\omega_{\Q,\nu}$ is consistently evaluated within
DFPT. In the latter case, the phonon frequencies are implicitly computed
using a static inverse dielectric matrix that incorporates the \xc kernel,
$f_{xc}$. Conversely, $\braket{\eta^2_{\Q,\nu}}^{\text{RPA}}_F$ neglects \xc
effects entirely.

Figure \ref{fig:bse_convergence} highlights the convergence of
$\braket{g^2_{\Q,\nu}}_F$ with respect to the number of conduction states
included in the two-particle Hamiltonian within the BSE framework.
The rapid convergence in terms of the BZ sampling enables
us to utilize a coarse mesh of 6x6x1 wave vectors.
We observe that full convergence requires up to 40 states.
\begin{figure}[htbp]
    \centering
    \includegraphics[width=0.8\columnwidth]{plot_bse_convergence.eps}
    \caption{Convergence of $\braket{g^2_{\Q,\nu}}_F$ (expressed in
    eV$^2$) with the number of unoccupied bands used to build the
    two-particle Hamiltonian within the BSE scheme.
    Values are reported for the symmetry points
    $\Q =\GG$,$\KK$ and for the highest optical phonon branch.
    All calculations have been performed using a 6x6x1 sampling of the BZ}
    \label{fig:bse_convergence}
\end{figure}
At convergence, we find $\braket{g^2_{\GG,\text{LO}}}_F = 0.031 \,\text{eV}^2$
and $\braket{g^2_{\KK,\text{TO}}}_F = 0.078 \,\text{eV}^2$, as computed within
the QS$G\widehat{W}$ framework. These values are in excellent agreement with those obtained
at $\GG$ by interpolating the phonon dispersion
near the Kohn anomaly (0.031 eV$^2$) and at $\KK$ (0.080 eV$^2$)
using Eq. (\ref{eq:tb_relation}) with  $\omega_{\KK,\text{TO}}$
from Ref. \onlinecite{lazzeri08}, respectively.

The condition $\lambda = 2.05$ is satisfied in these calculations,
indicating that \eph vertex corrections were not included in the evaluation
of the inverse dielectric matrix. A field-theoretic framework incorporating
the static \eph contribution ($W_{ph}^\A$; Eq. \ref{eq:Wph_A})
to the screened Coulomb interaction into the kernel of the
two-particle Hamiltonian would enable the introduction of an infinite
sum of \eph vertex diagrams within the BSE-corrected inverse dielectric matrix
and account for the static exciton-phonon coupling.
Such a framework could theoretically
recover the result $\lambda = 5.19$ reported in Ref. \onlinecite{basko08}, but
this approach is beyond the scope of the present study and is proposed
as a direction for future work.

Using a frozen-phonon approach within the $G_0W_0$ formalism,
Lazzeri et al. reported $\lambda = 3.07$~\cite{lazzeri08}, while
more recent refinements by Faber et al. yielded a slightly lower value of
$\lambda = 2.95$~\cite{faber15}.
These values, however, fall significantly below the reference theoretical
value of $\lambda=5.19$. Potential explanations for this discrepancy
include: (i) incomplete convergence with respect to BZ sampling,
a well-known challenge as discussed
in \sectionp \ref{sec:results_convergence}, and (ii) the adoption of
the plasmon pole model (PPM) approximation, which relies on a single
plasmon-pole frequency, at $\Q =\GG$ ranging between 27 eV and 7 eV
in Ref \onlinecite{faber15}. While the PPM approximation
is effective for systems where the energy-loss function is dominated by a
single plasmon feature, it fails to capture the more complex structure
observed in graphene. Specifically, first-principles calculations within
the RPA reveal two major features in the energy-loss function,
$-\text{Im}\,\varepsilon_{e,11}^{-1}(\GG,\omega)$: the combined
$\pi+\sigma$ plasmon mode at approximately 15 eV and the $\pi$ plasmon
mode at around 5 eV~\cite{eberlein08}. Additionally, low-energy
$\pi\to\pi^*$ single-particle excitations contribute a shoulder near
the lowest energy range. Trevisanutto et al. demonstrated that both
the $\pi$ plasmon and these low-energy $\pi\to\pi^*$ excitations play a
critical role in accurately describing the correlation
energy\cite{trevisanutto08}.
A PPM with a plasmon-pole frequency of 5 eV yields results comparable to those
obtained via contour deformation (CD) frequency integration\cite{trevisanutto08}, suggesting
that the plasmon-pole frequency employed in Refs. \onlinecite{lazzeri08} and
\onlinecite{faber15} may inadequately represent the screening $W_e$.
This limitation directly affects the accuracy of the calculated
$\braket{g^2_{\GG,\text{LO}}}_F$ and $\braket{g^2_{\KK,\text{TO}}}_F$ values,
as these are derived from the gap opening
at the Dirac point induced by atomic displacements along the $E_{2g}$ and $A_1'$
phonon modes in the distorted structure, respectively.
An approximate treatment of the screening within the frozen-phonon
approach can therefore compromise the accurate characterization of the \eph
interaction in graphene.

Nevertheless, the incorporation of long-range non-local exchange interactions
within the framework of MBPT leads to a marked enhancement of the
\eph coupling at the $\KK$ point compared to results obtained using
LDA and GGA functionals\cite{lazzeri08,faber15,venanzi23,graziotto24}.
This enhancement is attributed to the implicit
inclusion of \eph vertex corrections through
Coulomb vertex diagrams in the frozen-phonon method, which gives
rise to electron coupling to multiphonon excitations.
Conversely, our results demonstrate that the
inclusion of non-local and long-range
exchange interactions in \abinitio MBPT does not inherently enhance
\eph coupling at $\KK$ unless \eph vertex diagrams are explicitly included
in the formulation of the inverse dielectric function.

Finally, it is noteworthy that for phonon modes other than the
highest optical modes at $\GG$ and $\KK$, the Fermi surface-averaged
matrix elements $\braket{g^2_{\Q,\nu}}_F$ are zero, except for the doubly
degenerate mode at $\KK$. For this mode, QS\gw yields
$\braket{g^2_{\KK,\nu}}_F = 0.00723\,\text{eV}^2$ using a 6x6x1 BZ sampling,
while QS$G\widehat{W}$ results in $\braket{g^2_{\KK,\nu}}_F=0.00336 \text{eV}^2$.
These results align with Ref. \onlinecite{piscanec04} and confirm the
absence of Kohn anomalies for these branches.

\section*{Conclusions}

In this work, we present a derivation of the \eph coupling matrix element
$g_{in,\nu}(\K,\Q)$ and the dynamical matrix within a field-theoretic framework,
drawing inspiration from the foundational works of Baym~\cite{baym61} and
Hedin and Lundqvist~\cite{hedin_lundqvist69}. The central quantity in our
approach is the static electronic dielectric
function, $\varepsilon_e(\R,\R';0)$, which eliminates the need to compute
the induced density via a Sternheimer equation.
Importantly, we demonstrate that when the formalism is appropriately
structured, Pulay-like \ibccs terms are unnecessary for evaluating the
\eph matrix elements.

We also provide detailed insights into the implementation of this formalism
within the QS\gw method using a MPB set in the \texttt{Questaal} package.
Within the MBPT framework, we decompose the \eph matrix elements into a
long-range, nonanalytic contribution and a short-range analytic remainder.
This decomposition provides also a first-principles expression for
the Born effective charge tensor. The long-range term extends
Fr\"ohlich’s model to include anisotropic lattices and multiple phonon modes.

Additionally, we show that within the LMTO-MPB formalism, the \eph
matrix elements can be expressed as a linear combination of projection
coefficients for the product of two wave functions. This formulation simplifies
the algorithmic treatment of the transformation rule $g_{in,\nu}(\K,\rot\Q)$
under phonon wave vector rotations.

The \texttt{Questaal} code leverages its capability to compute the
polarizability with ladder diagrams to go beyond the RPA in evaluating the
inverse dielectric matrix, which significantly influences $g_{in,\nu}(\K,\Q)$.
Extensive validation of this field-theoretic framework demonstrates its ability
to achieve excellent agreement with experimental data. Specifically,
the calculated Fermi surface-averaged \eph matrix elements, derived
from the slope of the Kohn anomalies for the highest optical phonon mode
in graphene, closely match experimental observations when
electron-multiphonon coupling is neglected.

\section*{Acknowledgements}

The NREL staff was supported by the National Renewable Energy Laboratory,
operated by Alliance for Sustainable Energy, LLC,
for the U.S. Department of Energy (DOE) under Contract No. DE-AC36-08GO28308,
funding from Office of Science, Basic
Energy Sciences, Division of Materials. We acknowledge the use of
the National Energy Research Scientific Computing
Center, under Contract No. DE-AC02-05CH11231 using NERSC award BES-ERCAP0021783
and we also acknowledge that a portion
of the research was performed using computational resources sponsored by the
Department of Energy's Office of Energy
Efficiency and Renewable Energy and located at the National Renewable Energy
Laboratory. S.L. thanks Francesco Macheda for his
assistance in providing modifications to the \texttt{QUANTUM ESPRESSO} routines
for calculating \eph matrix elements,
along with Herve Ness, Fulvio Paleari and Claudio Attaccalite for the insightful
discussions that contributed to this work.
S.L. also profoundly thanks the late Nicola Bonini for the time he
dedicated to discussions and analysis of this body of work, for his
encouragement and for his mentorship during our time at King's College London.
He sadly passed away prior to the writing of this paper.

\bibliographystyle{apsrev4-1}



\end{document}